%% file: sn-article.tex
\documentclass[pdflatex,sn-mathphys-num]{sn-jnl}

\usepackage{graphicx}%
\usepackage{multirow}%
\usepackage{amsmath,amssymb,amsfonts}%
\usepackage{amsthm}%
\usepackage{mathrsfs}%
\usepackage[title]{appendix}%
\usepackage[svgnames]{xcolor}%
\usepackage{textcomp}%
\usepackage{manyfoot}%
\usepackage{booktabs}%
\usepackage{algorithm}%
\usepackage{algorithmicx}%
\usepackage{algpseudocode}%
\usepackage{listings}%

\theoremstyle{thmstyleone}%
\theoremstyle{thmstyletwo}%

\theoremstyle{thmstylethree}%

\begin{document}

\title{Agentic Swarms Discover Enhanchment of Fractional Chern Insulator Gaps Beyond Ideal Quantum Geometry}

\author[1]{\fnm{Yitong} \sur{Zhang}}
\equalcont{These authors contributed equally to this work.}

\author[2]{\fnm{Changwen} \sur{Xu}}
\equalcont{These authors contributed equally to this work.}

\author[3]{\fnm{Andrew Boshi} \sur{Li}}

\author[4]{\fnm{Siddhartha} \sur{Sarkar}}

\author*[1]{\fnm{Kai} \sur{Sun}}\email{sunkai@umich.edu}

\author*[2,3]{\fnm{Venkatasubramanian} \sur{Viswanathan}}\email{venkvis@umich.edu}

\affil[1]{\orgdiv{Department of Physics}, \orgname{University of Michigan at Ann Arbor}, \orgaddress{\city{Ann Arbor}, \postcode{48109}, \state{MI}, \country{USA}}}

\affil[2]{\orgdiv{Department of Mechanical Engineering}, \orgname{University of Michigan at Ann Arbor}, \orgaddress{\city{Ann Arbor}, \postcode{48109}, \state{MI}, \country{USA}}}

\affil[3]{\orgdiv{Department of Aerospace Engineering}, \orgname{University of Michigan at Ann Arbor}, \orgaddress{\city{Ann Arbor}, \postcode{48109}, \state{MI}, \country{USA}}}

\affil[4]{\orgdiv{Condensed Matter Theory Center and Joint Quantum Institute}, \orgname{University of Maryland}, \orgaddress{\city{College Park}, \postcode{20742}, \state{MD}, \country{USA}}}

\abstract{Can artificial intelligence (AI) move beyond solving well-defined problems to contribute to open-ended discovery in fundamental physics? In particular, could AI systems propose what to explore, identify unexpected phenomena, and uncover their mechanisms? Here we investigate this question using MOSAIC (Multi-agent Open-ended Scientific Analysis and Investigation through Collaboration), a collaborative swarm of AI agents (Claude Opus~5.5), to tackle a longstanding bottleneck: the small energy gaps limiting the robustness of fractional topological order and the study of anyons at higher temperatures. Rather than prescribing a model, we ask frontier models (GPT-6 Astra) to propose a theoretical framework, which MOSAIC explores through diverse strategies and rapid exchange of ideas. This exploration reveals a striking finding: although ideal quantum geometry is conventionally considered favorable for fractional states, departing from this limit can dramatically enhance their robustness, with many-body gaps more than an order of magnitude larger than those of their ideal-geometry counterparts. This finding emerged from the collective contribution of agent swarm.  Without any centralized coordination, the agents explored autonomously, combined findings from different agents, and formed insights from individual runs, reaching gap enhancements that isolated agents with the same tools did not reach. Going beyond numerical optimization, AI-assisted theoretical analysis further develop simplified models with an analytical gap enhancement factor that leads to unbounded gap enhancement within single-band approximation. Remarkably, AI discovers a common physical mechanism underlying the gap enhancement. Unlike conventional approaches centered on the momentum-space quantum metric, AI identifies the real-space quantum metric as the key quantity governing the many-body gap here, revealing how flat-band wavefunctions shape two-electron interactions, yielding an index directly computable from single-particle wavefunctions without many-body ED to guide gap optimization. Our study demonstrates how collaborative AI agents can integrate exploration of uncharted territory, discovery of unexpected physics, and development of theoretical understanding, opening new possibilities for AI-driven research in fundamental physics.}

\maketitle

\section*{Introduction}

Recent advances in artificial intelligence (AI) have demonstrated remarkable capabilities in mathematical reasoning, from tackling challenging problems~\cite{romeraparedes2024funsearch,novikov2025alphaevolve,anthropic2026riemann,openai2026mathblog} to generating formally verifiable proofs~\cite{alpoge2026boussinesq,openai2026navierstokesblog,openai2026navierstokes,openai2026mathrepo}. In the natural sciences, large language model (LLM) agents now plan and execute chemical syntheses~\cite{bran2024chemcrow,boiko2023coscientist}, run quantum-chemistry calculations~\cite{wang2025dreams,zou2025elagente}, drive autonomous laboratories~\cite{szymanski2023alab} and extract insights from data~\cite{mitchener2025kosmos}. AI scientists automate the full research cycle~\cite{lu2024aiscientist,yamada2025ai,lu2026aiscientist}, teams of agents propose hypotheses and designs that are later confirmed in the laboratory~\cite{swanson2025virtuallab,gottweis2026coscientist,ghareeb2026multi}, and decentralized collectives of agents coordinate through a shared state rather than a central planner~\cite{wang2026scienceclaw,pal2026swarmworld,gao2026autoscientists,el2026physicsofagents}. These achievements invite an even broader question: what might AI discover when the questions themselves are still taking shape? Frontier agents now solve unpublished research-level problems in theoretical physics with near-perfect accuracy~\cite{chandra2026theophyseval}, whereas such evaluations grade an agent against a known answer and therefore cannot measure discovery where no answer exists in advance. At the frontiers of physics, research often begins not with a precisely formulated problem, but with a puzzle that we do not yet know how to approach or which tools might be appropriate to solve the puzzle. Finding the right questions can be as important as, if not more than, finding their answers. The path forward is equally uncertain: progress may require analytical theory, numerical simulations, experiments, or an entirely new conceptual framework. Further, progress does not necessarily mean reaching a definitive solution. Uncovering a new physical mechanism, understanding an important limiting case, or revealing an unexpected connection can transform our understanding and open entirely new directions of inquiry. Here, we explore whether AI can extend its remarkable problem-solving capabilities into this open-ended process of scientific discovery, helping to push the frontiers of fundamental physics.

How can we put this possibility to a meaningful test? The choice of scientific problem is crucial. We propose three criteria for such a test. (1) The problem should address a longstanding bottleneck in an active area of physics, where substantial efforts have yet to overcome important obstacles and even partial progress would matter. (2) The frontier of existing knowledge should be sufficiently clear that AI can build upon the best available understanding while allowing us to recognize when it ventures into genuinely new territory. (3) New discoveries should be independently verifiable within a practical timescale, without requiring months or years of experimental work to establish their validity. Problems with well-established numerical solutions are particularly attractive: they allow AI to test and refine its own hypotheses while giving researchers a reliable way to independently verify its findings. In this way, the direction of discovery can remain open-ended, while the evidence for progress is concrete and verifiable.

Following the philosophy discussed above, we identify a central scientific problem for this test: finding new theoretical principles to enlarge the many-body energy gaps of fractional quantum Hall (FQH) states. The fractional quantum Hall effect~\cite{stormer1999fractional,laughlin1999nobel} is one of the cornerstones of modern condensed matter physics, revealing revolutionary concepts such as anyons~\cite{halperin1984statistics,arovas1984fractional} and topological order~\cite{wen1990ground}, and topological quantum computation~\cite{nayak2008nonabelian}. The theoretical prediction~\cite{tang2011hightemp,sun2011nearlyflat,neupert2011fqh,sheng2011fqhe,regnault2011fci,parameswaran2013fractional,bergholtz2013review,li2021spontaneous} and recent experimental discovery~\cite{cai2023signatures,zeng2023thermodynamic,park2023observation,xu2023observation,lu2024fqah} of the fractional quantum anomalous Hall effect (FQAHE)/fractional Chern insulators (FCIs) have further expanded this field to topological flat bands in crystalline and moir\'e materials. Yet despite four decades of remarkable progress, a major bottleneck remains: the small many-body energy gap, often on the kelvin scale or below, which confines these exotic states to extremely low temperatures and exceptionally clean samples, and constrains the speed of adiabatic braiding and topological quantum computing~\cite{nayak2008nonabelian}. Developing new strategies toward overcoming this long-standing bottleneck is a challenge of profound fundamental and technological significance, satisfying our first criterion. Despite extensive research, the boundary of current knowledge is remarkably clear and quantifiable. Landau levels have long served as the benchmark for robust fractional quantum Hall states. For over a decade, engineered Chern bands failed to outperform this benchmark under otherwise identical conditions, until very recent theoretical breakthroughs revealed new possibilities~\cite{zhang2025beyond,sarkar2026moireenhanced}. Here, we focus on flat bands with single-component Bloch wavefunctions. Within this class, the strongest established gap enhancements arise from ideal quantum geometry~\cite{parameswaran2013fractional,roy2014band,ledwith2020fci,wang2021exact} combined with spatial inhomogeneity in electron density~\cite{sarkar2026moireenhanced}. However, these constructions rely on ideal quantum geometry, and their gap enhancement is limited to $\Delta\leq M_2^{(0)}\Delta_{\mathrm{LLL}}$, where $M_2^{(0)} \ge 1$ is the second moment of the electron-density distribution in a fully filled band, and $\Delta_{\mathrm{LLL}}$ is the corresponding many-body gap in the lowest Landau level~\cite{sarkar2026moireenhanced}. To our knowledge, no single-component band departing from ideal quantum geometry has demonstrated a larger enhancement. Can we break this benchmark and achieve $\Delta>M_2^{(0)}\Delta_{\mathrm{LLL}}$ by departing from ideal quantum geometry? This defines a concrete, quantitative frontier to challenge, satisfying the second criterion. Finally, FQH and FCI physics benefit from powerful, well-established numerical techniques, particularly exact diagonalization. These tools provide a practical means to test new hypotheses and quantitatively verify theory predictions, satisfying the third criterion.

\begin{figure}[p]
\centering
\includegraphics[width=\linewidth,height=0.72\textheight,keepaspectratio]{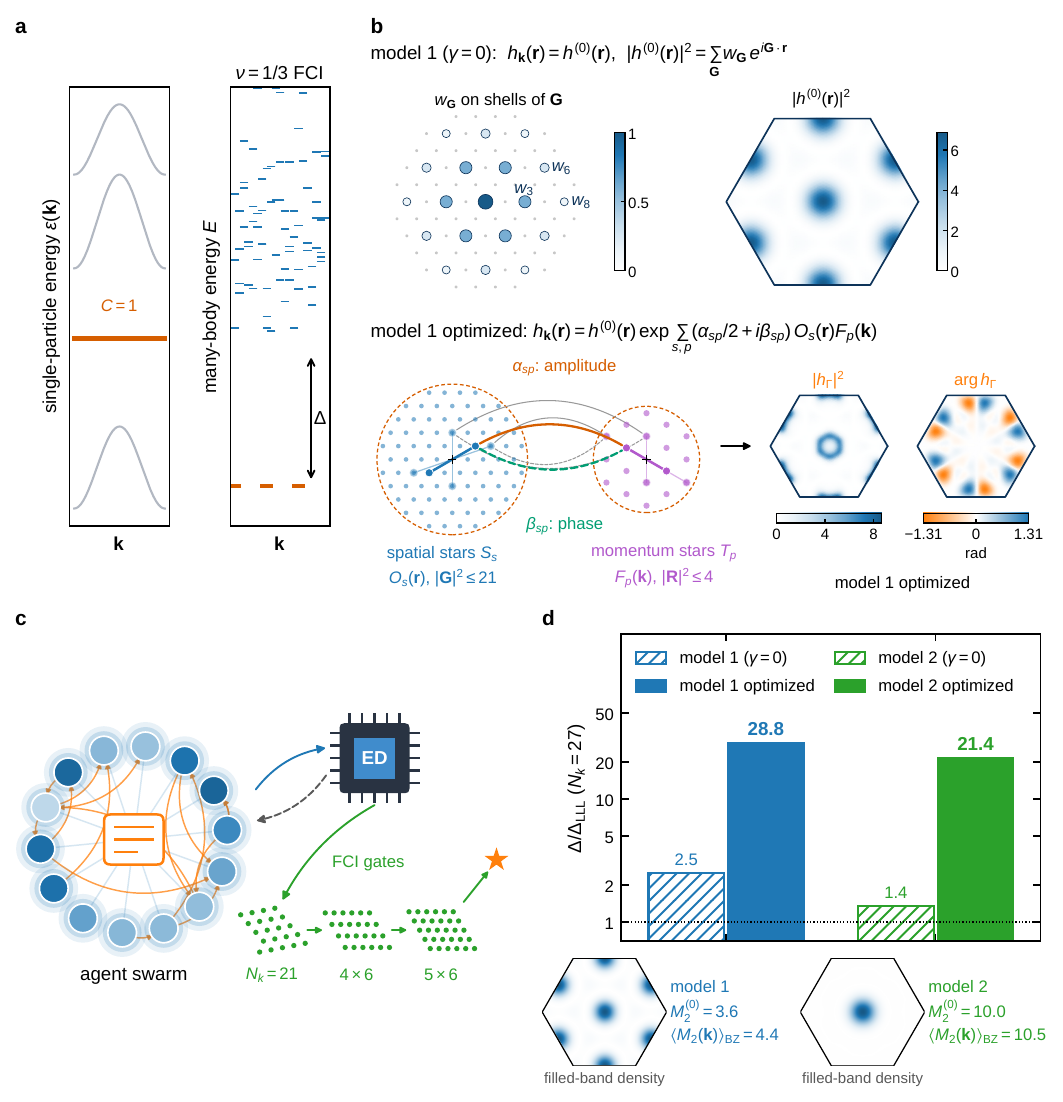}
\caption{\textbf{Agent-driven search for new pathways to enhance the fractional Chern insulator
gap.}
\textbf{a}, Schematic of an isolated topological flat band with Chern number $C=1$ (left), allowing interactions to stabilize a $\nu=1/3$ FCI (right). The many-body gap $\Delta$ sets the energy cost of excitations and an energy scale for the stability of the FCI.
\textbf{b}, Model 1 ($\gamma=0$) has ideal quantum geometry and a momentum-independent envelope $h^{(0)}(\mathbf r)$ whose intensity $|h^{(0)}(\mathbf r)|^2$ is set by the Fourier coefficients $w_{\mathbf G}$ (top). The optimized model adds momentum-dependent amplitude and phase patterns, controlled by $\alpha_{sp}$ and $\beta_{sp}$, which preserve $C=1$ but break ideal geometry and reshape the projected interaction (bottom; right: optimized envelope intensity and phase). All real-space plots show one unit cell.
\textbf{c}, Agents search for a larger gap while retaining Laughlin-type ground state. They share ideas and numerical results through a common message board, while exact diagonalization (ED) tests each candidate on the 21-site, $4\times6$ and $5\times6$ momentum clusters.
\textbf{d}, Many-body gaps calculated from the 27-site cluster ($N=9$ electrons), normalized by lowest Landau level $\Delta_{\mathrm{LLL}}$ under the same gate-screened Coulomb interaction $d_s/\ell_B = 0.1$. The optimized models far exceed the previous benchmark of $M_2\Delta_{\mathrm{LLL}}$ achieved by model 1 and 2 ($\gamma=0$). The two panels below show filled-band densities and envelope moments $M_2^{(0)}$ and $\langle M_2(\mathbf k)\rangle_{\rm BZ}$ for the $\gamma=0$ and optimized bands, respectively.
}
\label{fig:overview}
\end{figure}

Unlike solving a predefined problem, open-ended scientific inquiry has no fixed destination: the questions worth pursuing may evolve as new knowledge emerges. This makes some human guidance essential, even as we seek to give AI the freedom to explore independently. Therefore, we organized our study into three stages (Fig.~\ref{fig:overview}), with human involvement kept to a minimum. First, we introduced AI to the current frontier of knowledge and asked AI to propose a promising model for further investigation (Fig.~\ref{fig:overview}b).
Next, MOSAIC (Multi-agent Open-ended Scientific Analysis and Investigation through Collaboration), a decentralized swarm of up to 48 agents, explored this model, sharing hypotheses and results on a common
board, and building their work on top of each other, with a GPU ED referee that accepts only flat bands where the Laughlin state is a stable ground state across multiple cluster sizes and aspect ratios
(Fig.~\ref{fig:overview}c).
This effort uncovers results beyond what was previously known and substantially outperforms individual agents given the same task. The best flat bands raise the many-body gap of their parents more than tenfold (Fig.~\ref{fig:overview}d).
Finally, we employ AI to distill these discoveries into simpler models and physical explanations. Remarkably, AI found a unified physical understanding of the gap enhancement in both the swarm-discovered and simplified models, uncovering the central role of Bloch wavefunctions in shaping effective two-body interactions and identifying a single-particle indicator that could guide future optimization without costly many-body exact diagonalization.
Throughout this process, human researchers served primarily as guides and interpreters, clarifying the scientific context and what constitutes meaningful progress, rather than directing AI toward particular solutions.

\section*{Results}

\subsection*{Search space designed by AI}

\begin{figure}[!t]
\centering
\includegraphics[width=\linewidth,height=0.72\textheight,keepaspectratio]{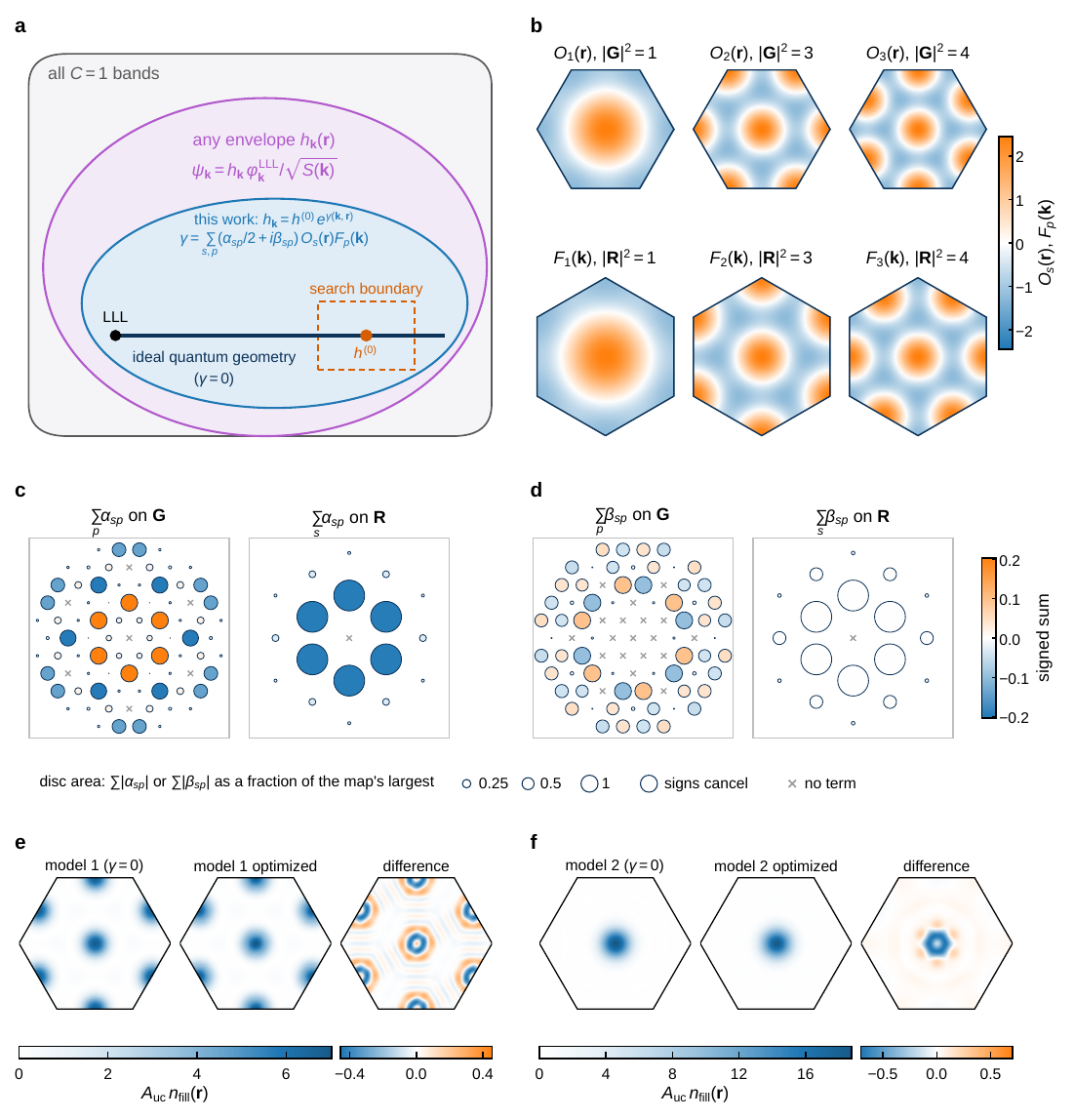}
\caption{\textbf{Search space for topological flat bands and properties of the optimized bands.}
\textbf{a}, Within the $C=1$ bands, the purple region contains states $\psi_{\mathbf k}=h_{\mathbf k}\varphi_{\mathbf k}^\text{LLL}/\sqrt{S(\mathbf k)}$ with general envelopes $h_{\mathbf k}$, and the blue region the exponential family $h_{\mathbf k}=h^{(0)}e^{\gamma(\mathbf k,\mathbf r)}$ of Eq.~(\ref{eq}). The black line marks ideal quantum geometry, including the LLL and models 1 and 2 ($\gamma=0$); the dashed box bounds the agents' search space.
\textbf{b}, Spatial modes $O_s(\mathbf r)$ (top; one unit cell) and momentum modes $F_p(\mathbf k)$ (bottom; Brillouin zone) in the truncated Fourier-product parametrization of $\gamma$, built from the first three shells of reciprocal vectors $\mathbf G$ and lattice vectors $\mathbf R$ (lengths in units of $|\mathbf G_1|$ and the lattice constant). Orange and blue are positive and negative.
\textbf{c,d}, Amplitude coefficients $\alpha_{sp}$ (\textbf{c}) and phase coefficients $\beta_{sp}$ (\textbf{d}) of model 1 optimized, summed over $p$ at each $\mathbf G$ (left) and over $s$ at each $\mathbf R$ (right). Color gives the signed sum and circle area the sum of magnitudes; hollow circles mark sign cancellation and crosses mark absent terms.
\textbf{e,f}, Filled-band densities $A_{\rm uc}n_{\rm fill}(\mathbf r)$ (cell average one) of model 1 (\textbf{e}) and model 2 (\textbf{f}) at $\gamma=0$ and optimized, with their difference. The modest changes show that density concentration alone cannot explain the larger gaps.
}
\label{fig:space}
\end{figure}

We begin by introducing the central scientific question of our study to AI: Can we find new mechanisms to enhance the many-body gap of fractional states in topological flat bands beyond the limits of existing theoretical knowledge? In this study, we focus on flat bands with single-component Bloch wavefunctions and screened Coulomb interactions. Relevant theoretical knowledge and the latest advances are provided to AI. In particular, we highlighted two key elements defining the current boundary of understanding~\cite{sarkar2026moireenhanced}: (1) ideal quantum geometry, $\mathrm{tr}(g(\mathbf{k})) = |F_{xy}(\mathbf{k})|$, where $g(\mathbf{k})$ and $F_{xy}(\mathbf{k})$ denote the quantum metric and Berry curvature, respectively, and (2) the empirical benchmark $\Delta \leq M_2^{(0)}\Delta_{\mathrm{LLL}}$, satisfied by previously known cases.
We emphasized to the agents that discovering new mechanisms capable of overcoming these limitations would represent an advance beyond existing knowledge.

Since the space of possible topological flat bands is effectively infinite-dimensional and impossible to explore exhaustively, we asked AI to propose a concrete model framework to initiate the investigation. The AI proposed exploring topological flat bands with Bloch wavefunctions of the form
\begin{equation}
\psi_{\mathbf k}(\mathbf r)=\frac{h^{(0)}(\mathbf r)\,e^{\gamma(\mathbf k,\mathbf r)}\,\varphi_{\mathbf k}^\text{LLL}(\mathbf r)}{\sqrt{S(\mathbf k)}}.
\label{eq}
\end{equation}
This construction represents a creative generalization of the established wavefunction framework for flat Chern bands with ideal quantum geometry.
Previous studies have shown that for bands with ideal quantum geometry, the Bloch wavefunctions can be expressed as $\psi_{\mathbf k}^{(0)}(\mathbf r)=h^{(0)}(\mathbf r)\varphi_{\mathbf k}^\text{LLL}(\mathbf r)/\sqrt{S_0(\mathbf k)}$~\cite{wang2021exact}. Here $\varphi_{\mathbf k}^\text{LLL}$ is a lowest Landau level (LLL) wavefunction defined on a torus, $h^{(0)}(\mathbf r)$ is a momentum-independent function describing the moir\'e-periodic modulation, and $S(\mathbf k)$ and $S_0(\mathbf k)$ ensure proper normalization. The key innovation proposed by AI is the introduction of an additional exponential factor, $e^{\gamma(\mathbf k,\mathbf r)}$, where $\gamma(\mathbf k,\mathbf r)$ is a smooth function periodic in both momentum and real space, respecting the corresponding moir\'e lattice periodicity. The momentum dependence of this additional factor allows the wavefunction to depart from ideal quantum geometry, while preserving the underlying lattice periodicity. Crucially, because the exponential factor is nonzero and smooth, it preserves momentum space zeros and corresponding vorticity of the parent band wavefunction $\psi_{\mathbf k}^{(0)}$ for any given $\mathbf{r}$, thus preserving its Chern number $C=1$~\cite{Kohmoto1985,Hatsugai1993}. This construction therefore opens up a vast, systematically tunable space of topological flat bands with highly diverse quantum geometries but unchanged topology, providing a controlled framework to search for previously unexplored mechanisms of many-body gap enhancement (Fig.~\ref{fig:space}a). Exploiting the periodicity of $\gamma(\mathbf k,\mathbf r)$ in both momentum and real space, the AI proposed a truncated Fourier-product parametrization $\gamma(\mathbf k,\mathbf r)=\sum_{s,p}\left(\tfrac12\alpha_{sp}+i\beta_{sp}\right)O_s(\mathbf r)F_p(\mathbf k)$,
where $O_s$ and $F_p$ are real-space and momentum-space Fourier modes, respectively (Fig.~\ref{fig:space}b), and $\alpha_{sp}$ and $\beta_{sp}$ provide tunable parameters for systematically exploring different quantum geometries.

\subsection*{An AI swarm finds gaps beyond ideal quantum geometry}

\begin{figure}[!t]
\centering
\includegraphics[width=\linewidth,height=0.72\textheight,keepaspectratio]{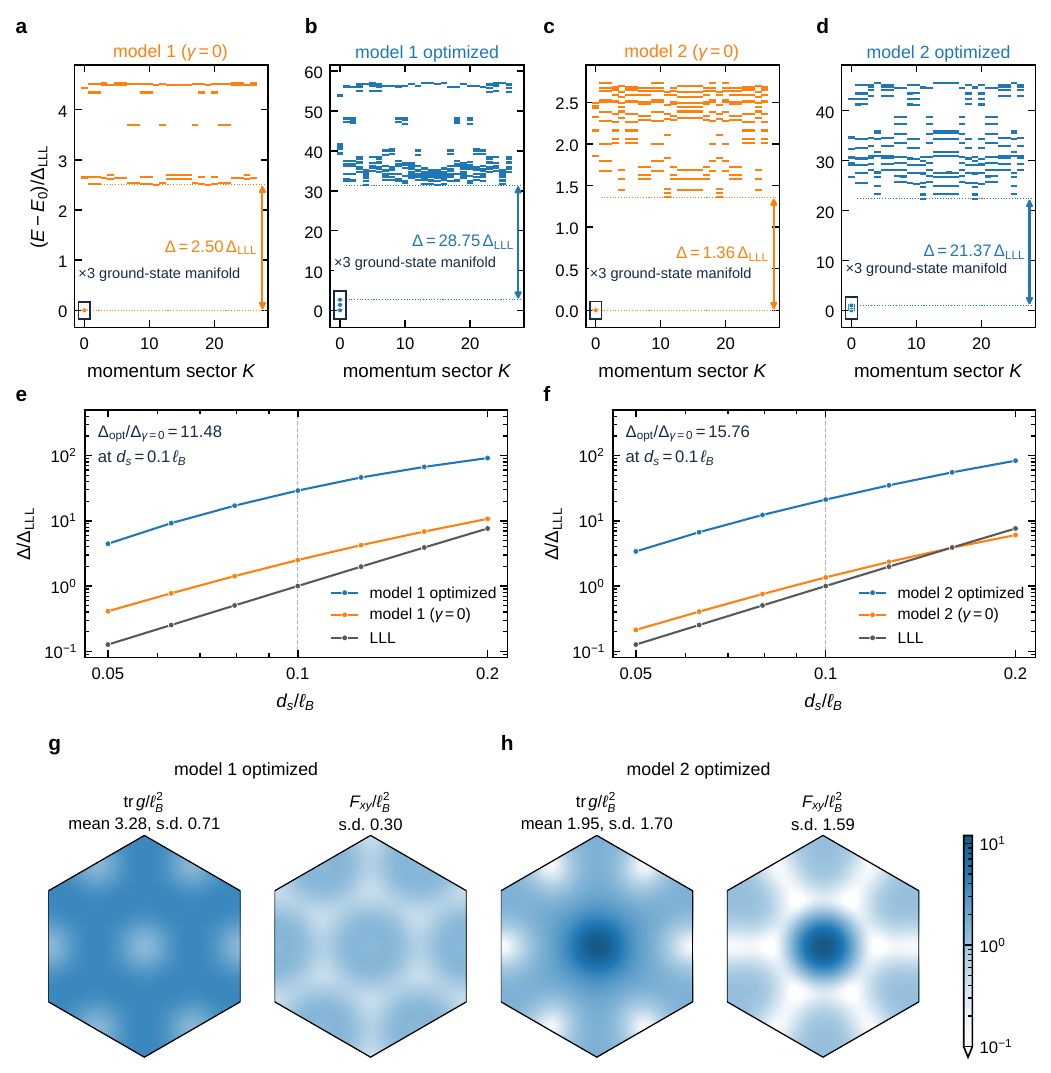}
\caption{\textbf{ED energy spectrum, gap enhancement, and quantum geometry of the optimized topological flat bands.}
\textbf{a--d}, ED spectra at $\nu=1/3$ on the 27-site cluster ($N=9$ electrons) for model 1 (\textbf{a},\textbf{b}) and model 2 (\textbf{c},\textbf{d}), at $\gamma=0$ and optimized (models defined in Methods). Energies $E-E_0$ in each total-momentum sector $K$ are in units of $\Delta_{\rm LLL}$; boxes enclose the three nearly degenerate Laughlin ground states and arrows mark the gap $\Delta$.
\textbf{e,f}, Gap against screening length $d_s/\ell_B$ for model 1 (\textbf{e}) and model 2 (\textbf{f}): optimized (blue), $\gamma=0$ (orange) and LLL (grey), normalized by $\Delta_{\rm LLL}$ at $d_s=0.1\ell_B$ (dashed line, the value used in the search; log--log axes). At small $d_s$ the LLL and optimized gaps follow the cubic law $\Delta\propto(d_s/\ell_B)^3$ of the leading short-range interaction.
\textbf{g,h}, Quantum metric trace ${\rm tr}\,g(\mathbf k)$ and Berry curvature $F_{xy}(\mathbf k)$ of the optimized model 1 (\textbf{g}) and model 2 (\textbf{h}) across the Brillouin zone, in units of $\ell_B^2$ (the mean Berry curvature of a $C=1$ band); labels give means and standard deviations. Unlike the $\gamma=0$ bands, which satisfy ${\rm tr}\,g=|F_{xy}|$, the optimized bands deviate substantially from ideal quantum geometry. Strikingly, large gap enhancement coexists with strong Berry-curvature fluctuations.
}
\label{fig:results}
\end{figure}

We next asked MOSAIC to maximize the $\nu=1/3$ many-body gap of the models in Eq.~(\ref{eq}) for a short-range screened Coulomb interaction with fixed screening length $d_s$ (see Methods). A key target was to exceed the density-based benchmark $M_2^{(0)}\Delta_{\mathrm{LLL}}$ of the ideal-geometry framework~\cite{sarkar2026moireenhanced}. Starting from model 1 ($\gamma=0$) and model 2 ($\gamma=0$), defined in Methods, the agents explored momentum-dependent amplitude and phase deformations (Fig.~\ref{fig:space}c,d). We retained only candidates whose exact-diagonalization spectra and topological diagnostics remained consistent with a Laughlin FCI across several finite-size clusters.

The outcome was striking. On a 27-site cluster with 9 electrons, the optimized gaps reached $28.75\,\Delta_{\mathrm{LLL}}$ and $21.37\,\Delta_{\mathrm{LLL}}$, respectively, representing enhancements by factors of 11.5 and 15.8 times over the corresponding parent ideal band many-body gaps $\Delta^{(0)}$ (Fig.~\ref{fig:results}a--d).
For models 1 and 2, $M_2^{(0)}=3.63$ and $10.04$, so the optimized gaps exceed $M_2^{(0)}\Delta_{\mathrm{LLL}}$ by factors of 7.9 and 2.1. Even compared with the momentum-averaged envelope moments of the optimized models, $\langle M_2(\mathbf k)\rangle_{\mathrm{BZ}}\simeq4.4$ and $10.5$, the gaps are larger by factors of 6.5 and 2.0. These results show that density concentration, which was the main mechanism in~\cite{sarkar2026moireenhanced}, alone cannot explain the enhancement. The optimized models also depart substantially from the ideal quantum geometry condition $\mathrm{tr}\,g(\mathbf k)=|F_{xy}(\mathbf k)|$ (Fig.~\ref{fig:results}g,h), and uniform Berry curvature. Earlier studies of graphene Landau levels and multicomponent bands have demonstrated how orbital structure can enhance fractional state gaps~\cite{tHoke2006fractional,zhang2025beyond}.
In the multicomponent setting of Ref.~\cite{zhang2025beyond}, the enhancement relies on the contact-potential part \(\propto d_s\) of the short-range interaction, which is not eliminated by fermionic antisymmetry for multicomponent wavefunctions and leads to the scaling \(\Delta\propto d_s\). By contrast, for the single-component wavefunctions considered here, fermionic antisymmetry removes this contribution, so that the leading behavior is \(\Delta\propto d_s^3\). Previous single-component examples showed only reduced gaps upon departing from ideal quantum geometry~\cite{zhang2025beyond}. To our knowledge, this is the first demonstration of such a large enhancement of the fractional-state gap in a single-component topological flat band with strongly non-ideal quantum geometry and large Berry-curvature fluctuations.

\begin{figure}[!htp]
\centering
\includegraphics[width=\linewidth,height=0.8\textheight,keepaspectratio]{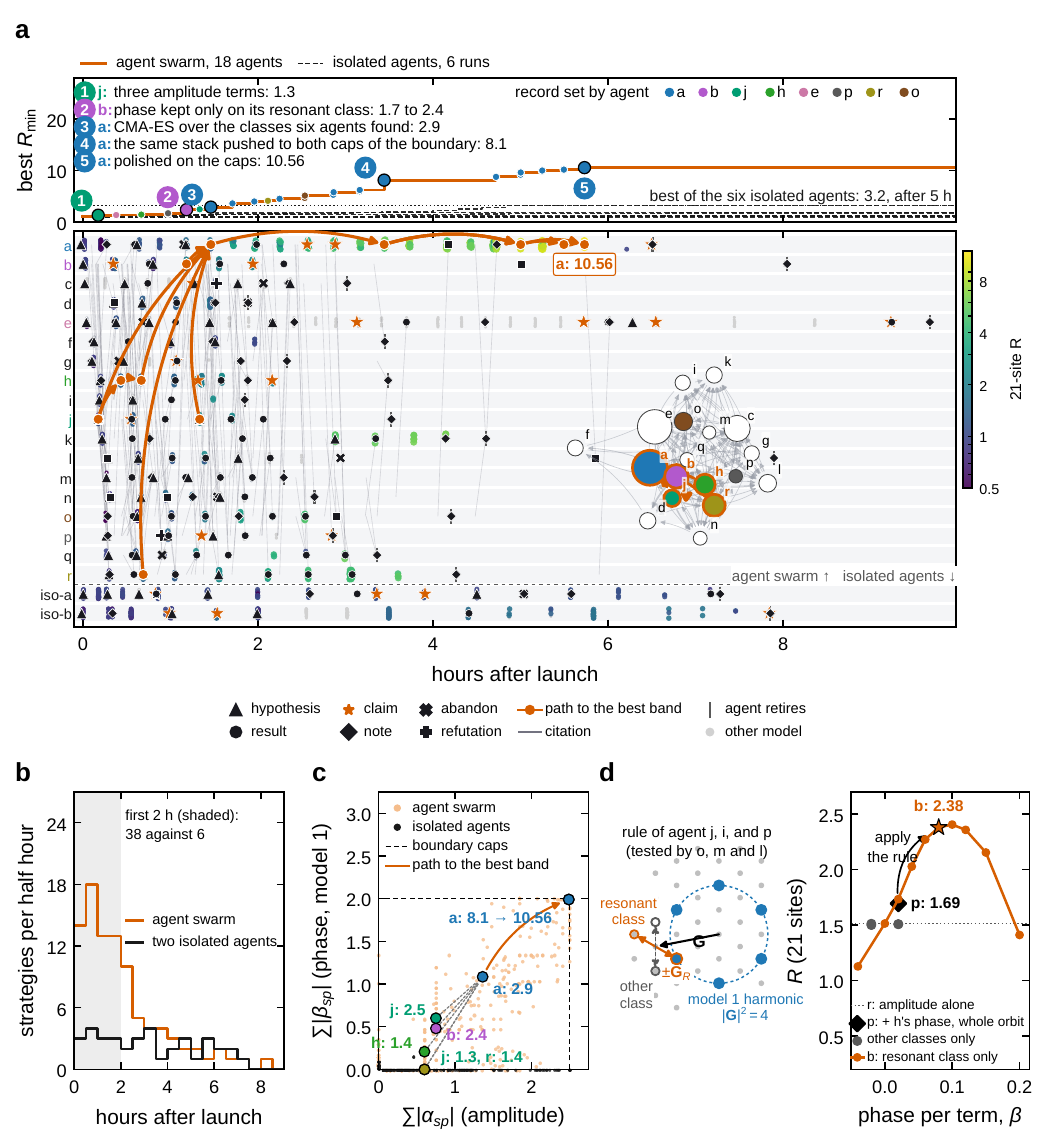}
\caption{
\textbf{Investigation of agent behavior behind the discovery.}
\textbf{a}, Agent behavior in finding the best band of model 1 (defined in Methods). Decentralized agent swarm collectively improves gap enhancement $R_{\min}$, the smallest ratio of the gap to its value at $\gamma=0$ over the meshes used in the search, with key findings marked by 1--5. The lower plot depicts the interactions between agents, with the path to the best band highlighted. The agent interactions are also visualized in a standard force-directed layout in the inset plot.
\textbf{b}, Number of distinct search strategies in use per half hour. The two isolated agents, given the same task without access to the swarm, are counted together.
\textbf{c}, Summed amplitude and phase coefficients of every candidate band submitted. Labeled points mark pieces found by individual agents and the bands that combined them.
\textbf{d}, A finding assembled by several agents. A subset of the agent swarm observed that a term $(\mathbf{G},\mathbf{R})$ of $\gamma$ can raise the gap only if $\mathbf{G}\pm\mathbf{G}_R$ is a Fourier component of $|h^{(0)}|^2$, where $\mathbf{G}_R=\hat{\mathbf z}\times\mathbf R/\ell_B^2$ is the reciprocal lattice vector perpendicular to $\mathbf R$ (on a LLL orbital, the factor $e^{i\mathbf k\cdot\mathbf R}$ is a magnetic translation by $\mathbf R$, i.e.\ a guiding-centre density wave of wavevector $\mathbf G_R$), which is turned into a selection rule for later runs (left). Applying the rule to the swarm's best band raised $R$ from 1.69 to 2.38 (right).
}
\label{fig:agents}
\end{figure}

The collective search mattered. On the clusters used during optimization, the strongest swarms raised the gaps of the two starting bands by factors of $R_{\min}=10.8$ and $16.3$, where $R_{\min}$ is the smallest gain $\Delta/\Delta^{(0)}$ over these clusters. The six isolated agents on model 1 reached only 1.1 to 3.2. On model 2, three of the six never found a band that passed all the tests, and the others reached 3.6, 7.5 and 11.5 (Supplementary Table~S1). Each swarm agent also spent fewer ED evaluations and tokens than an isolated agent, so the advantage did not come from a larger token budget.

We followed the 18-agent swarm on model 1, whose agents had no assigned roles and coordinated only through a shared message board (Fig.~\ref{fig:agents}a). The best band developed by this swarm was built by several agents in turns. 
Agent \textit{j} found the first gain with three amplitude terms ($R_{\min}=1.3$), agent \textit{b} raised it to 2.4 by keeping the phase terms on only one class of $(\mathbf{G},\mathbf{R})$ pairs, and agent \textit{a} combined the classes found by six agents in one covariance matrix adaptation evolution strategy (CMA-ES)~\cite{hansen2001cmaes} search (2.9), which it pushed to the limits of the search range (8.1) and refined to 10.56. Three features made this possible. First, the swarm searched more broadly: in the first two hours, the agents in the swarm ran 38 distinct search strategies, against 6 for the two isolated agents together (Fig.~\ref{fig:agents}b). Secondly, it found a way to combine the partial results effectively. The large gains required tuning amplitude and phase terms together, but different agents found the useful terms of each kind. Swarm agents reused each other's terms, so 45\% of the swarm's requests changed amplitude and phase together. Meanwhile, isolated agents changed one kind at a time, mostly amplitude, and fewer than 1\% of their requests changed both (Fig.~\ref{fig:agents}c). 
Third, agents built on each another's findings, so that a single improvement often passed through several agents, each contributing a different step: observation, explanation, testing and application (Fig.~\ref{fig:agents}d). By writing the deformation of Eq.~(\ref{eq}) as $\gamma(\mathbf k,\mathbf r)=\sum_{\mathbf G,\mathbf R}\gamma_{\mathbf G,\mathbf R}\,e^{i\mathbf G\cdot\mathbf r}e^{i\mathbf k\cdot\mathbf R}$, with $\gamma_{\mathbf G,\mathbf R}=\sum_{\mathbf G\in\mathcal S_s,\,\mathbf R\in\mathcal T_p}(\alpha_{sp}/2+i\beta_{sp})/\sqrt{|\mathcal S_s||\mathcal T_p|}$, where $\mathcal S_s$ and $\mathcal T_p$ are the sets of $\mathbf G$ and $\mathbf R$ that define $O_s$ and $F_p$. Each term $(\mathbf G,\mathbf R)$ pairs a spatial pattern $\mathbf G$ in the unit cell with a variation $\mathbf R$ across the Brillouin zone. Agent \textit{b} observed that, for $\mathbf G$ on the first shell and $\mathbf R$ on the first star, only terms with $\mathbf G\perp\mathbf R$ raised the gap. Agents \textit{j}, \textit{i} and \textit{p} explained this: on LLL orbitals $e^{i\mathbf k\cdot\mathbf R}$ is a guiding-center density wave of wavevector $\mathbf G_R=\hat{\mathbf z}\times\mathbf R/\ell_B^2$, so a term can raise the gap only if $\mathbf G\pm\mathbf G_R$ is a Fourier component of $|h^{(0)}|^2$. Agents \textit{o}, \textit{m} and \textit{l} tested this rule. They also found the limits: the rule held on the first momentum star but failed on some outer cases. Meanwhile, agents \textit{g} and \textit{h} had tested phase terms on whole orbits, only finding gains below 1\% and moving on. Agent \textit{b} then summarized that only one of the three classes in such an orbit satisfies the rule, and applied it to the swarm's best band, restricting its phase terms to the selected class, which raised $R$ from 1.69 to 2.38, a gain no single agent had found; this band became step 2 of the path. The other campaigns show the same pattern where decentralized agents collectively contribute to physical insights which eventually led to discovery (Supplementary Section~S-V). 

\subsection*{From a numerical surprise to a simpler physical picture}

\begin{figure}[p]
\centering
\includegraphics[width=\linewidth,height=0.72\textheight,keepaspectratio]{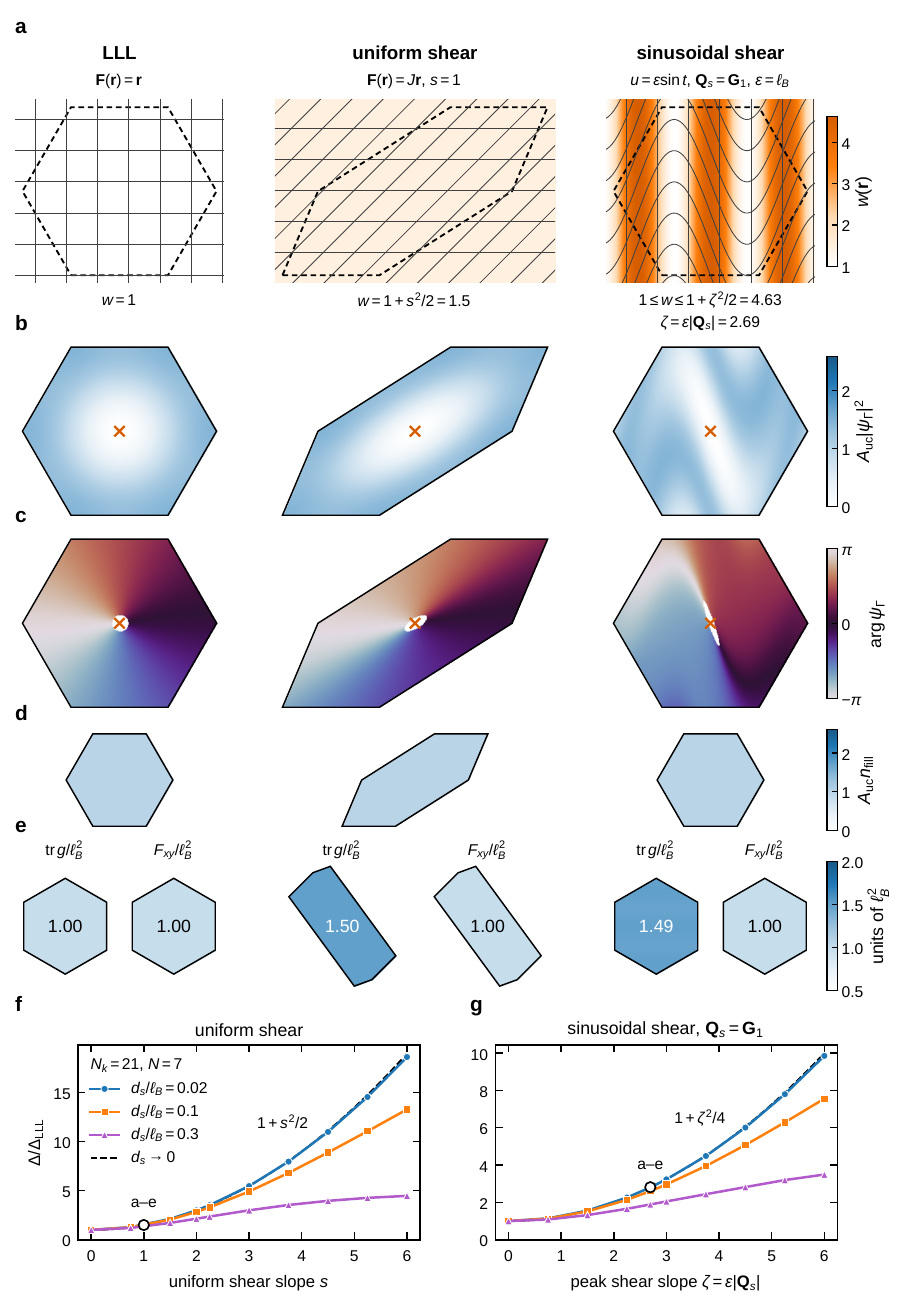}
\caption{\textbf{Gap enhancement by shear at fixed filled-band density.}
\textbf{a--e}, Comparison of the lowest Landau level (LLL, left), uniform shear ($s=1$, middle), and sinusoidal shear ($\epsilon=\ell_B$, $\mathbf Q_s=\mathbf G_1$, right). The area-preserving map is $\mathbf F(x,y)=(x,y+\xi(x))$, with $\xi(x)=sx$ or $\epsilon\sin(|\mathbf Q_s|x)$, where $x$ is along $\mathbf Q_s$ and $\ell_B$ is the magnetic length. The sheared orbitals are $\psi_{\mathbf k}^F=e^{i\chi_F}\varphi_{\mathbf k}^{\rm LLL}(\mathbf F(\mathbf r))$, with $\chi_F$ enforcing magnetic boundary conditions.
\textbf{a}, Real-space shear deformation (dashed lines: unit-cell boundaries). The background shows the local two-electron interaction weight $w(\mathbf r)=\operatorname{tr}(JJ^T)/2=1+[\xi'(x)]^2/2$, with $J=\partial\mathbf F/\partial\mathbf r$, which amplifies the leading short-range two-electron interaction.
\textbf{b,c}, Density $A_{\rm uc}|\psi_\Gamma|^2$ and phase $\arg\psi_\Gamma$ at $\Gamma$ point; crosses mark its zeros.
\textbf{d}, Filled-band density $A_{\rm uc}n_{\rm fill}$ on the 21-site cluster, remaining uniform despite orbital deformation.
\textbf{e}, Brillouin-zone-averaged trace quantum metric $\operatorname{tr}g$ and Berry curvature $F_{xy}$, in units of $\ell_B^2$. Shear increases the metric but preserves the curvature.
\textbf{f,g}, Gap enhancement $\Delta/\Delta_{\rm LLL}$ at $\nu=1/3$, relative to the LLL at the same screening length $d_s$, versus uniform shear slope $s$ (\textbf{f}) and peak sinusoidal shear slope $\zeta=\epsilon|\mathbf Q_s|$ (\textbf{g}), with $\mathbf Q_s=\mathbf G_1$. Blue, orange, and purple denote $d_s/\ell_B=0.02$, $0.1$, and $0.3$. Black dashed curves show theoretical predicted values in the $d_s\to 0$ limit for the two shear models: $1+s^2/2$ (exact) and $1+\zeta^2/4$ (up to Gaussian-suppressed corrections). The gap enhancement persists at longer screening lengths, although the enhancement factor decreases with increasing $d_s$.}
\label{fig:shear}
\end{figure}

The large gaps posed a more interesting question than how to improve the search further: \emph{what are the physical mechanism and theory principles behind this surprising gap enhancement revealed by the agents?} In Eq.~(\ref{eq}), varying $\gamma(\mathbf k,\mathbf r)$ changes several properties at once, including the electron-density distribution in real space, the Berry curvature and the quantum metric across momentum space. With all such effects tangled together, the numerical result alone could not easily give us a clear answer. We therefore asked AI to find a simpler setup where the mechanism could be understood directly.

Working with a researcher, AI proposed applying an area-preserving deformation to the lowest-Landau-level wavefunctions~\cite{haldane2011geometrical}. The simplest example is a uniform shear, $(x,y)\mapsto(x,y+sx)$, applied to every orbital with the phase needed to preserve its boundary conditions. Here $s$ controls the deformation, and $s=0$ gives the original lowest Landau level. Unlike the optimized bands of Eq.~(\ref{eq}), this construction keeps the filled-band electron density homogeneous in real space. Its Berry curvature remains unchanged, while the quantum metric grows with $s$ uniformly across the Brillouin zone (Fig.~\ref{fig:shear}a--e). 
Exact diagonalization with screened Coulomb interactions reveals substantial gap enhancement at small screening lengths (Fig.~\ref{fig:shear}f,g). For uniform shear with $s=6$ and $d_s=0.02\ell_B$, the gap reaches $18.6\,\Delta_{\mathrm{LLL}}$. A sinusoidal shear similarly enhances the gap. This enhancement persists for longer-range interactions, although the value of the enhancement factor is reduced there.

For comparison, the AI's construction is not the only smooth pathway for topological flat bands to depart from ideal quantum geometry while maintaining uniform Berry curvature and quantum metric in momentum space, and uniform electron density in real space. Previous studies have demonstrated alternative constructions with these same properties, for example, using weighted sums of lowest and higher Landau level wavefunctions~\cite{zhang2025beyond}. However, in those constructions, the many-body gap decreases upon departing from ideal quantum geometry, distinguishing them from the approach proposed here by AI.

To test the AI's ability to understand real physics, not just numerical results, we let AI search for a common physical mechanism for gap enhancement in the swarm-optimized models (models 1 and 2) and the simplified shear models. Remarkably, the AI-assisted analysis traced the enhancement to how flat-band Bloch wavefunctions reshape the effective two-body interactions: For the shear model, the shear increases the short-distance variation of the two-electron wavefunction, making close encounters more costly. In the small $d_s$ limit, the Laughlin gap increases by the factor $1+s^2/2$, while retaining its exact zero-energy Laughlin ground space. Since $s$ is unbounded, the enhancement can, in principle, also be unbounded, provided that mixing with remote bands remains negligible. The more complicated swarm-optimized models also turns out to follow this gap enhancement mechanism through strengthening the short-range two-electron interactions, although they do not necessarily preserve exact Laughlin zero modes~\cite{haldane1983hierarchy,trugman1985exact}. This AI-proposed physical mechanism agrees well and gives \emph{exactly} the same prediction as a complementary picture independently suggested by one of the researchers that maps the projected problem to an equivalent lowest Landau level (LLL) with a modified interaction. Further, AI also identified a two-electron interaction index $\mathcal I_{\mathrm 2e}$ (Methods), which can be evaluated directly from single-particle wavefunctions, without many-body ED. This indicator measures the total short-range two-electron interaction strength directly from the Bloch wavefunctions and could help guide FCI gap enhancement optimization without repeated many-body ED.

\section*{Discussion}

The exploring process demonstrated here closely mirrors a longstanding pattern in physics. Many important advances begin not with a complete theoretical understanding, but with the exploration of uncharted territory and the discovery of unexpected phenomena that exceed existing knowledge framework. These observations then motivate the development of new theoretical ideas and simplified models to uncover the underlying physical mechanisms. Our study follows this same trajectory. MOSAIC first explored a vast space of topological flat bands and identified anomalously large many-body gaps beyond the expectations of the existing ideal-geometry framework. Building on these discoveries, AI subsequently developed simplified models to isolate and understand the mechanisms responsible for the enhancement. Thus, the AI went beyond numerical optimization, progressing from exploration and discovery to physical understanding, following a central paradigm of research in physics.

A particularly important aspect of MOSAIC is its ability to explore large, previously uncharted parameter spaces with remarkable efficiency. In such problems, the optimal research strategy is itself unknown: one cannot predict in advance which theoretical ideas, computational methods, or exploration pathways will lead to a discovery. This sets such problems apart from the well-posed ones that frontier AI agents now solve at expert level. With no answer to check against, agents must decide which directions to pursue and when to pause and change, both of which are achieved in our study. This uncertainty makes the diversity of approaches especially valuable. In MOSAIC, individual agents adopt different research styles and strategies, independently pursuing promising directions while exchanging results and ideas with one another. This process closely resembles scientific interactions among human researchers, where different perspectives, independent investigations, and discussions at conferences and within collaborations collectively drive progress. Importantly, MOSAIC reproduces this collective exploration and communication on dramatically shorter timescales. Rather than relying on a single agent following one exploration strategy, the multi-agent framework combines diverse approaches with rapid information exchange, making the search for unexpected physics substantially more effective.

Finally, our results highlight that the role of AI need not end with discovering anomalous numerical results. In physics, identifying an unexpected phenomenon is only the beginning; a deeper goal is to construct a simplified theoretical description that reveals its essential mechanism. The subsequent development of simplified models and the deep underlying physical mechanism in our study demonstrates that AI, drawing on their mathematical and analytical capabilities, can contribute to this crucial stage of scientific understanding as well. Together, these results suggest a broader opportunity: AI-driven scientific research can integrate the exploration of unknown regimes, the identification of phenomena beyond existing knowledge, and the development of new physical explanations into a unified, iterative discovery process.

\section*{Methods}

\subsection*{Topological flat bands with momentum-dependent envelopes}

We use a triangular lattice with one flux quantum per unit cell,
$A_{\rm uc}=2\pi\ell_B^2$, and set $\ell_B=1$ in numerical calculations. The reciprocal vectors obey
$\mathbf a_i\cdot\mathbf b_j=2\pi\delta_{ij}$, so $A_{\rm BZ}=2\pi/\ell_B^2$.
For the optimization, we multiply lowest-Landau-level (LLL) Bloch orbitals by a momentum-dependent envelope,
\begin{equation}
\psi_{\mathbf k}(\mathbf r)=\frac{h_{\mathbf k}(\mathbf r)\,\varphi_{\mathbf k}^\text{LLL}(\mathbf r)}{\sqrt{S(\mathbf k)}},\qquad
S(\mathbf k)=\big\langle u^{\rm LLL}_{\mathbf k}\big|\,|h_{\mathbf k}|^2\,\big|u^{\rm LLL}_{\mathbf k}\big\rangle,
\label{eq:methods_band}
\end{equation}
where the full orbital is normalized over $N_{\rm{uc}}$ unit cells and
$u_{\mathbf k}=\sqrt{N_{\rm{uc}}}e^{-i\mathbf k\cdot\mathbf r}\psi_{\mathbf k}$ is normalized to one over a single cell.
We write $\langle X\rangle_{\rm uc}=A_{\rm uc}^{-1}\int_{\rm uc}X\,d^2r$ and use the analogous normalized average over the
Brillouin zone. A momentum-independent envelope $h^{(0)}$ gives ideal quantum geometry,
$\operatorname{tr}g=F_{xy}$~\cite{wang2021exact,ledwith2023vortexability}. Its intensity is
$\rho^{(0)}=|h^{(0)}|^2=\sum_{\mathbf G}w^{(0)}_{\mathbf G}e^{i\mathbf G\cdot\mathbf r}$, with
$w^{(0)}_{\mathbf0}=1$, and its density moment is
$M_2^{(0)}=\langle(\rho^{(0)})^2\rangle_{\rm uc}=\sum_{\mathbf G}|w^{(0)}_{\mathbf G}|^2$~\cite{sarkar2026moireenhanced}.
We use two such starting bands. Model 1 has shell weights
$(w^{(0)}_1,\ldots,w^{(0)}_8)=(0,0,3/5,0,0,1/4,0,1/8)$ and $M_2^{(0)}=3.63$.
Model 2 has $w^{(0)}_1=0.80$ and nine additional shells chosen to keep $\rho^{(0)}$ positive, giving
$M_2^{(0)}=10.04$. Their $\nu=1/3$ gaps on the 27-site cluster are
$\Delta^{(0)}/\Delta_{\rm LLL}=2.50$ and $1.36$, respectively.

We deform these bands through $h_{\mathbf k}=h^{(0)}e^\gamma$, with $\gamma=a/2+i\theta$ and smooth, real fields
$a$ and $\theta$ periodic in both $\mathbf r$ and $\mathbf k$. The amplitude changes the intensity to $\rho^{(0)}e^a$;
varying only the phase leaves the normalization and every orbital density unchanged. Periodicity preserves the LLL
boundary phases across the Brillouin zone, while
$S(\mathbf k)\ge e^{\min_{\mathbf r}a(\mathbf k,\mathbf r)}S_0(\mathbf k)>0$ ensures a well-defined normalized state.
The band therefore retains $C=1$~\cite{Kohmoto1985,Hatsugai1993}. Its departure from the ideal trace condition is
\begin{equation}
{\rm tr}\,g(\mathbf k)-F_{xy}(\mathbf k)={\rm Var}_{\mathbf k}\big(\partial_+\gamma\big)\geq0,
\qquad \partial_+=\partial_{k_x}+i\partial_{k_y},
\label{eq:methods_ideality}
\end{equation}
where $\operatorname{Var}_{\mathbf k}f=\langle|f|^2\rangle_{\mathbf k}-|\langle f\rangle_{\mathbf k}|^2$ and
$\langle f\rangle_{\mathbf k}=\int_{\rm uc}|u_{\mathbf k}|^2f\,d^2r$. Thus a nonzero $\gamma$ breaks the ideal trace
condition only where $\partial_+\gamma$ has nonzero variance within the orbital.

For the search we use the separable Fourier parametrization
\begin{equation}
\begin{gathered}
\gamma(\mathbf k,\mathbf r)=\sum_{s,p}\left(\frac{\alpha_{sp}}2+i\beta_{sp}\right)O_s(\mathbf r)F_p(\mathbf k),\\
O_s=\frac{\sum_{\mathbf G\in\mathcal S_s}e^{i\mathbf G\cdot\mathbf r}}{\sqrt{|\mathcal S_s|}},\qquad
F_p=\frac{\sum_{\mathbf R\in\mathcal T_p}e^{i\mathbf k\cdot\mathbf R}}{\sqrt{|\mathcal T_p|}}.
\end{gathered}
\label{eq:methods_fourier}
\end{equation}
The inversion-symmetric sets $\mathcal S_s$ and $\mathcal T_p$ contain nonzero reciprocal and real-space lattice
vectors, respectively. The functions are real, have zero mean and unit mean square, and the coefficients
$\alpha_{sp}$ and $\beta_{sp}$ are real. The agents could use individual inversion pairs or combine vectors related
by lattice rotations. Spatially uniform terms change only orbital normalization or overall phase; momentum-independent
terms can be absorbed into $h^{(0)}$. We exclude both. The boundary in Table~\ref{tab:boundary} was proposed by the
agents, approved by the researchers, and fixed before the campaigns. It allows 756 real coefficients and limits their
magnitudes, their absolute sums, and the average envelope moment to
$\langle M_2(\mathbf k)\rangle_{\rm BZ}\le3M_2^{(0)}$. This last constraint limits density concentration without
fixing the filled-band density. The envelope moment and filled-band density are defined below; derivations are given in
Supplementary Section~S-I.

\input{tables/tab_boundary}

\subsection*{Projected interaction and exact diagonalization}

We take the band to be exactly flat and project the interaction using the form factors
$\lambda_{\mathbf q}(\mathbf k,\mathbf k')=\langle\psi_{\mathbf k}|e^{-i\mathbf q\cdot\mathbf r}|\psi_{\mathbf k'}\rangle$,
with $\mathbf q=\mathbf k'-\mathbf k+\mathbf G$. They follow from the exact convolution
\begin{equation}
\lambda_{\mathbf q}(\mathbf k,\mathbf k')=\frac{\sum_{\mathbf G}w_{\mathbf G}(\mathbf k,\mathbf k')\,
\lambda^{\rm LLL}_{\mathbf q-\mathbf G}(\mathbf k,\mathbf k')}{\sqrt{S(\mathbf k)S(\mathbf k')}},\qquad
w_{\mathbf G}(\mathbf k,\mathbf k')=\big\langle h^*_{\mathbf k}h_{\mathbf k'}e^{-i\mathbf G\cdot\mathbf r}\big\rangle_{\rm uc},
\label{eq:methods_form_factor}
\end{equation}
with $w_{\mathbf G}(\mathbf k,\mathbf k')=w^{(0)}_{\mathbf G}$ at $\gamma=0$. The interaction is the
dual-gate-screened Coulomb potential $V(q)=2\pi e^2\tanh(qd_s)/(\varepsilon q)$, with
$E_C=e^2/(\varepsilon\ell_B)$ and $d_s=0.1\ell_B$. We diagonalize the projected Hamiltonian in each
total-momentum sector with a GPU-accelerated solver built on the open-source package
MomentumED.jl~\cite{zou2025momentumed} (Supplementary Section~S-IV). Searches use a $48\times48$ real-space mesh, envelope-kernel
cutoff $|m|,|n|\le8$, and transfer cutoff $q_{\max}=8$ shells. The diagnostics of
Figs.~\ref{fig:space}e,f and \ref{fig:results}g,h use envelope cutoff 12.
The screening scans in Fig.~\ref{fig:results}e,f use seven logarithmically spaced values from
$d_s=0.05$ to $0.2\ell_B$ and $q_{\max}=10$.

We diagonalize $N=N_{\rm{uc}}/3$ electrons in every total-momentum sector by Krylov iteration, using a precomputed two-body
vertex and an H100 GPU. Ordering the levels across sectors as $E_0\le E_1\le\cdots$, we define
$\Delta=E_3-E_2$ and the ground-state splitting as $E_2-E_0$. The validation ladder in Table~\ref{tab:ladder} checks
the threefold manifold, particle entanglement spectrum (PES) counting~\cite{li2008entanglement,sterdyniak2011pes}, and the
many-body Chern number averaged over the three lowest states, evaluated by twisting the boundary conditions in both
torus directions. The PES is calculated from $\rho_{\rm GS}=\frac13\sum_{\alpha=0}^{2}|\Psi_\alpha\rangle\langle\Psi_\alpha|$, where \(|\Psi_\alpha\rangle\) are the three lowest-energy eigenstates spanning the ground-state manifold,
by tracing out $N-N_A$ particles. We use $N_A=2$ at all stages except the 21-site confirmation stage, where $N_A=3$.
A strong claim
requires $R=\Delta/\Delta^{(0)}\ge1.10$ on $4\times6$, $5\times6$, and the higher-cutoff rerun
($q_{\max}=10$, a $64\times64$ mesh, envelope cutoff 10); candidates are ranked by the smallest ratio $R_{\min}$
over the validation ladder. Reported gaps use confirmation accuracy on the symmetric 27-site cluster ($N=9$), which was
excluded from the search, and are divided by the LLL gap on the same cluster at the same screening length.

\input{tables/tab_ladder}

\subsection*{Band diagnostics}

The normalized envelope intensity and its second moment are
\begin{equation}
\rho_{\mathbf k}(\mathbf r)=\frac{|h_{\mathbf k}(\mathbf r)|^2}{\langle|h_{\mathbf k}|^2\rangle_{\rm uc}},\qquad
M_2(\mathbf k)=\langle\rho_{\mathbf k}^2\rangle_{\rm uc}
=\sum_{\mathbf G}\left|\frac{w_{\mathbf G}(\mathbf k,\mathbf k)}{w_{\mathbf0}(\mathbf k,\mathbf k)}\right|^2.
\label{eq:methods_density_moment}
\end{equation}
We average $M_2(\mathbf k)$ over the sampled momenta. It equals one for a uniform envelope and reduces to
$M_2^{(0)}$ at $\gamma=0$. The physical filled-band density is
$n_{\rm fill}(\mathbf r)=\sum_{\mathbf k}|\psi_{\mathbf k}(\mathbf r)|^2$, evaluated on the 27-site cluster.
It also depends on the LLL orbital profiles and normalizations $S(\mathbf k)$. Phase-only deformations leave the
envelope moment and filled-band density unchanged but can change the off-diagonal form factors.

We use the quantum geometric tensor
$\mathcal Q_{ij}=\langle\partial_i u_{\mathbf k}|(1-|u_{\mathbf k}\rangle\langle u_{\mathbf k}|)|\partial_j u_{\mathbf k}\rangle$,
with $g_{ij}=\operatorname{Re}\mathcal Q_{ij}$, $F_{xy}=2\operatorname{Im}\mathcal Q_{xy}$, and Cartesian momentum
derivatives $\partial_i$. This gives $F_{xy}^{\rm LLL}=\ell_B^2$ and $C=(2\pi)^{-1}\int_{\rm BZ}F_{xy}\,d^2k$.
Numerically, $g_{xx}$ and $g_{yy}$ follow from
$g_{ii}\simeq[1-|\langle u_{\mathbf k}|u_{\mathbf k+\delta k\hat{\mathbf e}_i}\rangle|^2]/\delta k^2$,
with $\delta k=10^{-3}\ell_B^{-1}$. Berry curvature is the phase of the oriented overlap product around each
plaquette of a $24\times24$ grid, divided by its momentum-space area and using the magnetic Bloch boundary phases.
Both are reported in units of $\ell_B^2$.

\subsection*{Sheared Landau levels}

We construct a second family by applying a common area-preserving coordinate shear to every complete LLL orbital,
\begin{equation}
\psi^F_{\mathbf k}(\mathbf r)=e^{i\chi_F(\mathbf r)}\varphi_{\mathbf k}^{\rm LLL}(\mathbf F(\mathbf r)),\qquad
J=\frac{\partial\mathbf F}{\partial\mathbf r},\quad\det J=1.
\label{eq:methods_shear}
\end{equation}
The momentum-independent phase $\chi_F$ enforces magnetic boundary conditions, and area preservation makes the
transformation unitary. The interaction and the Euclidean metric remain fixed in laboratory coordinates.
These bands need not lie in the smooth exponential family above, because shear moves orbital zeros.

We use a uniform shear $\mathbf F_s(x,y)=(x,y+sx)$ and a periodic shear
$\mathbf F(\mathbf r)=\mathbf r+\mathbf d\,\epsilon\sin(\mathbf Q_s\cdot\mathbf r)$, where
$\mathbf d=\hat{\mathbf z}\times\mathbf Q_s/|\mathbf Q_s|$ and $\mathbf Q_s$ is a reciprocal-lattice vector.
The periodic shear acts on the original torus. For uniform shear, the period lattices obey
$\Lambda_F=J_s^{-1}\Lambda_{\rm LLL}$, and laboratory momenta enter the LLL form factors as $J_s^{-T}\mathbf q$.
For the periodic shear, the form factors are
$\lambda^F_{\mathbf q}=\sum_{n\in\mathbb Z}J_n[-\epsilon(\mathbf q\cdot\mathbf d)]
\lambda^{\rm LLL}_{\mathbf q+n\mathbf Q_s}$, where $J_n$ is the $n$-th Bessel function of the first kind.
Both families retain $C=1$, $F_{xy}=\ell_B^2$, and uniform filled-band density, while their metric trace increases.

Figure~\ref{fig:shear}a--e uses $s=1$ or $\mathbf Q_s=\mathbf G_1$ with $\epsilon=\ell_B$.
The gap scans in Fig.~\ref{fig:shear}f,g use $N=7$ electrons on the 21-site cluster in every momentum sector, varying
$s$ or varying $\epsilon$ at fixed $\mathbf Q_s=\mathbf G_1$ ($|\mathbf Q_s|\ell_B=2.69$).
Each gap is divided by the LLL gap at the same $d_s$ and transfer cutoff.
We increase the transfer cutoff until the relative gap change is below $2\times10^{-4}$.
Derivations are given in Supplementary Sections~S-II and S-III.

\subsection*{Two-electron interactions and short-range gap enhancement}

We asked AI to identify a unified mechanism for both of the optimized $h_{\mathbf k}$ bands and the shear model. AI proposed examining the short-distance slope of a two-electron wavefunction. Independently, a researcher proposed absorbing the band deformation into an effective LLL interaction. These pictures describe the same projected interaction physics and have the same quantitative predictions.

For normalized band orbitals $\psi_i(\mathbf r)$, labeled by Bloch momentum $i$, define $\mathbf D_{ij}=\psi_j\nabla\psi_i-\psi_i\nabla\psi_j$. At center coordinate $\mathbf r=(\mathbf r_1+\mathbf r_2)/2$ and relative separation $\boldsymbol\delta=\mathbf r_1-\mathbf r_2$, the antisymmetric two-electron wavefunction is $\Phi_{ij}(\mathbf r_1,\mathbf r_2)=\left[\psi_i(\mathbf r_1)\psi_j(\mathbf r_2)-\psi_j(\mathbf r_1)\psi_i(\mathbf r_2)\right]/{\sqrt2}$. To relate this slope to the interaction energy, write $d=d_s/\ell_B$ and $E_C=e^2/(\varepsilon\ell_B)$. At small $d$, $V(q)=2\pi E_C\ell_B^2[d-d^3(q\ell_B)^2/3+O(d^5(q\ell_B)^4)]$. The contact term vanishes because an antisymmetric two-electron wavefunction is zero when the electrons coincide. The remaining $-q^2$ term Fourier transforms to $\nabla_{\boldsymbol\delta}^2\delta^{(2)}(\boldsymbol\delta)$ and probes $\left.\nabla_{\boldsymbol\delta}^2(\Phi_{ij}^*\Phi_{kl})\right|_{\boldsymbol\delta=0}=\mathbf D_{ij}^*\cdot\mathbf D_{kl}$. Define $\widehat{\mathbf D}(\mathbf r)=\sum_{i<j}\mathbf D_{ij}(\mathbf r)c_jc_i$, where $c_i$ annihilates an electron in orbital $\psi_i$. The full projected Hamiltonian is thus $H(d)=d^3H_{\rm 2e}+O(d^5)$, with the leading short-range two-electron interaction Hamiltonian
\begin{equation}
H_{\rm 2e}=\frac{2\pi}{3}E_C\ell_B^4\int_A d^2r\,
\widehat{\mathbf D}^{\dagger}(\mathbf r)\cdot\widehat{\mathbf D}(\mathbf r).
\label{eq:methods_short_pair}
\end{equation}

For the agent optimized $h_{\mathbf k}$ bands, write $h_i=h^{(0)}e^{a_i/2+i\theta_i}$, with real amplitude and phase fields $a_i$ and $\theta_i$. Then $\mathbf D_{ij}=(h_i h_j/\sqrt{S_iS_j})\mathbf D_{ij}^{\rm LLL}+\psi_i\psi_j[\frac12\nabla(a_i-a_j)+i\nabla(\theta_i-\theta_j)]$, where $S_i$ and $S_j$ are the normalization factors of orbitals $i$ and $j$, defined by $S(\mathbf k)$ in Eq.~\eqref{eq:methods_band}. At $\gamma=0$, all orbitals share the envelope $h^{(0)}$, so the relative amplitude and phase gradients vanish. For $S_i\simeq1$, the interaction is weighted by $|h^{(0)}|^4$. Replacing this weight by its unit-cell average, $M_2^{(0)}=\langle|h^{(0)}|^4\rangle_{\rm uc}$, gives the approximation $H_{\rm 2e}^{(0)}\simeq M_2^{(0)}H_{\rm 2e}^{\rm LLL}$, accounting for the gap enhancement. For general $h_{\mathbf k}$, the relative amplitude and phase gradients can interfere constructively, strengthening short-range two-electron repulsion and favoring a larger FCI gap. 
The same mechanism applies to the shear model. For uniform shear $s$, the chain rule gives the two-electron wavefunction slopes $\mathbf D_{ij}^s(\mathbf r)=e^{2i\chi_F(\mathbf r)}J_s^T\mathbf D_{ij}^{\rm LLL}(J_s\mathbf r)$, as derived in Supplementary Section S-III. Every overlap $\mathbf D_{ij}^*\cdot\mathbf D_{kl}$ is then multiplied by $w=\tfrac12\operatorname{tr}(J_sJ_s^T)=1+s^2/2$. Since $\det J_s=1$, Eq.~\eqref{eq:methods_short_pair} therefore gives $H_{\rm 2e}^s=wH_{\rm 2e}^{\rm LLL}$, accounting for the gap enhancement in the shear model.

The effective-interaction picture proposed by the researcher gives the same result. Use the same fermion operators for corresponding sheared and LLL orbitals. The common shear phase cancels in the density matrix elements. Changing variables to $\mathbf y=J_s\mathbf r$, with $\det J_s=1$, gives $\langle\psi_i^s|e^{-i\mathbf q\cdot\mathbf r}|\psi_j^s\rangle=\langle\varphi_i^{\rm LLL}|e^{-i(J_s^{-T}\mathbf q)\cdot\mathbf y}|\varphi_j^{\rm LLL}\rangle$, giving $\widehat\rho^s_{\mathbf q}=\widehat\rho^{\rm LLL}_{J_s^{-T}\mathbf q}$. The full Hamiltonian therefore has the equivalent forms
\begin{equation}
H(d)=\frac{1}{2A}\sum_{\mathbf q}V(q)
:\widehat\rho^s_{\mathbf q}\widehat\rho^s_{-\mathbf q}:
=\frac{1}{2A}\sum_{\mathbf p}V_{\rm eff}(\mathbf p)
:\widehat\rho^{\rm LLL}_{\mathbf p}\widehat\rho^{\rm LLL}_{-\mathbf p}:.
\label{eq:methods_interaction_equivalence}
\end{equation}
Here colons denote normal ordering and $\mathbf p=J_s^{-T}\mathbf q$, comparing the Hamiltonian gives $V_{\rm eff}(\mathbf p)=V(|J_s^T\mathbf p|)$. Apply the 
small-$d$ expansion, $V_{\rm eff}(\mathbf p)=2\pi E_C\ell_B^2[d-d^3\ell_B^2\mathbf p^T J_sJ_s^T\mathbf p / 3+O(d^5)]$. 
Since the two-electron density is isotropic for LLL, the order-$d^3$ interaction therefore depends only on the angular average: 
$\langle\mathbf p^T J_sJ_s^T\mathbf p\rangle_{\rm angle}=\tfrac12p^2\operatorname{tr}(J_sJ_s^T)=wp^2$. 
This is exactly the factor obtained from Eq.~\eqref{eq:methods_short_pair}, so both descriptions give the same leading short-range 
gap enhancement. 

The two-electron interacting Hamiltonian in Eq.~\eqref{eq:methods_short_pair} also gives a single-particle measure of total interaction strength. Define the filled-band density $n_{\rm fill}=\sum_i|\psi_i|^2$ and the normalized orbital vector $|\omega\rangle=(\psi_1,\psi_2,\ldots)^T/\sqrt{n_{\rm fill}}$. Its real-space quantum metric is $g^r_{\mu\nu}=\operatorname{Re}\langle\partial_\mu\omega|(1-|\omega\rangle\langle\omega|)|\partial_\nu\omega\rangle$. Summing over all distinct choices of two band orbitals gives $\sum_{i<j}|\mathbf D_{ij}|^2=n_{\rm fill}^2\operatorname{tr}g^r$. Therefore, define the dimensionless index of two-electron interaction strength
\begin{equation}
\mathcal I_{\rm{2e}}=\frac{3}{N_{\mathrm{uc}}E_C}\sum_{i<j}\langle ij|H_{\rm 2e}|ij\rangle=\ell_B^2A_{\rm uc}^2
\big\langle n_{\rm fill}^2\operatorname{tr}g^r\big\rangle_{\rm uc},
\label{eq:methods_short_indicator}
\end{equation}
where $|ij\rangle=c_i^\dagger c_j^\dagger|0\rangle$ is a normalized two-electron state, $|0\rangle$ is the vacuum, 
and $N_{\mathrm{uc}}$ is the number of band orbitals. The sum includes each distinct choice $i<j$ once. In the shear model, this index 
is 1 without shear and equals $\langle w\rangle_{\rm uc}$ after shear. So this index provides a measure of the interaction strength in the sheared system.  However, the many-body gap also depends on how the ground 
and excited states occupy the interacting two-electron modes. The index therefore provides a useful guide to potential gap enhancement, though it does not guarantee a larger gap. Full derivations are given in Supplementary Sections S-II and S-III.

\subsection*{Usage of agents and design of MOSAIC}

LLM agents took part in all three stages of the study: search space definition, numerical study with agent swarm, and shear LLL theory proposal, while the researchers provided the background knowledge and checked the results. In the first stage, a Claude Code agent (Opus 5.5) was given the question of this work together with the theory of flat bands with ideal quantum geometry and the bound $\Delta\le M_2^{(0)}\Delta_{\rm LLL}$; it proposed the momentum-dependent envelope of Eq.~(\ref{eq}), its Fourier parametrization, and the search boundary. In the second stage, swarms of Claude Code agents (Opus 5.5) searched this space for large gaps, as described in the sections below. In the third stage, a separate AI (GPT-6 Pro), which was not part of the swarm, read the numerical results of the search and developed the theory of sheared Landau levels and its gap bounds. The search space definition and shear LLL theory proposal were achieved by human-AI discussions, while the numerical search for larger FCI gap was autonomously conducted by the agent swarm without human interference.

Our agent swarm framework MOSAIC consists of identical, autonomous agents, a shared message board through which they communicate, and an ED broker that evaluates every candidate band. There is no central planner in the agent team, thus no agent is assigned a role, a subspace to search, or a method to use, and coordination emerges only from what the agents read on the board. 

\bmhead{Agents}

In the agent swarm. each agent is an instance of Claude Opus~5.5 (Anthropic) running in Claude Code. An agent works in its own copy of the project repository, which contains the solver, the project's notes on the physics of the search space, and a private notebook. Agents can read files, write and run their own analysis scripts and optimizers, but they act on the shared state only through one command-line tool, which validates a candidate, submits jobs, reports results, and reads and writes the board. All agents of a campaign receive the same prompts and differ only in their names.

\bmhead{Board}

The board is the swarm's shared memory. A post is an immutable, time-stamped record of one of seven kinds: (i) a hypothesis, which an agent must post before spending more than ten evaluations on a family of candidates; (ii) a result; (iii) a claim; (iv) a refutation of another agent's claim; (v) an abandonment, which records what was learned from a dropped direction; (vi) a request to the human researchers; (vi) or a note. Agents can also mail each other directly, and a new request can declare the earlier request it builds on. Every number on the board must be quoted from a broker result by its request identifier, so that agents build on verified results rather than on each other's estimations. The prompts ask agents to read the open hypotheses before starting a direction, to critique other agents' claims in about one post in five, and to watch for herding onto a single family.

\bmhead{ED broker}

The broker is a deterministic process and the only writer of the result store. It manages the queue, the clusters, the accuracy settings, and the solver version; it rejects any request outside the search boundary (Table~\ref{tab:boundary}), runs accepted requests on persistent CPU and GPU solver lanes, applies the phase gates and promotes a request up the ladder of Table~\ref{tab:ladder} automatically when its gates pass and its smallest $R$ so far reaches the next threshold. 

\bmhead{Work cycle}

Agents run in Claude Code's \texttt{/loop} mode and set their own pace. At each wake-up, an agent reviews what has changed since its last visit, such as new posts and mail, the results of its own requests, the leaderboard and its remaining budget. Consequently, the agent would append to its notebook what it tried, what it learned, and what it would do next. The notebook carries the agent's memory across restarts of its session. An agent stops by posting a final summary when its quota is spent or when it judges that nothing worthwhile is left to try.

\bmhead{Prompts}

Each agent receives two texts (Supplementary Section~S-VI). A charter is appended to the Claude Code system prompt. It contains the campaign brief, which states the question: whether a momentum-dependent envelope can raise the Laughlin gap above that of the $\gamma=0$ band robustly across cluster sizes, and whether such a gain goes beyond the changed density. The charter also includes the physical setup, the search space and its boundary, the scoring, the phase gates, and when to accept the result. The charter then lists the tools and the rules of conduct, and leaves the choice of direction to the agent: ``Nobody assigns you a role or a region; the swarm organizes itself through the board.'' A short wake-up prompt is sent at the start of the session and again after every restart: ``Continue your swarm work as your charter describes: run \texttt{swarm digest}, decide, act, append to NOTEBOOK.md, run \texttt{swarm heartbeat}, then pace your next wakeup. When your quota is spent or nothing is left to try, run \texttt{swarm done} and stop the loop.''

\bmhead{Campaigns and isolated agents}

We launched six campaigns inside the search boundary, on models 1 and 2 with swarms of 5, 18 and 48 agents (Supplementary Table~S1). The agents of a campaign were started 60\,s apart and stopped at a 48\,h deadline. Every campaign also ran two isolated agents with the same brief, quota, and tools but without the board, the mail, or other agents' requests, giving six independent single-agent runs per model; their prompts differ from those of swarm agents only in the passages on coordination (Supplementary Section~S-VI). 
In each campaign, the analysis folders and the development history were hidden from the agents by file-system access controls. Token usage is counted from the agents' transcripts. The quantities used to analyze the agents' search are defined in Supplementary Section~S-V.

\section*{Data Availability}

The band parameters, spectra and agent records that support the findings of this study are available at \url{https://github.com/SciFM/MOSAIC}.

\section*{Code Availability}

The code for developing and launching the agent swarm MOSAIC and the implementation of the exact-diagonalization solver are available at \url{https://github.com/SciFM/MOSAIC}.

\section*{Acknowledgments}
VV, CX, and ABL acknowledge the support from Los Alamos National Laboratory under the grant number AWD026741 at the University of Michigan.
YZ and KS acknowledge support from the Office of Naval Research (Grant No. N00014-26-1-2565), the Air Force Office of Scientific Research MURI program (Grant No. FA9550-23-1-0334), and the National Science Foundation through the University of Michigan Materials Research Science and Engineering Center (Award No. DMR-2309029). SS is supported by the Laboratory for Physical Sciences through the Condensed Matter Theory Center.
The authors thank Advanced Research Computing at the University of Michigan for providing computing resources.

\section*{Author Contributions}

VV and KS conceived the idea; CX designed and implemented MOSAIC and ran the swarm campaigns; YZ formulated the flat-band model parametrization and its exact-diagonalization implementation; CX, YZ, and ABL monitored the progress of agent swarm campaigns; CX led the investigation into agent behavior; YZ led the analysis of gap enhancement mechanism and the sheared Landau level; YZ and SS reviewed and interpreted the physics; all authors modified and approved the manuscript.

\section*{Competing Interests}

The authors declare no competing interests.

\bibliography{sn-bibliography}

\end{document}


\title{Supplementary Information for:\\ AI Agent Swarms for Open-Ended Discovery in Fundamental Physics ---
Enhancing Fractional Chern Insulator Gaps Beyond Ideal Quantum Geometry}
\date{\today}
\maketitle

\vspace{-2\baselineskip}
\tableofcontents
\clearpage

A flat band can change the interaction between electrons through its orbital wavefunctions, even
when the bare repulsion is the same~\cite{wang2021exact}. We study different ways of deforming the orbital wavefunctions away from the
lowest Landau level (LLL). In a topological flat band with ideal
quantum geometry, every LLL orbital is multiplied by the same momentum-independent
envelope $h^{(0)}(\mathbf r)$. The modified Laughlin states remain zero modes
of the leading short-range interaction because their wavefunctions vanish
cubically as two electrons approach one another. The leading short-range
two-electron interacting energy penalty is renormalized by $|h^{(0)}|^4$.
With nearly momentum-independent orbital normalization and negligible
nonuniform two-electron center corrections, this gives an approximate enhancement
of the leading short-range gap by
$M_2^{(0)}=\langle|h^{(0)}|^4\rangle_{\rm uc}$, with
$\langle|h^{(0)}|^2\rangle_{\rm uc}=1$. This mechanism within ideal quantum geometry
was studied in Ref.~\cite{Sarkar2026GLL}.

We extend this construction to a broader family of $C=1$ topological flat bands by allowing the
envelope $h_{\mathbf k}(\mathbf r)$ to vary across the Brillouin zone.
In the exponential family $h_{\mathbf k}=h^{(0)}e^{\gamma(\mathbf k,\mathbf r)}$,
$\gamma=0$ denotes the band with ideal quantum geometry, while
$\gamma\ne0$ allows departures from ideal quantum geometry.
Smooth deformations with nonvanishing band norms preserve the Chern
number, while the quantum geometry can deviate from the ideal trace
condition. In our calculations for these deformed bands, the gap-enhancement ratio relative
to the LLL can be much larger than the envelope intensity moment
$M_2(\mathbf k)$ defined in Sec.~\ref{sec:band_density_diagnostics}. This indicates that
density concentration alone cannot solely account for the observed enhancement.
The momentum dependence changes both the strength of a two-electron energy penalty and the linear combinations of antisymmetrized two-particle
orbitals that the interaction penalizes. This model lets us
study gap enhancement in topological flat bands beyond ideal quantum geometry.

To study this enhancement in a more controlled setting, we consider a
shear model, which gives a distinct family of topological flat bands beyond
ideal quantum geometry. Applying a common area-preserving
coordinate shear to all LLL orbitals produces $C=1$ Chern bands with
nonideal quantum geometry. In the leading short-range interaction, this
construction preserves the Laughlin zero modes while increasing the
interaction cost of two-electron excitations.

The Supplemental Material is organized as follows.
Section~\ref{sec:setting} defines the topological flat bands, their Fourier
parametrizations, projected interactions, density diagnostics, and quantum
geometry. Section~\ref{sec:shear} gives the shear construction on a torus,
its form factors, and its topology and geometry.
Section~\ref{sec:pair_mechanism} brings the $\gamma=0$ bands,
their momentum-dependent deformations, and the shear models into a common discussion of two-electron interactions and gap
enhancement, and relates their two-electron wavefunction slopes to correlations and the local
real-space quantum metric.
Section~\ref{sec:engine} describes the GPU exact-diagonalization engine and the two-particle spectra,
Section~\ref{sec:agent_dynamics} the search history and behaviour of the agents, and
Section~\ref{sec:si_prompts} reproduces the prompts given to the agents.

\section{Topological flat bands}
\label{sec:setting}
\label{sec:sm_conventions}
\label{sec:momentum_gll}
We first define the normalized wavefunctions for the band with ideal
quantum geometry ($\gamma=0$), and then introduce momentum-dependent
envelope deformations ($\gamma\ne0$) that allow departures from this limit.
We introduce symmetry-preserving Fourier parametrizations for the exponential
family of topological flat bands.
The exact density matrix elements (form factors) of these states give the
projected many-body Hamiltonian. We then show some envelope and filled-band density 
diagnostics extended from the $\gamma=0$ band. Finally, we demonstrate how the envelope preserves
the Chern number while changing the quantum geometry.

\subsection{Conventions and bands with ideal quantum geometry}
Choose positively oriented primitive lattice vectors $\mathbf a_1,\mathbf a_2$
and define the reciprocal vectors $\mathbf b_i$ by
$\mathbf a_i\cdot\mathbf b_j=2\pi\delta_{ij}$. With one flux
quantum per cell, the magnetic length $\ell_B$, cell area $A_{\rm uc}$,
and Brillouin-zone area $A_{\rm BZ}$ satisfy
\begin{equation}
 A_{\rm uc}=\mathbf a_1\times\mathbf a_2=2\pi\ell_B^2,
 \qquad
 A_{\rm BZ}=\frac{(2\pi)^2}{A_{\rm uc}}=\frac{2\pi}{\ell_B^2}.
 \label{eq:sm_areas}
\end{equation}
The total area of a system with $N_{\rm uc}$ cells is $A=N_{\rm uc}A_{\rm uc}$.
The full Bloch orbital $\psi_{\mathbf k}$ is normalized to one over this
area. The corresponding Bloch amplitude, normalized over one unit cell, is
\begin{equation}
 u_{\mathbf k}(\mathbf r)=\sqrt{N_{\rm uc}}\,
 e^{-i\mathbf k\cdot\mathbf r}\psi_{\mathbf k}(\mathbf r),
 \qquad
 \int_{\rm uc}d^2r\,|u_{\mathbf k}(\mathbf r)|^2=1.
 \label{eq:sm_cell_normalization}
\end{equation}
We define the averages in the unit cell and Brillouin zone by
$\langle X\rangle_{\rm uc}=A_{\rm uc}^{-1}\int_{\rm uc}X\,d^2r$ and
$\langle X\rangle_{\rm BZ}=A_{\rm BZ}^{-1}\int_{\rm BZ}X\,d^2k$, respectively.
For magnetic Bloch functions, $u_{\mathbf k}$ obeys magnetic boundary
conditions. These conditions are
independent of $\mathbf k$ in a fixed gauge. Momentum derivatives act on
the Bloch amplitudes $u_{\mathbf k}$.

We define the Berry connection,
curvature, and Chern number by
\begin{equation}
 \mathcal A_i(\mathbf k)=-i\langle u_{\mathbf k}|\partial_i u_{\mathbf k}\rangle,
 \qquad
 F_{xy}=\partial_x\mathcal A_y-\partial_y\mathcal A_x,
 \qquad
 C=\frac{1}{2\pi}\int_{\rm BZ}d^2k\,F_{xy}.
 \label{eq:sm_berry_convention}
\end{equation}
The derivative $\partial_i=\partial/\partial k_i$ is taken in Cartesian
coordinates, and the LLL has $F^{\rm LLL}_{xy}=\ell_B^2$.

Let $\varphi_{n\mathbf k}$ denote an orbital of the $n$th
ordinary LL, with Bloch amplitude
$u^{\rm LL}_{n\mathbf k}=\sqrt{N_{\rm uc}}e^{-i\mathbf k\cdot\mathbf r}
\varphi_{n\mathbf k}$. For $n=0$, we write the LLL orbital as $\varphi_{\mathbf k}^\text{LLL}$ and its
Bloch amplitude as $u^{\rm LLL}_{\mathbf k}$. Multiplying the LLL orbital by an envelope and
normalizing the result gives the $\gamma=0$ wavefunction~\cite{wang2021exact}, with normalization factor $S_0(\mathbf k)$:
\begin{equation}
 \psi^{(0)}_{\mathbf k}(\mathbf r)
 =\frac{h^{(0)}(\mathbf r)\varphi_{\mathbf k}^\text{LLL}(\mathbf r)}{\sqrt{S_0(\mathbf k)}},
 \qquad
 S_0(\mathbf k)
 =\langle u^{\rm LLL}_{\mathbf k}|\,|h^{(0)}|^2\,|u^{\rm LLL}_{\mathbf k}\rangle.
 \label{eq:sm_static_band}
\end{equation}
The superscript $(0)$ labels the band with ideal quantum geometry
($\gamma=0$). The
envelope $h^{(0)}$ is the same at every momentum. Its modulus squared is periodic,
\begin{equation}
 \rho^{(0)}(\mathbf r)\equiv |h^{(0)}(\mathbf r)|^2
 =\sum_{\mathbf G}w^{(0)}_{\mathbf G}e^{i\mathbf G\cdot\mathbf r},
 \qquad
 w^{(0)}_{-\mathbf G}=w^{(0)*}_{\mathbf G}.
 \label{eq:sm_static_density}
\end{equation}
and we fix its overall scale by setting $w^{(0)}_{\mathbf0}=1$.
Positivity of $\rho^{(0)}$ constrains the allowed
Fourier coefficients $w^{(0)}_{\mathbf G}$.

The $\gamma=0$ density moment
\begin{equation}
 M_2^{(0)}=\frac{\langle(\rho^{(0)})^2\rangle_{\rm uc}}{\langle\rho^{(0)}\rangle_{\rm uc}^2}
 =\frac{\langle|h^{(0)}|^4\rangle_{\rm uc}}{\langle|h^{(0)}|^2\rangle_{\rm uc}^2}
 =\sum_{\mathbf G}|w^{(0)}_{\mathbf G}|^2.
 \label{eq:sm_static_m2}
\end{equation}
measures how strongly the envelope intensity is concentrated within a
cell and equals one for a uniform envelope~\cite{Sarkar2026GLL}. It approximately rescales the leading short-range spectrum when
orbital normalization is nearly $k$-independent and nonuniform
two-electron center (COM) contributions after projection are small
(Sec.~\ref{sec:ge_static}).

We characterize the band geometry through the quantum geometric
tensor~\cite{provost1980riemannian,roy2014band}
\begin{equation}
\mathcal Q_{ij}(\mathbf k)=\langle\partial_i u_{\mathbf k}|
(1-P_{\mathbf k})|\partial_j u_{\mathbf k}\rangle,\qquad g_{ij}=\operatorname{Re}\mathcal Q_{ij},\quad
F_{xy}=2\operatorname{Im}\mathcal Q_{xy},
\label{eq:sm_qgt}
\end{equation}
where $P_{\mathbf k}=|u_{\mathbf k}\rangle\langle u_{\mathbf k}|$.
The quantum metric $g_{ij}$ measures how rapidly the Bloch state
changes with momentum, while $F_{xy}$ describes the corresponding
Berry curvature. The $\gamma=0$ band retains the ideal trace condition
$\operatorname{tr}g^{(0)}(\mathbf k)=F_{xy}^{(0)}(\mathbf k)$~\cite{roy2014band,ledwith2023vortexability},
where $g^{(0)}$ is its quantum metric. Its Berry curvature is
\begin{equation}
F_{xy}^{(0)}(\mathbf k)
=\ell_B^2+\frac12\nabla_{\mathbf k}^2
\log S_0(\mathbf k).
\label{eq:sm_static_curvature}
\end{equation}
The momentum dependence of the normalization can therefore produce
nonuniform Berry curvature while preserving the ideal trace condition.

\subsection{Momentum-dependent envelopes and Fourier parametrizations}
\label{sec:sm_kdependent}
\label{sec:sm_parametrization}
\label{sec:numerics}

In the construction at $\gamma=0$, a single spatial envelope $h^{(0)}(\mathbf r)$
multiplies every lowest-Landau-level (LLL) orbital. To represent a broader family of $C=1$ Chern bands, we now allow this
envelope to change with crystal momentum $\mathbf k$. This adds two kinds
of freedom: the electron density can move within a unit cell as $\mathbf k$
changes, and the wavefunction can acquire a momentum-dependent spatial
phase. We first give the general scalar envelope within this construction
and then introduce the smaller family used in our calculations.

Write $h_{\mathbf k}(\mathbf r)=h^{(0)}(\mathbf r)g_{\mathbf k}(\mathbf r)$,
where $g_{\mathbf k}$ is smooth and periodic in both real space and the
Brillouin zone. Because both variables are periodic, its general Fourier
expansion is
\begin{equation}
 h_{\mathbf k}(\mathbf r)=h^{(0)}(\mathbf r)\sum_{\mathbf G}c_{\mathbf G}(\mathbf k)
      e^{i\mathbf G\cdot\mathbf r}=h^{(0)}(\mathbf r)\sum_{\mathbf G,\mathbf R}
      c_{\mathbf G,\mathbf R}
      e^{i\mathbf G\cdot\mathbf r+i\mathbf k\cdot\mathbf R},\qquad
 c_{\mathbf G}(\mathbf k)=\sum_{\mathbf R}c_{\mathbf G,\mathbf R}
      e^{i\mathbf k\cdot\mathbf R}.
\label{eq:sm_linear_envelope}
\end{equation}
Here $\mathbf G$ and $\mathbf R$ are reciprocal and real space lattice
vectors, which resolve the spatial structure within a unit cell and
describe how that structure varies across the Brillouin zone, respectively.
The coefficients $c_{\mathbf G,\mathbf R}$ can be complex. With all
Fourier components included, Eq.~\eqref{eq:sm_linear_envelope} represents
any sufficiently regular periodic scalar multiplier; keeping finitely
many components gives a numerical approximation. The periodicity
conditions are
\begin{equation}
 g_{\mathbf k}(\mathbf r+\mathbf a_i)=g_{\mathbf k}(\mathbf r),
 \qquad
 g_{\mathbf k+\mathbf G_i}(\mathbf r)=g_{\mathbf k}(\mathbf r),
\label{eq:sm_multiplier_periodicity}
\end{equation}
where $\mathbf a_i$ and $\mathbf G_i$ are primitive real space and
reciprocal lattice vectors.

The corresponding normalized topological flat-band states are
\begin{equation}
\begin{aligned}
 \psi_{\mathbf k}(\mathbf r)
 &=\frac{h_{\mathbf k}(\mathbf r)\varphi_{\mathbf k}^\text{LLL}(\mathbf r)}
        {\sqrt{S(\mathbf k)}},\\
 |u_{\mathbf k}\rangle
 &=\frac{h_{\mathbf k}|u^{\rm LLL}_{\mathbf k}\rangle}
        {\sqrt{S(\mathbf k)}},\\
 S(\mathbf k)
 &=\langle u^{\rm LLL}_{\mathbf k}|\,|h_{\mathbf k}|^2\,
       |u^{\rm LLL}_{\mathbf k}\rangle.
\end{aligned}
\label{eq:sm_kdependent_band}
\end{equation}
Here $\varphi_{\mathbf k}^\text{LLL}$ is the LLL orbital and
$|u^{\rm LLL}_{\mathbf k}\rangle$ its Bloch amplitude. The inner
product is over one unit cell. We require $S(\mathbf k)>0$ throughout
the Brillouin zone so that the state is well defined. Since the
multiplier has the periodicities in
Eq.~\eqref{eq:sm_multiplier_periodicity}, the normalized states retain
the boundary phases of the LLL and hence its Chern number
$C=1$, no matter whether their quantum geometry remains ideal
(Sec.~\ref{sec:sm_geometry}). Thus the Fourier expansion changes the
wavefunctions within this $C=1$ family without changing their topology.

The unrestricted coefficients in Eq.~\eqref{eq:sm_linear_envelope}
provide more freedom than is needed to examine the effects of amplitude
and phase separately. They also allow zeros of $g_{\mathbf k}$, which
make a continuous spatial phase difficult to define. We therefore use
a more restricted family of envelope function,
\begin{equation}
 h_{\mathbf k}(\mathbf r)
 =h^{(0)}(\mathbf r)e^{\gamma(\mathbf k,\mathbf r)},
 \qquad
 \gamma(\mathbf k,\mathbf r)
 =\frac12a(\mathbf k,\mathbf r)+i\theta(\mathbf k,\mathbf r),
\label{eq:sm_log_envelope}
\end{equation}
with smooth, real, periodic fields $a$ and $\theta$. This choice makes
the two changes explicit:
$|h_{\mathbf k}(\mathbf r)|^2=|h^{(0)}(\mathbf r)|^2e^{a(\mathbf k,\mathbf r)}$
while $\theta$ supplies the additional spatial phase. Setting $a=0$
gives a phase-only deformation, which leaves the local density weight
and $S(\mathbf k)$ unchanged; setting $\theta=0$ gives an amplitude-only
deformation. An exponential also keeps the multiplier nonzero so that the zeros
of $\psi_{\mathbf k}$ are unchanged. The positivity of $S(\mathbf k)$
is guaranteed by $e^{\min_{\mathbf r}a(\mathbf k,\mathbf r)}S_0(\mathbf k)
 \leq S(\mathbf k)\leq e^{\max_{\mathbf r}a(\mathbf k,\mathbf r)}S_0(\mathbf k).$

We expand this logarithm in the same real- and momentum-space Fourier
basis,
\begin{equation}
 \gamma(\mathbf k,\mathbf r)=\sum_{\mathbf G,\mathbf R}\gamma_{\mathbf G,\mathbf R}
    e^{i\mathbf G\cdot\mathbf r+i\mathbf k\cdot\mathbf R},\qquad \gamma_{\mathbf G,\mathbf R}=\frac12 a_{\mathbf G,\mathbf R}
+i\theta_{\mathbf G,\mathbf R}
\label{eq:sm_double_fourier}
\end{equation}
Because $a$ and $\theta$ are real,
their Fourier coefficients separately obey
$a_{-\mathbf G,-\mathbf R}=a_{\mathbf G,\mathbf R}^{*}$ and
$\theta_{-\mathbf G,-\mathbf R}=\theta_{\mathbf G,\mathbf R}^{*}$.
A term with $\mathbf G=0$ is spatially uniform and changes only the
normalization or overall phase of a given orbital. A term with
$\mathbf R=0$ is independent of momentum and can instead be included
in the envelope $h^{(0)}$ of the $\gamma=0$ state. We fix these choices and retain
$\mathbf G\neq0$ and $\mathbf R\neq0$ for the new deformation.

Lattice rotation symmetry reduces the number of remaining coefficients.
For a proper rotation $O$ that is also a symmetry of the $\gamma=0$
state, we require
\begin{equation}
 \gamma(O\mathbf k,O\mathbf r)=\gamma(\mathbf k,\mathbf r),
 \qquad
 \gamma_{O\mathbf G,O\mathbf R}
   =\gamma_{\mathbf G,\mathbf R}.
\label{eq:sm_joint_symmetry}
\end{equation}
The rotation acts on $\mathbf G$ and $\mathbf R$ together. Thus a
rotation-symmetric deformation can still distinguish different
relative directions between the spatial and momentum harmonics.
For sixfold rotation, let $\mu$ label the six-element orbit generated
from one nonzero pair $(\mathbf G_\mu,\mathbf R_\mu)$. Its real,
rotation-invariant basis function is
\begin{equation}
 \mathcal O_\mu(\mathbf k,\mathbf r)
 =\frac{1}{\sqrt6}\sum_{j=0}^{5}
 e^{i(O^j\mathbf G_\mu)\cdot\mathbf r
   +i\mathbf k\cdot(O^j\mathbf R_\mu)}.
\label{eq:sm_orbit_function}
\end{equation}
The terms at $j$ and $j+3$ are complex conjugates, so
$\mathcal O_\mu$ is real. Writing
$\gamma=\sum_\mu(\alpha_\mu/2+i\beta_\mu)\mathcal O_\mu$
then gives independent real parameters $\alpha_\mu$ and $\beta_\mu$
for amplitude and phase on each joint orbit.

For the calculations in this work, we use a smaller set of
rotation-symmetric functions that separates the spatial pattern from
its momentum dependence. Let $\mathcal S_s$ be an inversion-symmetric
set of nonzero reciprocal vectors related by lattice symmetries, and
$\mathcal T_p$ a corresponding set of real-space lattice vectors. Define
\begin{equation}
 O_s(\mathbf r)
 =\frac{1}{\sqrt{|\mathcal S_s|}}
     \sum_{\mathbf G\in\mathcal S_s}e^{i\mathbf G\cdot\mathbf r},
 \qquad
 F_p(\mathbf k)
 =\frac{1}{\sqrt{|\mathcal T_p|}}
     \sum_{\mathbf R\in\mathcal T_p}e^{i\mathbf k\cdot\mathbf R}.
\label{eq:sm_star_functions}
\end{equation}
Here $s$ and $p$ label the spatial and momentum sets, and the bars
count the vectors in each set. These functions are real, have zero
mean, and have unit mean square over their respective unit cells.
Using their products, we take
\begin{equation}
 h_{\mathbf k}(\mathbf r)
 =h^{(0)}(\mathbf r)\exp\!\left[
 \sum_{s,p}\left(\frac{\alpha_{sp}}2+i\beta_{sp}\right)
 O_s(\mathbf r)F_p(\mathbf k)
 \right],
\label{eq:sm_separable_envelope}
\end{equation}
with real $\alpha_{sp}$ and $\beta_{sp}$. For each pair $(s,p)$,
$O_s(\mathbf r)$ fixes the spatial pattern and $F_p(\mathbf k)$ controls
how it varies across the Brillouin zone. The coefficient $\alpha_{sp}$
changes the envelope intensity, and $\beta_{sp}$ changes its spatial
phase. For a fixed pair of sets $(s,p)$, the product $O_sF_p$ includes
every $(\mathbf G,\mathbf R)\in\mathcal S_s\times\mathcal T_p$ with equal
Fourier weight. Different pairs of sets have independent amplitude and
phase coefficients $\alpha_{sp}$ and $\beta_{sp}$. This preserves sixfold rotation symmetry and uses two
real parameters per pair of sets. 

For example, choose the three spatial sets
$\mathcal S_1,\mathcal S_2,\mathcal S_3$ and one momentum set
$\mathcal T_1$, with $|\mathcal S_s|=|\mathcal T_1|=6$ for $s=1,2,3$.
The three products $O_1F_1$, $O_2F_1$, and $O_3F_1$ have amplitude
coefficients $(\alpha_{11},\alpha_{21},\alpha_{31})$ and phase
coefficients $(\beta_{11},\beta_{21},\beta_{31})$, giving six real
parameters. A general
simultaneous-$C_6$ expansion on the same Fourier support, using the
basis functions in Eq.~\eqref{eq:sm_orbit_function}, contains
18 independent complex orbit coefficients (36 real
parameters). The restricted family retains momentum-dependent
intracell structure while keeping the comparison between amplitude-
only, phase-only, and combined deformations transparent.

\subsection{The projected interaction}
\label{sec:sm_form_factors}

The projected electron interaction depends on the overlap between Bloch
wavefunctions at different momenta. We first recall how this overlap
is calculated in the $\gamma=0$ band~\cite{wang2021exact,Sarkar2026GLL}, where a single set of Fourier
coefficients $w_{\mathbf G}$ completely describes the effect of the
spatial envelope. We then extend the same calculation to the
deformed topological flat-band model.

In the $\gamma=0$ band, every LLL orbital is multiplied by the same
envelope $h^{(0)}(\mathbf r)$. Its spatial modulation is described by
\begin{equation}
 |h^{(0)}(\mathbf r)|^2
 =\sum_{\mathbf G}w_{\mathbf G}e^{i\mathbf G\cdot\mathbf r},
 \qquad
 w_{\mathbf G}
 =\left\langle |h^{(0)}(\mathbf r)|^2
 e^{-i\mathbf G\cdot\mathbf r}\right\rangle_{\rm uc},
\end{equation}
where $\langle\cdots\rangle_{\rm uc}$ denotes the average over one
unit cell. The Fourier coefficients satisfy
$w_{-\mathbf G}=w_{\mathbf G}^*$ because the density is real. These coefficients enter the interaction through the form factor
\begin{equation}
 \lambda_{\mathbf q}(\mathbf k,\mathbf k')
 =\langle\psi_{\mathbf k}|
 e^{-i\mathbf q\cdot\mathbf r}
 |\psi_{\mathbf k'}\rangle.
 \label{eq:sm_form_factor_definition}
\end{equation}
It measures the matrix element of a density fluctuation carrying
momentum $\mathbf q$ between two normalized Bloch states.
Crystal-momentum conservation requires
\begin{equation}
 \mathbf q=\mathbf k'-\mathbf k+\mathbf G,
 \label{eq:sm_transfer}
\end{equation}
where $\mathbf G$ is a reciprocal-lattice vector. The physical
momentum transfer $\mathbf q$ can extend beyond the first Brillouin
zone even though $\mathbf k$ and $\mathbf k'$ are chosen within it.

Let $\lambda_{\mathbf q}^{\rm LLL}(\mathbf k,\mathbf k')$ denote
the corresponding form factor between LLL orbitals.
Substituting the momentum-independent envelope into
Eq.~\eqref{eq:sm_form_factor_definition} gives
\begin{equation}
 \lambda_{\mathbf q}^{(0)}(\mathbf k,\mathbf k')
 =\frac{1}{\sqrt{S_0(\mathbf k)S_0(\mathbf k')}}
 \sum_{\mathbf G}w_{\mathbf G}
 \lambda_{\mathbf q-\mathbf G}^{\rm LLL}
 (\mathbf k,\mathbf k'),
 \label{eq:sm_static_convolution}
\end{equation}
where
$S_0(\mathbf k)$ normalizes the orbital $\psi^{(0)}_{\mathbf k}$.
The coefficients $w_{\mathbf G}$ are independent of $\mathbf k$ and
$\mathbf k'$ because $h^{(0)}(\mathbf r)$ is momentum independent.

For completeness, in the magnetic Bloch convention of
Ref.~\cite{Sarkar2026GLL}, the LLL form factor appearing above
is
\begin{equation}
 \lambda_{\mathbf q}^{\rm LLL}(\mathbf k,\mathbf k')
 =\eta_{\mathbf G}\exp\bigg[-\frac{\ell_B^2|\mathbf q|^2}{4}
 +\frac{i\ell_B^2}{2}\mathbf k\times\mathbf k'+\frac{i\ell_B^2}{2}
 (\mathbf k+\mathbf k')\times\mathbf G
 \bigg],
\label{eq:sm_ll_form_factor}
\end{equation}
where $\mathbf q=\mathbf k'-\mathbf k+\mathbf G$ and
$\ell_B$ is the magnetic length. For
$\mathbf G=n_1\mathbf b_1+n_2\mathbf b_2$, with integer
$n_1,n_2$,
$\eta_{\mathbf G}=(-1)^{n_1+n_2+n_1n_2}$. The normalization $S_0(\mathbf k)$ can also be obtained from this LLL form
factor. Setting $\mathbf q=0$ and $\mathbf k'=\mathbf k$ in the numerator
of Eq.~\eqref{eq:sm_static_convolution} gives
\begin{equation}
 S_0(\mathbf k)=
 \sum_{\mathbf G}w_{\mathbf G}
 \eta_{\mathbf G}
 \exp\left[
 i\ell_B^2\mathbf G\times\mathbf k
 -\frac{\ell_B^2|\mathbf G|^2}{4}
 \right].
 \label{eq:sm_static_norm}
\end{equation}
The Gaussian factor in Eq.~\eqref{eq:sm_static_norm}
suppresses the contribution of large reciprocal vectors to the
normalization. These same harmonics can still contribute strongly to
the form factors, because the physical transfer $\mathbf q$ can approach
$\mathbf G$.

We now turn to the projected interaction of the deformed
topological flat-band model, where different Bloch orbitals have different spatial envelopes
$h_{\mathbf k}(\mathbf r)$. When calculating their overlap, the
spatial factor is now
$h_{\mathbf k}^*(\mathbf r)h_{\mathbf k'}(\mathbf r)$
rather than $|h^{(0)}(\mathbf r)|^2$. To describe this overlap, we
introduce the full Fourier coefficients
\begin{equation}
 w_{\mathbf G}(\mathbf k,\mathbf k')=\left\langle
 h_{\mathbf k}^*(\mathbf r)h_{\mathbf k'}(\mathbf r)
 e^{-i\mathbf G\cdot\mathbf r}
 \right\rangle_{\rm uc},\qquad
 h_{\mathbf k}^*(\mathbf r)h_{\mathbf k'}(\mathbf r)
 =\sum_{\mathbf G}w_{\mathbf G}(\mathbf k,\mathbf k')
 e^{i\mathbf G\cdot\mathbf r}.
\label{eq:sm_cross_kernel}
\end{equation}
These coefficients describe the spatial overlap between two
envelopes. Their product is periodic in position because all
$h_{\mathbf k}(\mathbf r)$ share the same spatial boundary conditions.

Equation~\eqref{eq:sm_cross_kernel} includes the $\gamma=0$ band as
a special case. If $h_{\mathbf k}=h^{(0)}$ for every momentum, then
$w_{\mathbf G}(\mathbf k,\mathbf k')=w_{\mathbf G}$.
For a momentum-dependent envelope, the diagonal coefficients
$w_{\mathbf G}(\mathbf k,\mathbf k)$ describe the spatial
intensity of an individual orbital, while the off-diagonal
coefficients contain the relative spatial phase between two
orbitals.

This distinction is particularly clear for the exponential
parametrization introduced in Eq.~\eqref{eq:sm_log_envelope},
\begin{equation}
 h_{\mathbf k}(\mathbf r)
 =h^{(0)}(\mathbf r)
 e^{a(\mathbf k,\mathbf r)/2+i\theta(\mathbf k,\mathbf r)}.
\end{equation}
Substitution into Eq.~\eqref{eq:sm_cross_kernel} gives
\begin{equation}
 w_{\mathbf G}(\mathbf k,\mathbf k')
 =\Big\langle |h^{(0)}(\mathbf r)|^2
 e^{[a(\mathbf k,\mathbf r)+a(\mathbf k',\mathbf r)]/2}\times e^{i[\theta(\mathbf k',\mathbf r)
                  -\theta(\mathbf k,\mathbf r)]}
 e^{-i\mathbf G\cdot\mathbf r}
 \Big\rangle_{\rm uc}.
\label{eq:sm_cross_amplitude_phase}
\end{equation}
The amplitude fields determine how strongly the two orbitals
overlap at each position, while the spatial variations of the phase fields difference
can change the matrix element for electron scattering. For example, in a phase-only
deformation, $a=0$, we have
$w_{\mathbf G}(\mathbf k,\mathbf k)=w_{\mathbf G}$ for every
$\mathbf k$. Thus every orbital retains the same spatial density
as the $\gamma=0$ state. Nevertheless,
$w_{\mathbf G}(\mathbf k,\mathbf k')$ generally changes when
$\mathbf k\neq\mathbf k'$, because the relative spatial phase is
different. A momentum-dependent phase can therefore modify the
projected interaction without changing the density.

We now derive the form factor for the topological flat bands. Using the normalized
topological flat-band wavefunction defined in
Eq.~\eqref{eq:sm_kdependent_band}, we obtain
\begin{equation}
 \lambda_{\mathbf q}(\mathbf k,\mathbf k')=\frac{1}{\sqrt{S(\mathbf k)S(\mathbf k')}}
 \big\langle\varphi_{\mathbf k}^\text{LLL}\big|
 h_{\mathbf k}^*e^{-i\mathbf q\cdot\mathbf r}
 h_{\mathbf k'}
 \big|\varphi_{\mathbf k'}^\text{LLL}\big\rangle=\frac{\displaystyle\sum_{\mathbf G}
 w_{\mathbf G}(\mathbf k,\mathbf k')
 \lambda_{\mathbf q-\mathbf G}^{\rm LLL}
 (\mathbf k,\mathbf k')}
 {\sqrt{S(\mathbf k)S(\mathbf k')}}.
\label{eq:sm_exact_convolution}
\end{equation}
This is the exact form factor of the topological flat-band
construction. Its structure is the same as that of the $\gamma=0$ band in Eq.~\eqref{eq:sm_static_convolution}, except that each
Fourier coefficient now depends on the initial and final Bloch
momenta. The normalization $S(\mathbf k)$ is obtained by setting $\mathbf q=0$
and $\mathbf k'=\mathbf k$ in the numerator of
Eq.~\eqref{eq:sm_exact_convolution}. It has the same form as the $S_0(\mathbf k)$
normalization in Eq.~\eqref{eq:sm_static_norm}, with
$S_0(\mathbf k)\to S(\mathbf k)$ and
$w_{\mathbf G}\to w_{\mathbf G}(\mathbf k,\mathbf k)$.
Thus only the diagonal intensity coefficients enter the normalization.

Having obtained the form factors, we can now construct the
projected interaction. We use the dual-gate-screened Coulomb potential~\cite{Sarkar2026GLL}
\begin{equation}
 V(q)=\frac{2\pi e^2}{\varepsilon q}
 \tanh(qd_s),
 \qquad
 E_C=\frac{e^2}{\varepsilon\ell_B},
 \label{eq:sm_interaction}
\end{equation}
where $q=|\mathbf q|$, $\varepsilon$ is the dielectric
constant, and $d_s$ is the screening length. The Coulomb scale
$E_C$ provides a common energy unit for comparing the deformed topological
flat bands with their $\gamma=0$ bands and the LLL.
Projecting this interaction onto the selected flat band, the density
operator becomes
\begin{equation}
 \widehat\rho_{\mathbf q}
 =\sum_{\mathbf k,\mathbf k'}
 \lambda_{\mathbf q}(\mathbf k,\mathbf k')
 c_{\mathbf k}^\dagger c_{\mathbf k'},
 \qquad
 \widehat\rho_{\mathbf q}^\dagger
 =\widehat\rho_{-\mathbf q},
 \label{eq:sm_projected_density}
\end{equation}
where $c_{\mathbf k}^\dagger$ creates an electron in the
normalized Bloch state at momentum $\mathbf k$. The projected two-body Hamiltonian is then
\begin{equation}
\begin{aligned}
 H_{\rm int}
 &=\frac{1}{2A}\sum_{\mathbf q}V(q)
 :\widehat\rho_{\mathbf q}\widehat\rho_{-\mathbf q}:\\
 &=\frac{1}{2A}
 \sum_{\mathbf q}
 \sum_{\mathbf k_1,\mathbf k_2,\mathbf k_3,\mathbf k_4}
 V(q)\,
 \lambda_{\mathbf q}(\mathbf k_1,\mathbf k_3)
 \lambda_{-\mathbf q}(\mathbf k_2,\mathbf k_4)\times
 c_{\mathbf k_1}^\dagger c_{\mathbf k_2}^\dagger
 c_{\mathbf k_4}c_{\mathbf k_3},
\end{aligned}
\label{eq:sm_projected_hamiltonian}
\end{equation}
where $A$ is the total system area and the colons denote normal
ordering. The kinetic energy is zero since we take the selected band to be exactly flat.

In numerical ED calculations, we construct the envelope first
and then evaluate the corresponding Fourier coefficients and
form factors. Products of spatial harmonics generate
contributions at their difference wavevectors, while the
exponential parametrization generally produces additional
higher harmonics. We therefore check convergence separately
with the envelope Fourier cutoff, the real-space integration
grid, and the physical momentum-transfer cutoff.

\subsection{Envelope intensity and filled-band diagnostics}
\label{sec:band_density_diagnostics}

In the $\gamma=0$ band, the moment $M_2^{(0)}$ measures how strongly
its common envelope is concentrated within a unit cell. When the
orbital normalization $S_0(\mathbf k)\simeq1$, the filled-band density
follows approximately the same spatial profile, so $M_2^{(0)}$ also
measures the concentration of the physical electron density. This
provides a natural starting point for the topological flat-band model, where the
envelope depends on momentum. We consider two extensions: the
Brillouin-zone average $\langle M_2(\mathbf k)\rangle_{\rm BZ}$,
which measures the typical concentration of an individual envelope,
and $M_2^{\rm fill}$, which measures the density modulation of the filled band.

We first make the connection in the $\gamma=0$ band explicit. With the
convention $w^{(0)}_{\mathbf0}=1$ introduced in
Eq.~\eqref{eq:sm_static_density}, the common envelope intensity is
$\rho^{(0)}(\mathbf r)=|h^{(0)}(\mathbf r)|^2$, and its moment is
$M_2^{(0)}=\langle(\rho^{(0)})^2\rangle_{\rm uc}$. The physical density
also contains the normalized LLL orbitals. Filling the $\gamma=0$ band
gives
\begin{equation}
 A_{\rm uc}n_{\rm fill}^{(0)}(\mathbf r)
 =\rho^{(0)}(\mathbf r)\frac{A_{\rm uc}}{N_{\rm uc}}
 \sum_{\mathbf k}\frac{|u^{\rm LLL}_{\mathbf k}(\mathbf r)|^2}
 {S_0(\mathbf k)}.
 \label{eq:filled_static_exact}
\end{equation}
Here $N_{\rm uc}$ is the number of unit cells, and
$\langle n_{\rm fill}^{(0)}\rangle_{\rm uc}=1/A_{\rm uc}$.
For any filled band, we use the dimensionless second moment
$M_2^{\rm fill}\equiv A_{\rm uc}^2
\langle n_{\rm fill}^2\rangle_{\rm uc}$ to measure its density
modulation. It equals one for a uniform density. If the normalization
satisfies the additional condition $S_0(\mathbf k)\simeq1$
throughout the Brillouin zone, the denominators in
Eq.~\eqref{eq:filled_static_exact} can be approximated by one.
Using the uniform density of the filled LLL then gives
\begin{equation}
 A_{\rm uc}n_{\rm fill}^{(0)}(\mathbf r)\simeq\rho^{(0)}(\mathbf r),
 \qquad M_2^{\rm fill}\simeq M_2^{(0)}.
 \label{eq:filled_static_approx}
\end{equation}
Thus $M_2^{(0)}$ characterizes both the envelope and, approximately,
the filled-band density when normalization variations (lower harmonics) are small.

For the topological flat-band model, different momenta have different envelopes
$h_{\mathbf k}(\mathbf r)$. We use the diagonal Fourier coefficients
$w_{\mathbf G}(\mathbf k)\equiv
w_{\mathbf G}(\mathbf k,\mathbf k)$ from
Eq.~\eqref{eq:sm_cross_kernel}.
We retain the log-Fourier
parametrization of Eq.~\eqref{eq:sm_log_envelope}; removing its spatially
uniform logarithmic component does not impose
$w_{\mathbf0}(\mathbf k)=1$. The normalized envelope intensity is
$\rho_{\mathbf k}(\mathbf r)
=|h_{\mathbf k}(\mathbf r)|^2/w_{\mathbf0}(\mathbf k)$,
with $\langle\rho_{\mathbf k}\rangle_{\rm uc}=1$, and its second moment is
$M_2(\mathbf k)=\langle\rho_{\mathbf k}^2\rangle_{\rm uc}
=1+\sum_{\mathbf G\ne\mathbf0}
\left|
\frac{w_{\mathbf G}(\mathbf k)}{w_{\mathbf0}(\mathbf k)}\right|^2$.
The excess $M_2(\mathbf k)-1
=\langle(\rho_{\mathbf k}-1)^2\rangle_{\rm uc}$ measures the
spatial variation of the envelope at momentum $\mathbf k$. Averaging
it across the Brillouin zone gives the extension of $M_2^{(0)}$ to momentum-dependent envelopes,
\begin{equation}
 \langle M_2(\mathbf k)\rangle_{\rm BZ}
 =1+\sum_{\mathbf G\ne\mathbf0}
 \left\langle\left|
 \frac{w_{\mathbf G}(\mathbf k)}{w_{\mathbf0}(\mathbf k)}
 \right|^2\right\rangle_{\rm BZ}.
 \label{eq:sm_bz_averaged_m2}
\end{equation}
On a finite momentum grid, $\langle\cdots\rangle_{\rm BZ}$ is
evaluated as $N_{\rm uc}^{-1}\sum_{\mathbf k}$. In the $\gamma=0$ limit,
$M_2(\mathbf k)=M_2^{(0)}$ for every momentum, and therefore
$\langle M_2(\mathbf k)\rangle_{\rm BZ}=M_2^{(0)}$. In general,
this average measures how concentrated the envelopes are.

To compare this measure with the actual electron density, we define
the dimensionless density of each normalized Bloch orbital,
\begin{equation}
 n_{\mathbf k}(\mathbf r)
 \equiv A_{\rm uc}|u_{\mathbf k}(\mathbf r)|^2
 =\frac{A_{\rm uc}w_{\mathbf0}(\mathbf k)}{S(\mathbf k)}\rho_{\mathbf k}(\mathbf r)
 |u^{\rm LLL}_{\mathbf k}(\mathbf r)|^2,
 \label{eq:sm_orbital_density}
\end{equation}
where $\langle n_{\mathbf k}\rangle_{\rm uc}=1$. The filled-band
density is the average over all occupied orbitals,
$A_{\rm uc}n_{\rm fill}(\mathbf r)
=N_{\rm uc}^{-1}\sum_{\mathbf k}n_{\mathbf k}(\mathbf r)$,
and its moment becomes
\begin{equation}
 M_2^{\rm fill}
 =\frac{1}{N_{\rm uc}^2}\sum_{\mathbf k,\mathbf k'}
 \langle n_{\mathbf k}n_{\mathbf k'}\rangle_{\rm uc}.
 \label{eq:filled_moment_cross}
\end{equation}
The two moments describe different aspects of the same band.
$\langle M_2(\mathbf k)\rangle_{\rm BZ}$ measures the average
concentration of the individual envelopes, while $M_2^{\rm fill}$
measures how much modulation remains in the total electron density.
For example, the individual orbitals can be strongly concentrated
but peak at different positions, producing a much smoother filled-band
density. The latter also depends on the LLL orbital profiles and the
normalizations $S(\mathbf k)$. Therefore, unlike in the $\gamma=0$ case,
the two moments need not agree even when $S(\mathbf k)\simeq1$.

A phase-only deformation makes their roles particularly clear.
For $h_{\mathbf k}=h^{(0)}e^{i\theta(\mathbf k,\mathbf r)}$, the envelope
intensity and every physical orbital density are unchanged. Thus
$\langle M_2(\mathbf k)\rangle_{\rm BZ}=M_2^{(0)}$ and
$M_2^{\rm fill}$ retains its $\gamma=0$ value. The phase can nevertheless
change the form factors between different momenta and hence the
projected interaction. These density moments are useful diagnostics
of spatial concentration, while the full form factors determine how
that concentration enters the many-body problem.

\subsection{Topology and geometry of the topological flat-band model}
\label{sec:sm_geometry}

We next try to understand the topology and geometry of this topological flat band. The momentum-dependent envelope changes how the Bloch wavefunctions
vary across the Brillouin zone. As a result, it can modify both the
Berry curvature and the quantum metric. However, the Chern number
remains $C=1$, inherited from the LLL. We first establish
this topological property and then derive how the envelope changes
the quantum geometry.

\paragraph*{Chern number.}

The Chern number is determined by how Bloch wavefunctions connect
across the Brillouin zone. For a Chern band, the wavefunctions cannot
be chosen smooth and periodic everywhere in momentum space.
Instead, the Bloch amplitudes at opposite Brillouin-zone
boundaries are related by a phase. For the LLL, this
relation takes the form
\begin{equation}
u^{\rm LLL}_{\mathbf k+\mathbf G}(\mathbf r)
=e^{i\chi_{\mathbf G}(\mathbf k)}
e^{-i\mathbf G\cdot\mathbf r}
u^{\rm LLL}_{\mathbf k}(\mathbf r),
\label{eq:sm_boundary_sewing}
\end{equation}
where $\mathbf G$ is a reciprocal-lattice vector and
$\chi_{\mathbf G}(\mathbf k)$ is a real phase.

The normalized topological flat-band state is
$|u_{\mathbf k}\rangle
=h_{\mathbf k}|u^{\rm LLL}_{\mathbf k}\rangle/\sqrt{S(\mathbf k)}$,
as defined in Eq.~\eqref{eq:sm_kdependent_band}.
Since $h_{\mathbf k+\mathbf G}=h_{\mathbf k}$, its norm satisfies
$S(\mathbf k+\mathbf G)=S(\mathbf k)$. The topological flat-band state
therefore obeys
\begin{equation}
u_{\mathbf k+\mathbf G}(\mathbf r)
=e^{i\chi_{\mathbf G}(\mathbf k)}
e^{-i\mathbf G\cdot\mathbf r}
u_{\mathbf k}(\mathbf r).
\end{equation}
Thus the momentum-dependent envelope $h(\mathbf k)$ preserves the boundary
phases of the LLL. The condition $S(\mathbf k)>0$
ensures that the normalized state remains smooth and well defined
throughout the Brillouin zone.

The Berry curvature can be calculated directly from the normalized wavefunction
in Eq.~\eqref{eq:sm_kdependent_band}. Writing the unnormalized state as
$|v_{\mathbf k}\rangle
=h_{\mathbf k}|u^{\rm LLL}_{\mathbf k}\rangle$,
with $S=\langle v_{\mathbf k}|v_{\mathbf k}\rangle$, we obtain
\begin{equation}
F_{xy}(\mathbf k)
=\frac{2}{S}\operatorname{Im}\bigg[\langle\partial_xv_{\mathbf k}|
\partial_yv_{\mathbf k}\rangle-\frac{
\langle\partial_xv_{\mathbf k}|v_{\mathbf k}\rangle
\langle v_{\mathbf k}|\partial_yv_{\mathbf k}\rangle}{S}\bigg].
\label{eq:envelope_curvature}
\end{equation}
Here $\partial_i=\partial/\partial k_i$. To obtain the Chern number, we rewrite the curvature as a total
derivative. Using
$\partial_iS=2\operatorname{Re}\langle v_{\mathbf k}|
\partial_i v_{\mathbf k}\rangle$, Eq.~\eqref{eq:envelope_curvature}
becomes
\begin{equation}
F_{xy}(\mathbf k)
={}\partial_x\left[
\frac{\operatorname{Im}\langle v_{\mathbf k}|
\partial_yv_{\mathbf k}\rangle}{S}\right]-\partial_y\left[
\frac{\operatorname{Im}\langle v_{\mathbf k}|
\partial_xv_{\mathbf k}\rangle}{S}\right]
={}\partial_x\mathcal A_y-\partial_y\mathcal A_x,
\end{equation}
where
$\mathcal A_i=\operatorname{Im}\langle v_{\mathbf k}|
\partial_i v_{\mathbf k}\rangle/S$
is the Berry connection. 

As established in Eq.~\eqref{eq:sm_boundary_sewing}, the deformed
and LLL states have the same boundary phases. Their
Berry connections therefore differ by a smooth, periodic function
$\delta\mathcal A_i=\mathcal A_i-\mathcal A_i^{\rm LLL}$.
To see this directly, the unnormalized state obeys
$v_{\mathbf k+\mathbf G}
=e^{i\chi_{\mathbf G}(\mathbf k)}e^{-i\mathbf G\cdot\mathbf r}v_{\mathbf k}$.
Since $e^{-i\mathbf G\cdot\mathbf r}$ is independent of momentum and
$S(\mathbf k+\mathbf G)=S(\mathbf k)$, substitution into the connection gives
\begin{equation*}
\mathcal A_i(\mathbf k+\mathbf G)=\frac{1}{S(\mathbf k)}\operatorname{Im}\Big[
\langle v_{\mathbf k}|\partial_i v_{\mathbf k}\rangle
+i\bigl(\partial_i\chi_{\mathbf G}(\mathbf k)\bigr)
\langle v_{\mathbf k}|v_{\mathbf k}\rangle\Big]=\mathcal A_i(\mathbf k)+\partial_i\chi_{\mathbf G}(\mathbf k).
\end{equation*}
The LLL Berry connection $\mathcal A_i^{\rm LLL}$ acquires the same phase derivative, so
$\delta\mathcal A_i(\mathbf k+\mathbf G)=\delta\mathcal A_i(\mathbf k)$. Since the LLL has flat Berry curvature $F_{xy}^{\rm LLL}=\ell_B^2$, the
curvature obtained above can be written as
\begin{equation}
F_{xy}(\mathbf k)
=\ell_B^2+\partial_x\delta\mathcal A_y
-\partial_y\delta\mathcal A_x.
\end{equation}
The additional curvature integrates to zero because
$\delta\boldsymbol{\mathcal A}$ is periodic across the Brillouin
zone. Therefore,
\begin{equation}
C=\frac{1}{2\pi}\int_{\rm BZ}d^2k\,F_{xy}(\mathbf k)=\frac{\ell_B^2A_{\rm BZ}}{2\pi}
+\frac{1}{2\pi}\oint_{\partial{\rm BZ}}
\delta\boldsymbol{\mathcal A}\cdot d\mathbf k=1,
\label{eq:envelope_chern}
\end{equation}
where $A_{\rm BZ}=2\pi/\ell_B^2$ for one flux quantum per unit
cell. Thus the momentum-dependent envelope changes the local Berry
curvature while the periodicity of the envelope ensures that its total contribution vanishes.
The topological flat band therefore retains Chern number $C=1$.

\paragraph*{Quantum geometry.}

Although the Chern number remains fixed, the momentum-dependent
envelope can change the quantum metric and the distribution of Berry
curvature. 

For the exponential family considered
here, the wavefunction is
$v_{\mathbf k}=h^{(0)}e^{\gamma(\mathbf k,\mathbf r)}
u^{\rm LLL}_{\mathbf k}$, with normalized state
$u_{\mathbf k}=v_{\mathbf k}/\sqrt{S(\mathbf k)}$.
With $D_i u=(1-P_{\mathbf k})\partial_i u_{\mathbf k}$, the quantum
geometric tensor in Eq.~\eqref{eq:sm_qgt} is
$\mathcal Q_{ij}=\langle D_i u|D_j u\rangle$.
Using $\mathcal Q_{yx}=\mathcal Q_{xy}^*$,
$g_{ij}=\operatorname{Re}\mathcal Q_{ij}$, and
$F_{xy}=2\operatorname{Im}\mathcal Q_{xy}$, we expand the squared norm as
\begin{equation}
\begin{aligned}
\|D_xu+iD_yu\|^2
&=\mathcal Q_{xx}+\mathcal Q_{yy}
+i(\mathcal Q_{xy}-\mathcal Q_{yx})\\
&=g_{xx}+g_{yy}-2\operatorname{Im}\mathcal Q_{xy}
=\operatorname{tr}g-F_{xy}.
\end{aligned}
\label{eq:sm_ideality_norm}
\end{equation}
The ideal trace condition $\operatorname{tr}g=F_{xy}$
is therefore equivalent to
$(D_x+iD_y)u=0$.

To see how the envelope affects this condition, define
$\partial_+=\partial_{k_x}+i\partial_{k_y}$.
The LLL satisfies
\begin{equation}
(1-P_{\rm LLL})\partial_+
u^{\rm LLL}_{\mathbf k}=0,
\end{equation}
where $P_{\rm LLL}$ projects onto the LLL Bloch state $u^{\rm LLL}_{\mathbf k}$. Thus $\partial_+u^{\rm LLL}_{\mathbf k}$
is proportional to $u^{\rm LLL}_{\mathbf k}$.
The $\gamma=0$ state inherits this property because its envelope
$h^{(0)}(\mathbf r)$ is independent of $\mathbf k$ so that the ideal trace condition is
preserved. We next examine how a nonzero deformation $\gamma$ can produce
departure from ideal quantum geometry.

Differentiating the unnormalized topological flat-band state gives
\begin{equation}
\label{eq:partial_plus_v}
\partial_+v_{\mathbf k}
=h_{\mathbf k}\partial_+u^{\rm LLL}_{\mathbf k}
+(\partial_+\gamma)v_{\mathbf k},
\end{equation}
where $h_{\mathbf k}=h^{(0)}e^\gamma$.
The first term is parallel to $v_{\mathbf k}$, but the second term
is not because
$\partial_+\gamma$ generally depends on position. Passing to the normalized state gives
$\partial_+u_{\mathbf k}=S^{-1/2}\partial_+v_{\mathbf k}
-\tfrac12(\partial_+\log S)u_{\mathbf k}$.
The normalization term is parallel to $u_{\mathbf k}$, so applying
$1-P_{\mathbf k}$ removes it together with the first term in Eq.~\eqref{eq:partial_plus_v}, giving
\begin{equation}
D_xu+iD_yu=(1-P_{\mathbf k})[(\partial_+\gamma)u_{\mathbf k}]=\left[\partial_+\gamma
-\langle\partial_+\gamma\rangle_{\mathbf k}\right]
u_{\mathbf k}.
\label{eq:sm_projected_plus}
\end{equation}
Here we define $\langle f\rangle_{\mathbf k}
=\int_{\rm uc}d^2r\,|u_{\mathbf k}(\mathbf r)|^2f(\mathbf r)$,
so $P_{\mathbf k}(fu_{\mathbf k})
=\langle f\rangle_{\mathbf k}u_{\mathbf k}$.

Taking the squared norm and using
Eq.~\eqref{eq:sm_ideality_norm}, we arrive at
\begin{equation}
\operatorname{tr}g(\mathbf k)-F_{xy}(\mathbf k)
=\left\langle|\partial_+\gamma|^2\right\rangle_{\mathbf k}
-\left|\langle\partial_+\gamma\rangle_{\mathbf k}\right|^2= \operatorname{Var}_{\mathbf k}(\partial_+\gamma).
\label{eq:sm_ideality_defect}
\end{equation}
The right-hand side is the variance of
$\partial_+\gamma$ over the orbital density.

This result identifies the origin of the nonideal geometry.
For the band with ideal quantum geometry ($\gamma=0$), $\partial_+\gamma=0$,
so the variance vanishes and the band remains ideal
\cite{mera2021engineering,ledwith2023vortexability}.
For $\gamma\ne0$, a spatially varying
$\partial_+\gamma$ produces a positive variance and hence
$\operatorname{tr}g>F_{xy}$. The variance quantifies the departure from
ideal quantum geometry; a spatially uniform $\partial_+\gamma$ gives no
departure even when $\gamma\ne0$.

\section{Sheared topological flat bands}
\label{sec:shear}
\label{sec:sm_shear}

A shear changes the shape of the band wavefunctions while preserving the area. 
It therefore provides a way to modify the projected interaction
while preserving the topology and integrated density moments. We apply the
same area-preserving coordinate transformation to every orbital, but
keep the interaction fixed in the laboratory frame. We first construct
sheared lowest-Landau-level (LLL) orbitals on a torus and derive their
form factors. The same construction can then be applied to a band with
an envelope function.

\subsection{Sheared topological flat bands on a torus}
\label{sec:shear_band_construction}
\label{sec:torus}
Let $\mathbf a_1,\mathbf a_2$ be primitive lattice vectors, with
$\mathbf a_1\times\mathbf a_2=A_{\rm uc}=2\pi\ell_B^2$.
The reciprocal vectors obey
$\mathbf a_i\cdot\mathbf b_j=2\pi\delta_{ij}$.
We put $N_i$ cells along $\mathbf a_i$, so the torus is
spanned by $\mathbf L_i=N_i\mathbf a_i$ and contains
$N_{\rm uc}=N_1N_2$ flux quanta. For example, a triangular lattice has
$\mathbf a_1=a(1,0)$ and
$\mathbf a_2=a(1/2,\sqrt3/2)$, with
$\sqrt3a^2/2=2\pi\ell_B^2$. Let $\varphi_{\mathbf k}^\text{LLL}$ be orthonormal magnetic Bloch orbitals of
the LLL, normalized over the full torus~\cite{haldane1985periodic,haldane1985many}.
For the auxiliary LLL Hamiltonian, we take charge $+e$ with $e>0$
and $B_0>0$, and use the Landau gauge $\mathbf A=(0,B_0x)$.
Thus $\boldsymbol\pi=-i\hbar\boldsymbol\nabla-e\mathbf A$,
consistent with the $C=+1$ and holomorphic conventions above.

We consider a periodic shear along a direction $\mathbf d$
perpendicular to a reciprocal-lattice wavevector
$\mathbf Q_s=m_1\mathbf b_1+m_2\mathbf b_2\ne0$.
For definiteness, choose
$\mathbf d=\hat{\mathbf z}\times\mathbf Q_s/|\mathbf Q_s|$.
With $t=\mathbf Q_s\cdot\mathbf r$, the shear map is
\begin{equation}
 \mathbf F(\mathbf r)=\mathbf r+\mathbf d\,u(t),\qquad J(\mathbf r)=\frac{\partial\mathbf F}{\partial\mathbf r}
 =\mathbb I+\mathbf d\otimes\mathbf Q_s\,u'(t).
 \label{eq:shear-hk-map}
\end{equation}
Here $u(t)$ is real, $2\pi$-periodic, and has zero mean.
Because $\mathbf d\cdot\mathbf Q_s=0$, we have
$\det J=1$ and the inverse map is exactly
$\mathbf F^{-1}(\mathbf r)=\mathbf r-\mathbf d\,u(t)$.
Also $\mathbf F(\mathbf r+\mathbf a_i)
=\mathbf F(\mathbf r)+\mathbf a_i$, so the shear
acts on the same torus for any such displacement $u(t)$.

We transform each complete LLL orbital by
\begin{equation}
 \psi_{\mathbf k}^{F}(\mathbf r)
 =(U_F\varphi_{\mathbf k}^\text{LLL})(\mathbf r)
 =e^{i\chi_F(\mathbf r)}
 \varphi_{\mathbf k}^\text{LLL}(\mathbf F(\mathbf r)).
 \label{eq:shear-hk-sheared-state}
\end{equation}
The $k$-independent phase $\chi_F$ ensures that the sheared state
satisfies the same magnetic boundary conditions as the
parent LLL state. Since the coordinate map preserves area, $U_F$ is unitary.
It therefore preserves the normalization, orthogonality,
and Chern number ($C=1$) of the deformed LLL band. In the gauge chosen above, one convenient
choice is
\begin{equation}
 \chi_F(\mathbf r)=-\frac{1}{\ell_B^2}\bigg[d_yx\,u(t)+\frac{d_xd_y}{2}u(t)^2-\frac{1}{|\mathbf Q_s|}\int_0^t u(s)\,ds\bigg].
 \label{eq:shear-hk-gauge}
\end{equation}
Indeed, the phase changes by
$\chi_F(\mathbf r+\mathbf a_i)-\chi_F(\mathbf r)
=-a_{i,x}d_yu(t)/\ell_B^2$, which cancels the extra
boundary phase from the displacement of $y$.
For $\mathbf Q_s=Q\hat{\mathbf x}$ and
$\mathbf d=\hat{\mathbf y}$, this reduces to the familiar
$\mathbf F(x,y)=(x,y+u(Qx))$ and
$\chi_F=-\ell_B^{-2}[xu(Qx)-\int_0^x u(Qx')\,dx']$.

Away from zeros of each parent LLL orbital, the shear admits
a momentum-dependent multiplicative representation
$\psi_{\mathbf k}^F=h_{\mathbf k}^F\varphi_{\mathbf k}^\text{LLL}$,
where
\begin{equation}
 h_{\mathbf k}^F(\mathbf r)=e^{i\chi_F(\mathbf r)}
 \frac{\varphi_{\mathbf k}^\text{LLL}(\mathbf F(\mathbf r))}
      {\varphi_{\mathbf k}^\text{LLL}(\mathbf r)}.
 \label{eq:shear-hk-effective-envelope}
\end{equation}
The ratio is generally singular because shear moves the orbital
zeros to their inverse images under $\mathbf F$. Thus the sheared band
need not belong to the smooth exponential family used in the optimization,
although the sheared wavefunction remains smooth. The smooth-logarithm
identity in Eq.~\eqref{eq:sm_ideality_defect} cannot generally be applied
to this ratio; we derive the shear geometry directly below.

A single-particle Hamiltonian that can realize this sheared band is
\begin{equation}
 H_F^{\rm sp}=U_F\frac{\boldsymbol\pi^2}{2m_*}U_F^\dagger
 =\frac{1}{2m_*}\sum_{\mu=x,y}
 \big[(J^{-T}\boldsymbol\pi)_\mu\big]^2,
 \label{eq:sm_shear_parent}
\end{equation}
Here $\boldsymbol\pi=-i\hbar\boldsymbol\nabla-e\mathbf A$,
$m_*>0$. The squares denote differential-operator composition,
including derivatives of the position-dependent coefficients in $J^{-T}$.
The bands are exactly flat and the Landau-level energies are unchanged.

\subsection{Form factors and projected interactions}
\label{sec:shear_form_factors}
\label{sec:sm_shear_form_geometry}

We next derive the form factors of the sheared orbitals, which
determine the projected interaction.
We use the density-matrix-element convention and allowed momentum
transfers of Eqs.~\eqref{eq:sm_form_factor_definition}
and~\eqref{eq:sm_transfer}, together with the LLL form factor
in Eq.~\eqref{eq:sm_ll_form_factor}.
Using $|\psi_{\mathbf k}^F\rangle=U_F|\varphi_{\mathbf k}^\text{LLL}\rangle$
from Eq.~\eqref{eq:shear-hk-sheared-state}, the density matrix element is
\begin{equation*}
 \lambda_{\mathbf q}^{F}(\mathbf k,\mathbf k')
 =\langle\psi_{\mathbf k}^{F}|e^{-i\mathbf q\cdot\mathbf r}
 |\psi_{\mathbf k'}^{F}\rangle
 =\langle\varphi_{\mathbf k}^\text{LLL}|U_F^\dagger
 e^{-i\mathbf q\cdot\mathbf r}U_F|\varphi_{\mathbf k'}^\text{LLL}\rangle.
\end{equation*}
Since $\det J=1$, $U_F^\dagger$ applies the inverse coordinate map
with the inverse phase. The phases cancel around the density operator,
so $U_F^\dagger e^{-i\mathbf q\cdot\mathbf r}U_F
=e^{-i\mathbf q\cdot\mathbf F^{-1}(\mathbf r)}$.
Using $\mathbf F^{-1}(\mathbf r)=\mathbf r-\mathbf d\,
u(\mathbf Q_s\cdot\mathbf r)$ then gives
\begin{equation}
 \lambda_{\mathbf q}^{F}(\mathbf k,\mathbf k')=\langle\varphi_{\mathbf k}^\text{LLL}|
 e^{-i\mathbf q\cdot\mathbf F^{-1}(\mathbf r)}
 |\varphi_{\mathbf k'}^\text{LLL}\rangle=\langle\varphi_{\mathbf k}^\text{LLL}|
 e^{-i\mathbf q\cdot\mathbf r}
 e^{i(\mathbf q\cdot\mathbf d)u(\mathbf Q_s\cdot\mathbf r)}
 |\varphi_{\mathbf k'}^\text{LLL}\rangle.
 \label{eq:shear-hk-form-factor}
\end{equation}

We can also shear-transform a $\gamma=0$ band whose LLL orbitals
carry a momentum-independent envelope function $h^{(0)}(\mathbf r)$.
Let $\rho^{(0)}=|h^{(0)}|^2$ and
$S_0(\mathbf k)=\langle\varphi_{\mathbf k}^\text{LLL}|\rho^{(0)}
|\varphi_{\mathbf k}^\text{LLL}\rangle$.
The normalized sheared state is
\begin{equation}
 \psi_{\mathbf k}^{(0,F)}(\mathbf r)
 =\frac{e^{i\chi_F(\mathbf r)}
 h^{(0)}(\mathbf F(\mathbf r))
 \varphi_{\mathbf k}^\text{LLL}(\mathbf F(\mathbf r))}
 {\sqrt{S_0(\mathbf k)}}.
 \label{eq:ge_combined_band}
\end{equation}
Its normalization is unchanged because the complete $\gamma=0$ state is sheared unitarily. Define the Fourier
coefficients, evaluated in the unsheared coordinates, by
\begin{equation}
 C_{\mathbf G}(\mathbf q)=
 \left\langle \rho^{(0)}(\mathbf r)
 e^{i(\mathbf q\cdot\mathbf d)u(\mathbf Q_s\cdot\mathbf r)}
 e^{-i\mathbf G\cdot\mathbf r}\right\rangle_{\rm uc},
 \label{eq:sm_shear_vertex_weights}
\end{equation}
where $\mathbf G$ is a reciprocal-lattice vector and
$\langle\cdots\rangle_{\rm uc}$ is a unit-cell average.
The combined-band form factor is
\begin{equation}
 \lambda_{\mathbf q}^{(0,F)}(\mathbf k,\mathbf k')
 =\frac{\displaystyle\sum_{\mathbf G}
 C_{\mathbf G}(\mathbf q)
 \lambda_{\mathbf q-\mathbf G}^{\rm LLL}
 (\mathbf k,\mathbf k')}
 {\sqrt{S_0(\mathbf k)S_0(\mathbf k')}}.
 \label{eq:sm_shear_convolution}
\end{equation}

Finally, with $A=N_{\rm uc}A_{\rm uc}$, the density and
projected two-body Hamiltonian are
\begin{equation}
 \bar\rho_F(\mathbf q)=\sum_{\mathbf k,\mathbf k'}
 \lambda_{\mathbf q}^{F}(\mathbf k,\mathbf k')
 c_{\mathbf k}^\dagger c_{\mathbf k'},\qquad
 H_{\rm int}^{F}
 =\frac{1}{2A}\sum_{\mathbf q}V(|\mathbf q|)
 :\bar\rho_F(\mathbf q)\bar\rho_F(-\mathbf q):\, .
 \label{eq:sm_shear_projected_hamiltonian}
\end{equation}
For the enveloped band, $\lambda^F$ is replaced by
$\lambda^{(0,F)}$. In both cases, $V(|\mathbf q|)$ is
the bare interaction in the fixed Euclidean laboratory frame.
The shear changes the band orbitals and their form factors;
the interaction kernel remains $V(|\mathbf q|)$.

\subsection{Topology and quantum geometry of sheared bands}
\label{sec:shear_geometry}

The shear preserves the Chern number, but changes how the Bloch
wavefunctions vary with momentum. We analyze this effect on the
quantum metric $g_{ij}(\mathbf k)$ and Berry curvature $F_{xy}(\mathbf k)$. Both follow
from the quantum geometric tensor defined in Eq.~\eqref{eq:sm_qgt}.
Using the cell normalization in
Eq.~\eqref{eq:sm_cell_normalization}, the sheared state is
\begin{equation}
 u_{\mathbf k}^{F}(\mathbf r)
 =e^{i\chi_F(\mathbf r)}
 e^{i\mathbf k\cdot[\mathbf F(\mathbf r)-\mathbf r]}
 u_{\mathbf k}^{\rm LLL}(\mathbf F(\mathbf r)).
 \label{eq:sm_shear_cell_phase}
\end{equation}

We obtain the quantum geometry by substituting
Eq.~\eqref{eq:sm_shear_cell_phase} into Eq.~\eqref{eq:sm_qgt}.
Choose $x$ along the shear wavevector $\mathbf Q_s$ and $y$ along
the displacement, so that $\mathbf F(x,y)=(x,y+\xi(x))$, with
$\xi(x)=u(Qx)$ and $Q=|\mathbf Q_s|$. Since $\mathbf F$ and $\chi_F$
are independent of momentum, differentiating the sheared state gives
\begin{equation}
\partial_{k_i}u_{\mathbf k}^{F}(\mathbf r)
=e^{i\chi_F(\mathbf r)+ik_y\xi(x)}
\left[(\partial_{k_i}u_{\mathbf k}^{\rm LLL})(\mathbf F(\mathbf r))
+i\delta_{iy}\xi(x)u_{\mathbf k}^{\rm LLL}(\mathbf F(\mathbf r))\right].
\label{eq:sm_shear_state_derivative}
\end{equation}
Thus the $k_x$ derivative acts on the parent LLL state, while the
$k_y$ derivative also contains the contribution $i\xi(x)u_{\mathbf k}^{F}$
from the shear phase. To evaluate these derivatives on a torus, we use the magnetic-image
representation. Integration along $y$ makes distinct images orthogonal
and allows the remaining $x$ integral to be unfolded onto the real line.
Up to momentum-independent factors and an overall phase, the
$x$-dependent part of each unsheared LLL image is proportional to
$e^{-(x-\ell_B^2k_y)^2/(2\ell_B^2)}e^{-ik_xx}$.
Its momentum derivatives therefore contribute the factors $-ix$ and
$x-\ell_B^2k_y$, respectively. Orbital averages at fixed $\mathbf k$
use the normalized weight
$p(x)=(\sqrt{\pi}\ell_B)^{-1}
e^{-(x-\ell_B^2k_y)^2/\ell_B^2}$:
$\langle f\rangle=\int_{-\infty}^{\infty}dx\,p(x)f(x)$.
In particular, $\langle x\rangle=\ell_B^2k_y$ and
$\langle(x-\langle x\rangle)^2\rangle=\ell_B^2/2$.

The projector in Eq.~\eqref{eq:sm_qgt} removes the component of each
derivative parallel to $u_{\mathbf k}^{F}$, subtracting the mean of
its derivative factor. Defining $\delta x=x-\langle x\rangle$ and
$\delta \xi=\xi-\langle \xi\rangle$, the projected derivatives are
\begin{equation}
D_xu_{\mathbf k}^{F}
\longleftrightarrow -i\delta x\,u_{\mathbf k}^{F},
\qquad
D_yu_{\mathbf k}^{F}
\longleftrightarrow(\delta x+i\delta \xi)u_{\mathbf k}^{F},
\end{equation}
where
$D_i=(1-|u_{\mathbf k}^{F}\rangle
\langle u_{\mathbf k}^{F}|)\partial_{k_i}$.

Substituting these projected derivatives into
Eq.~\eqref{eq:sm_qgt} gives the diagonal components
\begin{equation}
\mathcal Q_{xx}=\langle(\delta x)^2\rangle=\frac{\ell_B^2}{2},
\qquad
\mathcal Q_{yy}=\langle(\delta x)^2+(\delta \xi)^2\rangle
=\frac{\ell_B^2}{2}+\operatorname{Var}\xi,
\end{equation}
and the off-diagonal component
\begin{equation}
\mathcal Q_{xy}=\langle i\delta x(\delta x+i\delta \xi)\rangle
=-\operatorname{Cov}(x,\xi)+i\frac{\ell_B^2}{2},
\end{equation}
where $\operatorname{Var}\xi=\langle \xi^2\rangle-\langle \xi\rangle^2$ and
$\operatorname{Cov}(x,\xi)=\langle x\xi\rangle
-\langle x\rangle\langle \xi\rangle$.

Finally, taking $g_{ij}=\operatorname{Re}\mathcal Q_{ij}$ and
$F_{xy}=2\operatorname{Im}\mathcal Q_{xy}$ as in Eq.~\eqref{eq:sm_qgt}
gives
\begin{equation}
g_{xx}=\frac{\ell_B^2}{2},\qquad
g_{yy}=\frac{\ell_B^2}{2}+\operatorname{Var}\xi,\qquad
g_{xy}=-\operatorname{Cov}(x,\xi),\qquad
F_{xy}=\ell_B^2.
\label{eq:sm_shear_metric}
\end{equation}
The displacement variance increases $g_{yy}$ while the Berry curvature remains uniform.

\subsection{Sinusoidal and uniform shear}
\label{sec:shear_examples}
We next apply the general construction to two simple shear profiles.
The sinusoidal shear acts within the original torus; a uniform shear
instead maps it to a torus of a different shape.

\paragraph*{Sinusoidal shear.}
Take $u(t)=\epsilon\sin t$, where $\epsilon$ is a displacement.
Expanding the phase in Eq.~\eqref{eq:shear-hk-form-factor} as
$e^{iz\sin t}=\sum_nJ_n(-z)e^{-int}$ gives
\begin{equation}
 \lambda_{\mathbf q}^{F}(\mathbf k,\mathbf k')
 =\sum_{n\in\mathbb Z}
 J_n[-\epsilon(\mathbf q\cdot\mathbf d)]
 \lambda_{\mathbf q+n\mathbf Q_s}^{\rm LLL}(\mathbf k,\mathbf k').
 \label{eq:sm_shear_bessel}
\end{equation}
Here $J_n$ is the Bessel function of the first kind. Each term is a
LLL form factor at a shifted momentum. For a general periodic shear profile, the Bessel weights are
replaced by the corresponding Fourier coefficients of
$e^{i(\mathbf q\cdot\mathbf d)u(t)}$.

To evaluate the quantum metric for this profile, we need the moments
of the shear displacement $\xi(x)=u(Qx)=\epsilon\sin(Qx)$ within each
LLL orbital. At fixed momentum $\mathbf k$, the orbital
average $\langle \xi\rangle$ is the mean displacement along $y$, while
$\operatorname{Var}\xi=\langle \xi^2\rangle-\langle \xi\rangle^2$ measures
the variation of that displacement across the orbital.
Equation~\eqref{eq:sm_shear_metric} shows why this variation matters:
it adds directly to $g_{yy}$ and hence to the metric trace.
Using the LLL form factor in Eq.~\eqref{eq:sm_ll_form_factor}
gives
\begin{equation}
 \langle \xi\rangle
 =\epsilon\eta_{\mathbf Q_s}e^{-Q^2\ell_B^2/4}\sin(Q\ell_B^2k_y),
 \qquad
 \operatorname{Var}\xi
 =\frac{\epsilon^2}{2}(1-e^{-Q^2\ell_B^2/2})
 (1+e^{-Q^2\ell_B^2/2}\cos(2Q\ell_B^2k_y)).
 \label{eq:sm_sine_geometry_moments}
\end{equation}
Here $\eta_{\mathbf Q_s}$ is the reciprocal-lattice sign defined below
Eq.~\eqref{eq:sm_ll_form_factor}; it fixes the origin of the oscillation
in the mean displacement. Averaging the displacement variance over the
Brillouin zone and using Eq.~\eqref{eq:sm_shear_metric} gives
\begin{equation}
 \frac{\langle\operatorname{tr}g^F\rangle_{\rm BZ}}{\ell_B^2}
 =1+\frac{\epsilon^2}{2\ell_B^2}(1-e^{-Q^2\ell_B^2/2}).
 \label{eq:sm_sine_geometry}
\end{equation}
The Berry curvature remains constant with $F_{xy}=\ell_B^2$, while the average trace of the quantum metric 
grows quadratically in the shear amplitude $\epsilon$.

\paragraph*{Uniform shear.}
For comparison, take $\mathbf F(\mathbf r)=J\mathbf r$ with a constant
real $2\times2$ matrix $J$ and $\det J=1$. The laboratory coordinates
$\mathbf r$ and their Euclidean metric remain fixed; the map shears the
band orbitals by evaluating the parent LLL state at the mapped
position $\mathbf R=\mathbf F(\mathbf r)=J\mathbf r$. This is a unimodular
deformation of the Landau-orbit metric~\cite{haldane2011geometrical}.
Let $\mathbf a_{i,{\rm lab}}$, with $i=1,2$, be the fixed laboratory
real-space primitive vectors, and let $\mathbf b_{i,{\rm lab}}$ be their
reciprocal vectors, defined by
$\mathbf a_{i,{\rm lab}}\cdot\mathbf b_{j,{\rm lab}}=2\pi\delta_{ij}$.
The deformed LLL lattice then has
$\mathbf a_{i,{\rm def}}=J\mathbf a_{i,{\rm lab}}$ and
$\mathbf b_{i,{\rm def}}=J^{-T}\mathbf b_{i,{\rm lab}}$,
which obey $\mathbf a_{i,{\rm def}}\cdot\mathbf b_{j,{\rm def}}=2\pi\delta_{ij}$.
A laboratory momentum $\mathbf q$ therefore becomes $J^{-T}\mathbf q$
in the coordinates of the deformed LLL lattice, where $J^{-T}=(J^{-1})^T$. In the convention of
Eq.~\eqref{eq:sm_ll_form_factor}, the full form factor is
\begin{equation}
 \lambda_{\mathbf q}^{F}(\mathbf k,\mathbf k')
 =\eta_{\mathbf G}\exp\bigg[-\frac{\ell_B^2}{4}|J^{-T}\mathbf q|^2
 +\frac{i\ell_B^2}{2}\mathbf k\times\mathbf k'
 +\frac{i\ell_B^2}{2}(\mathbf k+\mathbf k')\times\mathbf G\bigg],
 \label{eq:sm_affine_form_factor}
\end{equation}
where $\mathbf q=\mathbf k'-\mathbf k+\mathbf G$ and
$\mathbf G=n_1\mathbf b_{1,{\rm lab}}+n_2\mathbf b_{2,{\rm lab}}$, with
$n_1,n_2\in\mathbb Z$.
The sign $\eta_{\mathbf G}=(-1)^{n_1+n_2+n_1n_2}$ uses the same integer
coefficients as in the deformed LLL lattice. All momenta in the phase are
laboratory momenta; their cross products are unchanged by $J^{-T}$
because $\det J=1$. For a pure LLL parent state, the geometry in Cartesian
laboratory momentum coordinates is
\begin{equation}
 g_J=\frac{\ell_B^2}{2}J^{-1}J^{-T},
 \qquad F_{xy}=\ell_B^2.
 \label{eq:sm_affine_geometry_general}
\end{equation}
For a dimensionless uniform-shear parameter $s$ and
$\mathbf F_s(x,y)=(x,y+sx)$, this gives
\begin{equation}
 g_s=\frac{\ell_B^2}{2}
 \begin{pmatrix}1&-s\\-s&1+s^2\end{pmatrix},
 \qquad
 \frac{\operatorname{tr}g_s}{F_{xy}}=1+\frac{s^2}{2}.
 \label{eq:sm_affine_geometry}
\end{equation}
A uniform shear therefore saturates $\det g=F_{xy}^2/4$, while its
trace exceeds $F_{xy}$ for $s\ne0$.

\section{Two-electron interactions and the mechanism of gap enhancement}
\label{sec:pair_mechanism}
\label{sec:short_static}
\label{sec:ge_short_range}

The excitation gap is the interaction cost of changing the pair
correlations of the fractional liquid~\cite{girvin1986magneto}. For short-range repulsion, this
cost is set by how rapidly the two-electron wavefunction rises from zero when
two electrons separate. This gives a common description of the three
band constructions. A momentum-independent envelope $h^{(0)}$ changes the spatial weight of
the two-electron interaction, and a sheared topological flat band changes the local distance over which
the two-electron wavefunction varies. Both preserve the cubic vanishing of the
Laughlin wavefunction at particle coincidence, described in the LLL by
the factor $(z_i-z_j)^3$. General $k$-dependent envelopes need not preserve
this cubic behavior. We first derive the two-electron interaction, then use this distinction to explain the gap
response in each model.

\subsection{The leading short-range two-electron interaction}
\label{sec:projected_gap}
\label{sec:ge_pair_gap}
\label{sec:sm_ed}

Let $\psi_i(\mathbf r)$ be an orthonormal scalar band orbital, with $i$
labeling its Bloch momentum. The projected field is
$\widehat\psi(\mathbf r)=\sum_i\psi_i(\mathbf r)c_i$. For a many-body
state $|\Psi\rangle$, the pair density is
\begin{equation}
G_\Psi(\mathbf r_1,\mathbf r_2)
=\big\langle\widehat\psi^\dagger(\mathbf r_1)
\widehat\psi^\dagger(\mathbf r_2)
\widehat\psi(\mathbf r_2)\widehat\psi(\mathbf r_1)\big\rangle_\Psi.
\label{eq:ge_pair_density}
\end{equation}
Weighting it by the repulsion gives the interaction energy. Order the
many-body levels over all momentum sectors as $E_0\leq E_1\leq\cdots$.
For a Laughlin-like ground-state triplet on the torus~\cite{haldane1985periodic,haldane1985many}, let $|g\rangle$ be the state
at its upper edge $E_2$ and $|e\rangle$ the next state, at $E_3$.
Their separation is
\begin{equation}
\Delta=E_3-E_2
=\frac12\int_A d^2r_1\,d^2r_2\,
V(\mathbf r_1-\mathbf r_2)
\bigl[G_e(\mathbf r_1,\mathbf r_2)-G_g(\mathbf r_1,\mathbf r_2)\bigr].
\label{eq:ge_realspace_gap}
\end{equation}
Thus the gap depends on the difference between the pairs present in
the excitation and in the ground state.

For the screened Coulomb interaction defined in Sec.~\ref{sec:sm_form_factors},
write $d=d_s/\ell_B$ and $E_C=e^2/(\varepsilon\ell_B)$. Its short-screening expansion is
\begin{equation}
V(q)=2\pi E_C\ell_B^2
\left[d-\frac{d^3(q\ell_B)^2}{3}+O(d^5q^4\ell_B^4)\right].
\label{eq:ge_screening_series}
\end{equation}
The constant term gives a contact repulsion, which acts only when the
two particles coincide. A scalar antisymmetric two-electron wavefunction
vanishes there, so this term contributes no energy. The next term,
proportional to $d^3q^2$, probes how rapidly the two-electron wavefunction grows
as the particles separate.
This contact cancellation need not hold in multicomponent bands
with internal antisymmetry. For example, in multicomponent higher-vortexable
bands formed from LL0 and LL1 components, the contact interaction can
contribute to the fermionic $V_1$ channel~\cite{Sarkar2026GLL,zhang2025beyond}.

To compute this energy cost, we need the short-distance behavior
of each two-electron pair. For $i<j$, introduce the normalized antisymmetric two-electron
wavefunction and its slope,
\begin{equation}
\Phi_{ij}(\mathbf r_1,\mathbf r_2)=\frac{\psi_i(\mathbf r_1)\psi_j(\mathbf r_2)
       -\psi_j(\mathbf r_1)\psi_i(\mathbf r_2)}{\sqrt2},\qquad
\mathbf D_{ij}(\mathbf r)=\psi_j(\mathbf r)\nabla\psi_i(\mathbf r)
 -\psi_i(\mathbf r)\nabla\psi_j(\mathbf r).
\label{eq:ge_pair_wavefunction}
\end{equation}
Here $\mathbf r=(\mathbf r_1+\mathbf r_2)/2$ is the center coordinate, and
$\nabla$ acts on position. With
$\boldsymbol\delta=\mathbf r_1-\mathbf r_2$, antisymmetry gives
\begin{equation}
\Phi_{ij}(\mathbf r+\boldsymbol\delta/2,
          \mathbf r-\boldsymbol\delta/2)
=\frac{\boldsymbol\delta\cdot\mathbf D_{ij}(\mathbf r)}{\sqrt2}
+O(|\boldsymbol\delta|^3).
\label{eq:ge_D}
\end{equation}
The factor $-q^2$ in Eq.~\eqref{eq:ge_screening_series} Fourier
transforms to $\nabla_{\boldsymbol\delta}^2\delta^{(2)}(\boldsymbol\delta)$ in real space.
Its matrix element is therefore determined by the Laplacian of
$\Phi_{ij}^*\Phi_{kl}$ at zero separation. Using the linear expansion
in Eq.~\eqref{eq:ge_D}, this Laplacian equals
$\mathbf D_{ij}^*\cdot\mathbf D_{kl}$. Integrating over the center
coordinate $\mathbf r$ then gives the two-electron interaction matrix element
\begin{equation}
\begin{aligned}
\mathcal V_{ij,kl}(d)
&\equiv\langle\Phi_{ij}|V|\Phi_{kl}\rangle
=d^3\bigl(\mathcal V_{\rm 2e}\bigr)_{ij,kl}+O(d^5),\\
\bigl(\mathcal V_{\rm 2e}\bigr)_{ij,kl}
&=\kappa_3\int_A d^2r\,\mathbf D_{ij}^*\cdot\mathbf D_{kl},
\qquad \kappa_3=\frac{2\pi}{3}E_C\ell_B^4.
\end{aligned}
\label{eq:ge_V3}
\end{equation}
The subscript 2e labels this leading two-electron term, the coefficient of $d^3$. The short-range interaction Hamiltonian is $H=\sum_{i<j,k<l}\mathcal V_{ij,kl}c_i^\dagger c_j^\dagger c_lc_k$.
Defining
$\widehat{\mathbf D}=\sum_{i<j}\mathbf D_{ij}c_jc_i$, we obtain
\begin{equation}
H(d)=d^3H_{\rm 2e}+O(d^5),\qquad
H_{\rm 2e}=\kappa_3\int_A d^2r\,
\widehat{\mathbf D}^{\dagger}\cdot\widehat{\mathbf D}.
\label{eq:ge_H3}
\end{equation}
The leading short-range interaction is therefore a positive sum of
squared two-electron wavefunction slopes. In the LLL, it is the fermionic relative-angular-momentum
$m=1$ pseudopotential parent interaction~\cite{haldane1983fractional,trugman1985exact}.
The same quantity appears in the local pair
correlation:
\begin{equation}
G_\Psi(\mathbf r+\boldsymbol\delta/2,
       \mathbf r-\boldsymbol\delta/2)
=\sum_{\mu,\nu=x,y}\delta_\mu\delta_\nu
\langle\widehat D_\mu^\dagger(\mathbf r)
       \widehat D_\nu(\mathbf r)\rangle_\Psi
+O(|\boldsymbol\delta|^4).
\label{eq:ge_short_pair_density}
\end{equation}
For the LLL Laughlin state~\cite{laughlin1983anomalous,trugman1985exact}, the linear two-electron wavefunction amplitude and hence this
quadratic term vanish. Its short-range excitations pay the cost of
restoring such a term.

To understand how the two-electron interaction determines the many-body gap,
we first examine its two-particle spectrum. Since the interaction
preserves total momentum, the two-electron matrix $\mathcal V_{\rm 2e}$ can be
diagonalized separately in each momentum sector
$\mathbf K=\mathbf k_i+\mathbf k_j$, defined modulo reciprocal vectors. Let $\varepsilon_\alpha(\mathbf K)$ and $v_{\mathbf K\alpha}$
be the eigenvalues and normalized eigenvectors of this matrix.
Each eigenvector describes a particular combination of two-particle
states with total momentum $\mathbf K$, while its eigenvalue gives
the interaction energy of that combination. These are precisely
the energies and states obtained from two-particle ED.

To connect these two-electron energy penalties to the many-body problem, define
the corresponding two-electron annihilation operator $Q_{\mathbf K\alpha}
=\sum_{(i,j)\in\mathbf K}(v_{\mathbf K\alpha})_{ij}^*c_jc_i$, 
where the sum runs over independent two-electron pairs $i<j$.
The leading interaction can then be written as
\begin{equation}
H_{\rm 2e}=\sum_{\mathbf K,\alpha}
\varepsilon_\alpha(\mathbf K)
Q_{\mathbf K\alpha}^\dagger Q_{\mathbf K\alpha}.
\label{eq:ge_pair_decomposition}
\end{equation}
This expression separates the interaction into independent
two-particle modes. The coefficient $\varepsilon_\alpha(\mathbf K)$
is the energy cost of the corresponding two-electron mode, while
$Q_{\mathbf K\alpha}^\dagger Q_{\mathbf K\alpha}$ measures its
weight in a many-body state. For a normalized many-body state $|\Psi\rangle$, define
$n_{\mathbf K\alpha}^{\Psi}=\langle\Psi|Q_{\mathbf K\alpha}^\dagger Q_{\mathbf K\alpha}|\Psi\rangle.$
Its interaction energy is therefore
\begin{equation}
E_\Psi^{\rm 2e}
=\sum_{\mathbf K,\alpha}
\varepsilon_\alpha(\mathbf K)n_{\mathbf K\alpha}^{\Psi}.
\end{equation}
Thus the two-particle spectrum provides the energy costs,
while the many-body wavefunction determines how strongly
each two-electron mode is occupied.

At filling $\nu=1/3$, let $|g\rangle$ be the highest-energy
state of the threefold ground-state manifold and $|e\rangle$
the lowest excited state. With the eigenenergies ordered
as $E_0^{\rm 2e}\leq E_1^{\rm 2e}\leq\cdots$, the leading gap is
\begin{equation}
\Delta_{\rm 2e}=E_3^{\rm 2e}-E_2^{\rm 2e}
=\sum_{\mathbf K,\alpha}\varepsilon_\alpha(\mathbf K)
\left(n_{\mathbf K\alpha}^{e}
-n_{\mathbf K\alpha}^{g}\right).
\label{eq:ge_mode_gap}
\end{equation}
This relation shows why a larger two-particle eigenvalue
does not necessarily produce a larger many-body gap:
the result also depends on the difference in two-electron mode
occupations between the excited and ground states.

For a Laughlin parent Hamiltonian, the ground states have
zero energy and contain no weight in any positive-energy
two-electron mode. The gap is then determined entirely by the
positive-energy two-electron modes present in the lowest excitation.
A deformation can enhance the gap by increasing the cost
of these modes. The two-particle ED spectrum
reveals which two-electron energy penalties increase, while the many-body
correlations determine their contribution to the gap.

Finally, in the short-screening limit, the interaction Hamiltonian
takes the form
$H(d)=d^3H_{\rm 2e}+O(d^5)$, which gives
\begin{equation}
\Delta(d)=d^3\Delta_{\rm 2e}+O(d^5).
\end{equation}
Here $\Delta_{\rm 2e}=E_3^{\rm 2e}-E_2^{\rm 2e}$ is the many-body gap
of $H_{\rm 2e}$ defined in Eq.~\eqref{eq:ge_mode_gap}. Leading $d^3$
behavior requires $\Delta_{\rm 2e}>0$ for the level separation considered.
Since $d=d_s/\ell_B$, this is consistent with the $d_s^3$ scaling
found in the main text at fixed $\ell_B$.

\subsection{Ideal quantum geometry and Laughlin zero modes}
\label{sec:ge_static}

For the $\gamma=0$ band, every LLL orbital $\varphi_i$ carries the same
envelope $h^{(0)}$. Substitution of
$\psi_i^{(0)}=h^{(0)}\varphi_i/\sqrt{S_i^{(0)}}$ into the two-electron wavefunction slope gives
\begin{equation}
\mathbf D_{ij}^{(0)}
=\frac{(h^{(0)})^2}{\sqrt{S_i^{(0)}S_j^{(0)}}}\mathbf D_{ij}^{\rm LLL},
\qquad
\mathbf D_{ij}^{\rm LLL}
=\varphi_j\nabla\varphi_i-\varphi_i\nabla\varphi_j.
\label{eq:ge_static_D}
\end{equation}
The derivatives of $h^{(0)}$ cancel between the two particle assignments.
Consequently, multiplying the Laughlin wavefunction by
$\prod_a h^{(0)}(\mathbf r_a)$ preserves its cubic vanishing as two particles
approach one another. The resulting state is an exact zero mode of
$H_{\rm 2e}^{(0)}$~\cite{wang2021exact,ledwith2023vortexability}.

For $S_i^{(0)}\simeq1$, substituting Eq.~\eqref{eq:ge_static_D}
into the two-electron matrix and replacing $(\rho^{(0)})^2=|h^{(0)}|^4$ by its
unit-cell average gives
\begin{equation}
\bigl(\mathcal V_{\rm 2e}^{(0)}\bigr)_{ij,kl}
\simeq\langle(\rho^{(0)})^2\rangle_{\rm uc}\,
\kappa_3\int_A d^2r\,
(\mathbf D_{ij}^{\rm LLL})^*\cdot\mathbf D_{kl}^{\rm LLL}
=M_2^{(0)}\bigl(\mathcal V_{\rm 2e}^{\rm LLL}\bigr)_{ij,kl}.
\label{eq:ge_static_pair_scaling}
\end{equation}
Here $M_2^{(0)}=\langle(\rho^{(0)})^2\rangle_{\rm uc}$ with
$\langle\rho^{(0)}\rangle_{\rm uc}=1$, as defined in
Eq.~\eqref{eq:sm_static_m2}.
This approximation requires both
$S_0(\mathbf k)\simeq1$ and small nonuniform two-electron center (COM)
contributions after projection. Define the convolution
$C_{\mathbf K}=\sum_{\mathbf G}
w^{(0)}_{\mathbf G}w^{(0)}_{\mathbf K-\mathbf G}$. One sufficient diagnostic is
\(
\epsilon_{\rm COM}
=\sum_{\mathbf K\ne0}
|C_{\mathbf K}|e^{-K^2\ell_B^2/8}/M_2^{(0)}\ll1
\)~\cite{Sarkar2026GLL}.
The same factor then
rescales the leading Hamiltonian and gap:
\begin{equation}
H_{\rm 2e}^{(0)}\simeq M_2^{(0)}H_{\rm 2e}^{\rm LLL},\qquad
\frac{\Delta_{\rm 2e}^{(0)}}{\Delta_{\rm 2e}^{\rm LLL}}
\simeq M_2^{(0)}.
\label{eq:ge_static_scaling}
\end{equation}
Thus the two-particle ED levels and the many-body excitation energies
are approximately rescaled by $M_2^{(0)}$, while the Laughlin zero modes
are preserved exactly.

\subsection{Shear preserves Laughlin zero modes and raises excitation energies}
\label{sec:ge_shear_theorem}
\label{subsec:shear-hk-clustering}

A shear also preserves the Laughlin zero modes. Apply the
common area-preserving map $\mathbf F$ of Sec.~\ref{sec:shear} to the
LLL orbitals, keeping the interaction fixed in laboratory coordinates.
To determine how shear changes the two-electron interaction, consider
the antisymmetric two-electron wavefunction slope
$\mathbf D_{ij}^{\rm LLL}=\varphi_j\nabla\varphi_i
-\varphi_i\nabla\varphi_j$.
Under the transformation
$\psi_i^F=e^{i\chi_F}\varphi_i(\mathbf F(\mathbf r))$,
the chain rule gives $\nabla\psi_i^F=e^{i\chi_F}
\left[i(\nabla\chi_F)\varphi_i
+J^T\nabla\varphi_i\right]$,
where the LLL orbitals on the right are evaluated at
$\mathbf F(\mathbf r)$ and $J$ is the Jacobian of the map.
Substituting into the antisymmetric combination, the terms
containing $\nabla\chi_F$ cancel, leaving
\begin{equation}
\mathbf D_{ij}^F(\mathbf r)
=e^{2i\chi_F(\mathbf r)}J^T
\mathbf D_{ij}^{\rm LLL}(\mathbf F(\mathbf r)).
\label{eq:ge_shear_D}
\end{equation}

The LLL orbitals have the local form
$\varphi_i=a(\mathbf r)f_i(z)$, where $z=x+iy$
and $f_i$ is holomorphic. Since
$\nabla f_i=f_i'(z)(1,i)^T$, the common factor
$a$ gives
\begin{equation}
\mathbf D_{ij}^{\rm LLL}
=\sqrt{2}a^2(f_j f_i'-f_i f_j')\mathbf e,
\label{eq:ge_lll_pair_direction}
\end{equation}
where $\mathbf e=(1,i)^T/\sqrt{2}$.
Thus all LLL two-electron wavefunction slopes share the same direction
$\mathbf e$.

Equation~\eqref{eq:ge_V3} then gives the full two-electron interaction matrix
to leading order in $d$:
\begin{equation}
\mathcal V_{ij,kl}^{F}(d)
=d^3\kappa_3\int_A d^2r\,
\bigl(\mathbf D_{ij}^{F}\bigr)^*
\cdot\mathbf D_{kl}^{F}+O(d^5)
=d^3\kappa_3\int_A d^2r\,w(\mathbf r)
\bigl(\mathbf D_{ij}^{\rm LLL}\bigr)^*
\cdot\mathbf D_{kl}^{\rm LLL}
+O(d^5),
\label{eq:ge_shear_pair_matrix}
\end{equation}
where the sheared slopes are evaluated at $\mathbf r$, the LLL slopes
at $\mathbf F(\mathbf r)$, and the two-electron interaction weight is
\begin{equation}
w(\mathbf r)=|J^T\mathbf e|^2
=\frac12\operatorname{tr}(JJ^T)\geq1,
\label{eq:ge_shear_w}
\end{equation}
where the inequality follows from $\det J=1$.

Physically, $w(\mathbf r)$ measures the local amplification of the
squared two-electron wavefunction slope relative to the parent LLL. It averages the squared
local stretching factors: a rotation gives $w=1$, while stretching at
fixed area gives $w>1$. The short-range interaction therefore penalizes
nonzero two-electron wavefunction slopes more strongly in regions of larger $w$. The
Laughlin states remain zero modes because their linear two-electron wavefunction amplitude
vanishes. For the one-directional shear
$\mathbf F(x,y)=(x,y+\xi(x))$, this weight is simply
$w(x)=1+[\xi'(x)]^2/2$.

Equation~\eqref{eq:ge_shear_pair_matrix} gives the two-particle matrix
elements. To bound the many-body spectrum, consider the leading
interaction energy of an arbitrary normalized LLL many-body state $|\Psi\rangle$
and its sheared counterpart $U_F|\Psi\rangle$, with $U_F$ acting on
every electron. Using Eq.~\eqref{eq:ge_H3} and changing variables to
$\mathbf y=\mathbf F(\mathbf r)$ gives
\begin{equation}
\langle\Psi|U_F^\dagger H_{\rm 2e}^F U_F|\Psi\rangle
=\kappa_3\int_A d^2y\,\widetilde w(\mathbf y)
\bigl\|\widehat{\mathbf D}^{\rm LLL}(\mathbf y)|\Psi\rangle\bigr\|^2,
\label{eq:ge_shear_master}
\end{equation}
where $\widetilde w(\mathbf y)=w(\mathbf F^{-1}(\mathbf y))$.
The squared norm sums over the two components of the two-electron wavefunction slope.
The parent LLL energy is the same expression with $\widetilde w=1$.
Since $\widetilde w>0$, either energy vanishes precisely when
$\widehat{\mathbf D}^{\rm LLL}(\mathbf y)|\Psi\rangle=0$ at every
two-electron center. Thus the two Hamiltonians have the same zero modes in
this LLL representation; in the sheared band, the corresponding
Laughlin state is $U_F\Psi_L$.

Let $w_{\min}$ and $w_{\max}$ be the minimum and maximum weights.
Because the squared norm in Eq.~\eqref{eq:ge_shear_master} is
nonnegative, replacing $\widetilde w$ by either extreme bounds the
energy of every state. Equivalently,
\begin{equation}
w_{\min}H_{\rm 2e}^{\rm LLL}
\leq U_F^\dagger H_{\rm 2e}^F U_F
\leq w_{\max}H_{\rm 2e}^{\rm LLL}.
\label{eq:ge_shear_operator_bounds}
\end{equation}
The common zero-energy subspace allows us to apply the variational
principle on its orthogonal complement, giving
\begin{equation}
w_{\min}\Delta_{\rm 2e}^{\rm LLL}
\leq\Delta_{\rm 2e}^F
\leq w_{\max}\Delta_{\rm 2e}^{\rm LLL}.
\label{eq:ge_shear_gap_bounds}
\end{equation}
The same comparison holds for every ordered positive eigenenergy.
Since $w_{\min}\geq1$, it also preserves any nonzero thermodynamic gap
of the corresponding LLL parent Hamiltonian.

For the uniform shear $\mathbf F_s(x,y)=(x,y+sx)$,
$w=1+s^2/2$ is constant, so the entire leading spectrum is rescaled:
\begin{equation}
U_s^\dagger H_{\rm 2e}^sU_s
=\left(1+\frac{s^2}{2}\right)H_{\rm 2e}^{\rm LLL}.
\label{eq:ge_uniform_operator}
\end{equation}
In particular, the gap satisfies
\begin{equation}
\Delta_{\rm 2e}^s(J_s^{-1}\Lambda)
=\left(1+\frac{s^2}{2}\right)
\Delta_{\rm 2e}^{\rm LLL}(\Lambda).
\label{eq:ge_uniform_torus}
\end{equation}
Here $U_s=U_{F_s}$, $J_s=\partial\mathbf F_s/\partial\mathbf r$, and
$\Lambda$ is the period lattice of the unsheared LLL torus. Because
the parent orbitals are evaluated at $J_s\mathbf r$, the sheared torus
has period lattice $J_s^{-1}\Lambda$, as in
Sec.~\ref{sec:shear_examples}.

\paragraph*{Equivalent effective interaction.}
The same factor follows from absorbing the uniform shear into the
interaction. Use the same fermion operators for corresponding sheared
and LLL orbitals. The common phase $e^{i\chi_F}$ cancels in every
density matrix element, and changing variables to
$\mathbf y=J_s\mathbf r$, with $\det J_s=1$, gives
\begin{equation}
\langle\psi_i^s|e^{-i\mathbf q\cdot\mathbf r}|\psi_j^s\rangle
=\langle\varphi_i^{\rm LLL}|e^{-i(J_s^{-T}\mathbf q)\cdot\mathbf y}|\varphi_j^{\rm LLL}\rangle,
\qquad
\widehat\rho^{\,s}_{\mathbf q}=\widehat\rho^{\rm LLL}_{J_s^{-T}\mathbf q}.
\label{eq:ge_uniform_density}
\end{equation}
The allowed laboratory momenta $\mathbf q$ of the sheared torus
$J_s^{-1}\Lambda$ map one to one onto the momenta
$\mathbf p=J_s^{-T}\mathbf q$ of the LLL torus $\Lambda$, and both tori
have area $A$. The projected Hamiltonian therefore takes the two
equivalent forms
\begin{equation}
H(d)=\frac{1}{2A}\sum_{\mathbf q}V(q)
:\widehat\rho^{\,s}_{\mathbf q}\widehat\rho^{\,s}_{-\mathbf q}:
=\frac{1}{2A}\sum_{\mathbf p}V_{\rm eff}(\mathbf p)
:\widehat\rho^{\rm LLL}_{\mathbf p}\widehat\rho^{\rm LLL}_{-\mathbf p}:,
\qquad
V_{\rm eff}(\mathbf p)=V(|J_s^T\mathbf p|),
\label{eq:ge_effective_interaction}
\end{equation}
where colons denote normal ordering. The sheared band is thus an LLL
with an anisotropic interaction. Its small-$d$ expansion is
\begin{equation}
V_{\rm eff}(\mathbf p)=2\pi E_C\ell_B^2
\left[d-\frac{d^3\ell_B^2}{3}\,\mathbf p^TJ_sJ_s^T\mathbf p+O(d^5)\right].
\label{eq:ge_effective_series}
\end{equation}
Write $J_sJ_s^T=w\,I+T$, with $I$ the identity, $w=\tfrac12\operatorname{tr}(J_sJ_s^T)$
and $T$ real, symmetric and traceless. The isotropic part is $w$ times
the order-$d^3$ term of the LLL. In real space, the $T$ part acts as
$T_{\mu\nu}\partial_\mu\partial_\nu\delta^{(2)}(\boldsymbol\delta)$ and,
by Eq.~\eqref{eq:ge_D}, has two-electron matrix elements proportional
to $T_{\mu\nu}(D^{\rm LLL}_{ij,\mu})^*D^{\rm LLL}_{kl,\nu}$. All LLL
slopes lie along $\mathbf e=(1,i)^T/\sqrt2$
[Eq.~\eqref{eq:ge_lll_pair_direction}], and
$\mathbf e^\dagger T\mathbf e=(T_{xx}+T_{yy})/2=0$. The anisotropic part
therefore drops out at order $d^3$: only the angular average
$\langle\mathbf p^TJ_sJ_s^T\mathbf p\rangle_{\rm angle}=w\,p^2$ enters,
and $H_{\rm 2e}^s$ equals $w H_{\rm 2e}^{\rm LLL}$, in agreement with
Eq.~\eqref{eq:ge_uniform_operator}. The same identity,
$|J_s^T\mathbf e|^2=\mathbf e^\dagger J_sJ_s^T\mathbf e=w$, gives the
weight in Eq.~\eqref{eq:ge_shear_w}.

The periodic shear $\mathbf F(\mathbf r)=\mathbf r+\mathbf d\,\epsilon\sin(\mathbf Q_s\cdot\mathbf r)$ acts on the
original torus. With $x$ along $\mathbf Q_s$ and $Q=|\mathbf Q_s|$, it reads $\mathbf F(x,y)=(x,y+\epsilon\sin Qx)$. Its weight and average are
\begin{equation}
w(x)=1+\frac{\epsilon^2Q^2}{4}[1+\cos(2Qx)],\qquad
\bar w=1+\frac{\epsilon^2Q^2}{4},
\label{eq:ge_periodic_w}
\end{equation}
where $\epsilon$ is the displacement amplitude; in terms of the peak shear slope
$\zeta=\epsilon|\mathbf Q_s|$ of main-text Fig.~5, $\bar w=1+\zeta^2/4$. Projection into the two-electron center LLL, with magnetic length
$\ell_B/\sqrt2$, suppresses the oscillation at $2Q$ by
$e^{-Q^2\ell_B^2/2}$, using the Gaussian form-factor suppression
discussed in Ref.~\cite{Sarkar2026GLL}. Defining
$\eta_w=(\epsilon Q)^2e^{-Q^2\ell_B^2/2}/4$, we obtain
\begin{equation}
(\bar w-\eta_w)H_{\rm 2e}^{\rm LLL}
\leq U_F^\dagger H_{\rm 2e}^F U_F
\leq(\bar w+\eta_w)H_{\rm 2e}^{\rm LLL}.
\label{eq:ge_periodic_operator_bound}
\end{equation}
The projection bound applies to the collision amplitude for each
fixed configuration of the other electrons; integration over those
coordinates gives the many-body inequality. Since the zero modes
are unchanged,
\begin{equation}
\left|\frac{\Delta_{\rm 2e}^F}{\Delta_{\rm 2e}^{\rm LLL}}
-\left(1+\frac{\epsilon^2Q^2}{4}\right)\right|
\leq\frac{\epsilon^2Q^2}{4}e^{-Q^2\ell_B^2/2}.
\label{eq:ge_periodic_gap}
\end{equation}
Thus rapid spatial oscillations leave an almost uniform enhancement
by $\bar w$, without requiring a small shear amplitude. The lower
bound exceeds one for any nonzero $\epsilon$ and $Q$.
For $\epsilon=\ell_B$ and $Q\ell_B=\sqrt{16\pi/\sqrt3}$,
$\bar w=8.2551974569$ and $\eta_w=3.6213\times10^{-6}$. For the shear of main-text Fig.~5, $\mathbf Q_s=\mathbf G_1$ ($|\mathbf Q_s|\ell_B=2.69$) and $\epsilon=\ell_B$, $\bar w=2.8138$ and $\eta_w=0.0482$, so the leading gap ratio lies between 2.766 and 2.862. 
Note that these bounds concern the leading two-electron Hamiltonian $H_{\rm 2e}$ only, $\eta_w$ is not an exact error bar
on a finite-screening ED gap.

\subsection{Momentum-dependent envelopes change the two-electron interacting energy penalty}
\label{sec:momentum_mechanism}
\label{sec:ge_amplitude_phase}
\label{sec:ge_kgll_response}

Unlike a common envelope or shear, a general momentum-dependent
envelope need not preserve the cancellations that make the Laughlin
wavefunction vanish cubically at particle coincidence. For
$h_i=h^{(0)}e^{a_i/2+i\theta_i}$ and
$\psi_i=h_i\varphi_i/\sqrt{S_i}$, the product rule gives
\begin{equation}
\mathbf D_{ij}
=\frac{h_ih_j}{\sqrt{S_iS_j}}\mathbf D_{ij}^{\rm LLL}
+\psi_i\psi_j\left[
\frac12\nabla(a_i-a_j)+i\nabla(\theta_i-\theta_j)\right].
\label{eq:ge_envelope_D}
\end{equation}
The first term dresses the LLL slope. The second survives because
the two orbitals have different amplitude and phase gradients.

The amplitude and phase terms also interfere in the two-electron matrix.
Write $\mathbf A_{ij}=\nabla(a_i-a_j)/2$ and
$\mathbf B_{ij}=\nabla(\theta_i-\theta_j)$. Their mixed contribution is
\begin{equation}
\bigl(\mathcal V_{\rm 2e}^{a\theta}\bigr)_{ij,kl}
=i\kappa_3\int_A d^2r\,\psi_i^*\psi_j^*\psi_k\psi_l
\bigl(\mathbf A_{ij}\cdot\mathbf B_{kl}
      -\mathbf B_{ij}\cdot\mathbf A_{kl}\bigr).
\label{eq:ge_mixed_term}
\end{equation}
Its diagonal elements vanish. 
Therefore, although this mixed term cannot universally raise all two-electron energy penalties togehter, 
constructive interference between the phase and amplitude terms can raise the energy cost of a given two-electron pair combination.

To understand the amplitude--phase interference in
Eq.~\eqref{eq:ge_mixed_term}, we separate the two-electron wavefunction slope into
two circular components,
$D_{\pm,ij}=(D_{ij,x}\pm iD_{ij,y})/\sqrt{2}$.
Their overlap becomes
$\mathbf D_{ij}^*\cdot\mathbf D_{kl}
=\sum_{\nu=\pm}D_{\nu,ij}^*D_{\nu,kl}$.
The amplitude and phase contributions are
$D^a_{\pm,ij}=\psi_i\psi_j(A_{ij,x}\pm iA_{ij,y})/\sqrt{2}$
and
$D^\theta_{\pm,ij}
=i\psi_i\psi_j(B_{ij,x}\pm iB_{ij,y})/\sqrt{2}$.
Thus the mixed interaction is obtained by integrating
$(D^a_{\nu,ij})^*D^\theta_{\nu,kl}
+(D^\theta_{\nu,ij})^*D^a_{\nu,kl}$
over the two-electron center position and summing over $\nu=\pm$.

For the LLL, $D_{ij,y}=iD_{ij,x}$, so $D_{+,ij}=0$
and only $D_{-,ij}$ is nonzero. We therefore examine
how the amplitude and phase gradients combine in the
$D_+$ component. Writing
$D_{\pm,ij}^{\rm add}=D^a_{\pm,ij}+D^\theta_{\pm,ij}$,
direct substitution gives
\begin{equation}
|D_{+,ij}^{\rm add}|^2
=\frac{|\psi_i\psi_j|^2}{2}
\left|\mathbf B_{ij}
-\hat{\mathbf z}\times\mathbf A_{ij}\right|^2.
\label{eq:ge_matching}
\end{equation}
When $\mathbf B_{ij}=\hat{\mathbf z}\times\mathbf A_{ij}$,
we have $D^\theta_{+,ij}=-D^a_{+,ij}$ and
$D^\theta_{-,ij}=D^a_{-,ij}$. The two contributions therefore
cancel in $D_+$ and add constructively in $D_-$, giving
$D_{-,ij}^{\rm add}=2D^a_{-,ij}$. The added two-electron wavefunction slope thus lies
entirely along the LLL direction $(1,i)^T$.
Physically, this enhances the added two-electron wavefunction slope weight in the
LLL chirality, with $|D_{-,ij}^{\rm add}|^2=4|D^a_{-,ij}|^2$.
When this contribution interferes constructively with the dressed LLL
slope, it can strengthen the short-range repulsion in the $V_1$-like
channel. This provides a mechanism for increasing the excitation gap
while preserving Laughlin-like correlations. Its effect on the many-body gap
still depends on the two-electron mode occupation differences in
Eq.~\eqref{eq:ge_mode_gap}.

For smooth periodic envelopes, this matching can occur
locally. Global matching for a given pair would imply
$\nabla^2(a_i-a_j)=0$, so periodicity forces both its amplitude and
phase differences to be spatially constant. When it occurs where the pair weight is large,
the interference can substantially change the interaction
energy. For the optimized band of model 1,
the full deformation has a gap about $5.15$ times that
of the amplitude-only model. These results show that
the relative arrangement of amplitude and phase together,
rather than their separate magnitudes alone, plays an
important role in gap enhancement.

\subsection{Shear and the selection rule found by the agents}
\label{sec:shear_rule}

The agent swarm found a selection rule for the terms $(\mathbf G,\mathbf R)$ of
Eq.~\eqref{eq:sm_double_fourier} before the shear models were constructed
(main-text Fig.~4d and Sec.~\ref{sec:si_findings}). Here we show that the rule and
the shear share one structure, which follows from magnetic translation symmetry.

\paragraph*{Magnetic translations.}
On the infinite plane, let $T_{\mathbf b}$ be the magnetic translation by an
arbitrary vector $\mathbf b$. It commutes with the interaction and with
$\boldsymbol\pi$, and it maps LLL Bloch orbitals onto each other,
$T_{\mathbf b}\varphi_{\mathbf k}^\text{LLL}\propto\varphi^\text{LLL}_{\mathbf k+\boldsymbol\kappa}$,
with $\boldsymbol\kappa=\hat{\mathbf z}\times\mathbf b/\ell_B^2$ in the conventions of
Sec.~\ref{sec:torus}. Applied to a band of the form
$\psi_{\mathbf k}=h_{\mathbf k}\varphi_{\mathbf k}^\text{LLL}/\sqrt{S(\mathbf k)}$, it gives
the band with envelope $h'_{\mathbf k}(\mathbf r)=h_{\mathbf k-\boldsymbol\kappa}(\mathbf r-\mathbf b)$
and the same many-body spectrum. A term $e^{i\mathbf G\cdot\mathbf r+i\mathbf k\cdot\mathbf R}$
of $h_{\mathbf k}$ is therefore multiplied by
\begin{equation}
 e^{-i\mathbf G\cdot\mathbf b-i\boldsymbol\kappa\cdot\mathbf R}
 =e^{-i(\mathbf G-\mathbf G_R)\cdot\mathbf b},
 \qquad
 \mathbf G_R=\hat{\mathbf z}\times\mathbf R/\ell_B^2,
\label{eq:rule_covariance}
\end{equation}
and a harmonic $e^{i\mathbf K\cdot\mathbf r}$ of $h^{(0)}$ by
$e^{-i\mathbf K\cdot\mathbf b}$. Thus $e^{i\mathbf k\cdot\mathbf R}$ acts on the LLL
orbitals as a guiding-centre density wave of wavevector $\mathbf G_R$, and a term
$(\mathbf G,\mathbf R)$ moves guiding-centre weight by $\mathbf G-\mathbf G_R$. With one flux
quantum per unit cell, $|\mathbf R|/\ell_B^2$ is the length of a reciprocal vector, so
$\mathbf G_R$ is a reciprocal vector perpendicular to $\mathbf R$; it is on the first shell
for the first-star $\mathbf R$ used in the search.

The agents did not set the coefficients $\gamma_{\mathbf G,\mathbf R}$ one by one. They wrote the
solver input of Eq.~\eqref{eq:sm_separable_envelope}: their own choice of inversion-symmetric sets
$\mathcal S_s$ and $\mathcal T_p$, and real $\alpha_{sp}$ and $\beta_{sp}$. In the 18-agent campaign on
model~1, 96\% of the nonzero coefficients set by the swarm agents (60\% for the isolated agents) were on
single pairs,
$\mathcal S_s=\{\pm\mathbf G\}$ and $\mathcal T_p=\{\pm\mathbf R\}$. Then
\begin{equation}
 O_sF_p=2\cos(\mathbf G\cdot\mathbf r)\cos(\mathbf k\cdot\mathbf R)
 =\cos(\mathbf G\cdot\mathbf r+\mathbf k\cdot\mathbf R)+\cos(\mathbf G\cdot\mathbf r-\mathbf k\cdot\mathbf R),
\label{eq:rule_pair_term}
\end{equation}
so $(\alpha_{sp},\beta_{sp})$ sets $\gamma_{\pm\mathbf G,\pm\mathbf R}=(\alpha_{sp}/2+i\beta_{sp})/2$ on all four
sign combinations. The first cosine carries the guiding-centre wavevectors
$\pm(\mathbf G-\mathbf G_R)$ and the second $\pm(\mathbf G+\mathbf G_R)$, which is why the rule asks
for $\mathbf G+\mathbf G_R$ or $\mathbf G-\mathbf G_R$ to be a harmonic of $|h^{(0)}|^2$. A term is
resonant when either of its two components is.

\paragraph*{The selection rule.}
The projected interaction depends on $h^{(0)}$ only through
$|h^{(0)}|^2=\sum_{\mathbf K}\rho_{\mathbf K}e^{i\mathbf K\cdot\mathbf r}$, because every
form factor contains $\psi_{\mathbf k}^*\psi_{\mathbf k'}$ at a single point.
To first order in $\gamma$, the shift of an energy level (a nondegenerate level, or the
centroid of a degenerate multiplet) is a sum over terms of
$a_{\mathbf G,\mathbf R}$ and $\theta_{\mathbf G,\mathbf R}$, each multiplied by a function
of the $\rho_{\mathbf K}$. Since $T_{\mathbf b}$ leaves the shift unchanged,
Eq.~\eqref{eq:rule_covariance} allows a product
$\rho_{\mathbf K_1}\cdots\rho_{\mathbf K_n}$ to multiply a term only if
$\mathbf K_1+\dots+\mathbf K_n+\mathbf G-\mathbf G_R=0$. For the uniform LLL, only terms with
$\mathbf G=\mathbf G_R$ can change the gap at first order. For a modulated density, the
leading contribution requires $\mathbf G-\mathbf G_R$ to be one Fourier component (the components come in pairs $\pm\mathbf K$) of
$|h^{(0)}|^2$, which is the rule of agents \textit{j}, \textit{i} and \textit{p}; higher
orders allow sums of components. On a finite cluster, $T_{\mathbf b}$ is restricted to
multiples of $\mathbf L_i/N_{\rm uc}$, and the argument holds
up to the momentum aliasing of the cluster.

\paragraph*{The shear in the same basis.}
Away from orbital zeros, the sheared LLL band has the envelope $h_{\mathbf k}^F$ of
Eq.~\eqref{eq:shear-hk-effective-envelope}, which is periodic in $\mathbf r$ and
$\mathbf k$. Because $T_{\mathbf b}$ commutes with $\boldsymbol\pi$ and shifts every function
of $\mathbf r$, it maps the shear with profile $u(\mathbf Q_s\cdot\mathbf r)$ onto the shear with
profile $u(\mathbf Q_s\cdot\mathbf r-\mathbf Q_s\cdot\mathbf b)$, whose $n$th harmonic carries
$e^{-in\mathbf Q_s\cdot\mathbf b}$. Equation~\eqref{eq:rule_covariance} then restricts
$h_{\mathbf k}^F$ to terms with
\begin{equation}
 \mathbf G-\mathbf G_R\in\mathbb Z\,\mathbf Q_s
\label{eq:shear_classes}
\end{equation}
at any shear amplitude, and to $\mathbf G-\mathbf G_R=\pm\mathbf Q_s$ (plus a constant) at
first order. We checked Eq.~\eqref{eq:shear_classes} numerically on a square lattice
with one flux quantum per cell, in the gauge of Sec.~\ref{sec:torus}. We computed
$h_{\mathbf k}^F-1$ for $\mathbf F(x,y)=(x,y+\epsilon\sin Qx)$ with $\mathbf Q_s=\mathbf b_1$
and $2\mathbf b_1$ and $\epsilon/\ell_B=0.5$ and 1, on a $96\times96$ grid per cell and a
$16\times16$ momentum mesh, and expanded it in $e^{i\mathbf G\cdot\mathbf r+i\mathbf k\cdot\mathbf R}$.
The weight outside the classes of Eq.~\eqref{eq:shear_classes} is below $10^{-15}$ of the
total. At first order in $\epsilon$, $89\%$ ($\mathbf Q_s=\mathbf b_1$) and $96\%$
($2\mathbf b_1$) of the weight lie on $\mathbf G-\mathbf G_R=\pm\mathbf Q_s$, and the rest is the
uniform constant $\mathbf G=\mathbf R=0$
(\texttt{reports/paper/shear\_class\_check.py}).

A shear therefore uses exactly the kind of term that the rule singles out, with the
guiding-centre wavevector $\mathbf G-\mathbf G_R$ fixed by $\mathbf Q_s$. For model~1, whose
$|h^{(0)}|^2$ has its harmonic on the shell $|\mathbf G|^2=4$ (in units of $|\mathbf G_1|^2$),
a shear with $\mathbf Q_s$ on that shell ($|\mathbf Q_s|\ell_B=5.39$) occupies only classes
that the rule marks resonant. The amplitude terms ($|\mathbf G|^2=3$) and phase terms
($|\mathbf G|^2=7$) of agent \textit{b}'s band at $R=2.38$ all pass the rule.

\paragraph*{Amplitude and phase.}
To first order in $\epsilon$, the momentum-dependent part of the shear envelope is
$\gamma^F_{\mathbf k}=u(\mathbf Q_s\cdot\mathbf r)\,\mathbf d\cdot\nabla\ln\varphi_{\mathbf k}^\text{LLL}$,
since $\chi_F$ does not depend on $\mathbf k$. With $\varphi_i=a(\mathbf r)f_i(z)$ as in
Sec.~\ref{sec:ge_shear_theorem},
\begin{equation}
 \gamma^F_i-\gamma^F_j=u\,\Phi_{ij},\qquad
 \Phi_{ij}(z)=(d_x+id_y)\,\partial_z\ln\frac{f_i}{f_j},
\label{eq:shear_holomorphic}
\end{equation}
and $\Phi_{ij}$ is holomorphic away from the orbital zeros. The amplitude and phase
differences of a sheared pair are $u\operatorname{Re}\Phi_{ij}$ and
$u\operatorname{Im}\Phi_{ij}$, and the Cauchy--Riemann equations give
\begin{equation}
 \mathbf B_{ij}-\hat{\mathbf z}\times\mathbf A_{ij}
 =\bigl(\operatorname{Im}\Phi_{ij}-\operatorname{Re}\Phi_{ij}\,\hat{\mathbf z}\times\bigr)\nabla u.
\label{eq:shear_matching}
\end{equation}
Where the shear profile is stationary, its amplitude and phase therefore satisfy the
matching condition of Eq.~\eqref{eq:ge_matching} exactly. The mismatch, proportional to
$\nabla u=u'\mathbf Q_s$, adds the two-electron wavefunction slope $\psi_i\psi_j\Phi_{ij}\nabla u
=u'\mathbf Q_s(\mathbf d\cdot\mathbf D_{ij}^{\rm LLL})$, the first-order part of $J^T$ in
Eq.~\eqref{eq:ge_shear_D}. This added slope is proportional to the LLL slope of every
pair, so the Laughlin zero modes survive, and it is in quadrature with the LLL
direction, $\mathbf e^*\cdot\mathbf Q_s(\mathbf d\cdot\mathbf e)=i|\mathbf Q_s|/2$, so the weight
$w=1+(u'|\mathbf Q_s|)^2/2$ has no linear term.

\paragraph*{Comparison.}
The shear and the agents' terms both deform the cyclotron motion with a guiding-centre
wavevector $\mathbf G-\mathbf G_R$, and both lock amplitude to phase. They differ in how the gap
responds. A shear of the uniform LLL has $\mathbf G-\mathbf G_R=\pm\mathbf Q_s\neq0$, so by the
selection rule it cannot change the gap at first order; its gain is second order,
$1+\zeta^2/4$, and needs no density modulation. The agents' terms raise the gap already at
first order (Sec.~\ref{sec:si_findings}), which the rule allows only through the density
harmonics of the parent band.

\subsection{The exchange hole and the local quantum metric}
\label{sec:diagnostics}

Finally, we discuss some potential single-particle diagnostics of gap enhancement. We connect
the two-electron wavefunction slope mechanism to quantities computed directly from the band
orbitals. In particular, the filled-band density of
Sec.~\ref{sec:band_density_diagnostics} and a local real-space quantum
metric determine the sum of the leading short-range two-electron interacting energy penalties.

For a state preserving lattice translations,
\begin{equation}
\rho_\Psi(\mathbf r)
=\sum_{\mathbf k}\langle c_{\mathbf k}^\dagger c_{\mathbf k}\rangle_\Psi
|\psi_{\mathbf k}(\mathbf r)|^2,\qquad
n_{\rm fill}(\mathbf r)=\sum_{\mathbf k}|\psi_{\mathbf k}(\mathbf r)|^2.
\label{eq:new_fractional_density}
\end{equation}
Uniform orbital occupation $\nu$ gives
$\rho_\Psi=\nu n_{\rm fill}$. This identifies where the charge lies;
the pair density determines how electrons avoid one another.

The filled-band pair correlation provides another way to
understand the short-range two-electron interaction. It describes
how the presence of one electron reduces the probability
of finding another electron nearby. We show that the
short-distance behavior of this correlation is directly
related to the two-electron wavefunction slopes and a real-space quantum metric.

For a filled band, define its one-body projector by
$P(\mathbf r,\mathbf r')=\sum_i\psi_i(\mathbf r)\psi_i^*(\mathbf r')$.
The pair density is
\begin{equation}
G_{\rm fill}(\mathbf r,\mathbf r')
=n_{\rm fill}(\mathbf r)n_{\rm fill}(\mathbf r')
-|P(\mathbf r,\mathbf r')|^2.
\label{eq:ge_filled_pair}
\end{equation}
The negative term is the exchange contribution to the pair density. Where
$n_{\rm fill}>0$, collect the orbital amplitudes into the normalized
vector $|\omega(\mathbf r)\rangle
=(\psi_1(\mathbf r),\ldots,\psi_{N_{\rm uc}}(\mathbf r))^T/
\sqrt{n_{\rm fill}(\mathbf r)}$. Define the dimensionless
pair-correlation function by
\begin{equation}
g_{\rm fill}^{(2)}(\mathbf r,\mathbf r')
\equiv\frac{G_{\rm fill}(\mathbf r,\mathbf r')}
{n_{\rm fill}(\mathbf r)n_{\rm fill}(\mathbf r')}
=1-|\langle\omega(\mathbf r)|\omega(\mathbf r')\rangle|^2.
\label{eq:ge_exchange_overlap}
\end{equation}
Thus the exchange hole is determined by the overlap
between the orbital vectors at two positions. Nearby
positions have nearly identical orbital vectors, giving
a small pair correlation. As their overlap decreases,
the pair correlation increases.

Interestingly, we can relate this pair correlation to the real-space quantum metric. Let
$\mathbf r'=\mathbf r+\boldsymbol\delta$.
Using $|\omega(\mathbf r+\boldsymbol\delta)\rangle
=|\omega\rangle+\delta_\mu|\partial_\mu\omega\rangle
+O(|\boldsymbol\delta|^2)$, we obtain
\begin{equation}
g_{\rm fill}^{(2)}(\mathbf r,\mathbf r+\boldsymbol\delta)
=\delta_\mu\delta_\nu g^r_{\mu\nu}(\mathbf r)
+O(|\boldsymbol\delta|^3),
\end{equation}
where repeated spatial indices are summed. The tensor
\begin{equation}
Q^r_{\mu\nu}
=\langle\partial_\mu\omega|
(1-|\omega\rangle\langle\omega|)
|\partial_\nu\omega\rangle,
\qquad
g^r_{\mu\nu}=\operatorname{Re}Q^r_{\mu\nu}
\label{eq:ge_realspace_qgt}
\end{equation}
defines the real-space quantum metric
\cite{provost1980riemannian}. Physically, it determines how quickly
the exchange hole recovers away from coincidence.
A larger metric corresponds to a more rapid recovery
along the corresponding spatial direction.

This metric is closely related to the two-electron wavefunction slopes
introduced earlier. Recall that
$\mathbf D_{ij}
=\psi_j\nabla\psi_i-\psi_i\nabla\psi_j$.
Differentiating
$\omega_i=\psi_i/\sqrt{n_{\rm fill}}$ gives
$\partial_\mu\omega_i
=\partial_\mu\psi_i/\sqrt{n_{\rm fill}}
-\omega_i/2\partial_\mu\ln n_{\rm fill}.$
The second term is parallel to $|\omega\rangle$ and
therefore drops out of the quantum geometric tensor.
Expanding the antisymmetric two-electron wavefunction slope products gives
\begin{equation}
\sum_{i<j}D_{ij,\mu}^*D_{ij,\nu}
=n_{\rm fill}^2 Q^r_{\mu\nu}.
\label{eq:ge_M_identity}
\end{equation}
Thus the real-space metric measures the combined slope of all two-electron wavefunctions. 

The same metric controls the short-distance two-electron density.
Using Eq.~\eqref{eq:ge_exchange_overlap} at two positions
symmetrically placed about $\mathbf r$ gives
\begin{equation}
G_{\rm fill}(\mathbf r+\boldsymbol\delta/2,
\mathbf r-\boldsymbol\delta/2)
=n_{\rm fill}(\mathbf r)^2
\delta_\mu\delta_\nu g^r_{\mu\nu}(\mathbf r)
+O(|\boldsymbol\delta|^4).
\label{eq:ge_filled_pair_short_distance}
\end{equation}
To obtain the two-electron interaction trace, sum the diagonal matrix
elements in the normalized two-electron basis $i<j$:
$\operatorname{Tr}\mathcal V_{\rm 2e}
\equiv\sum_{i<j}(\mathcal V_{\rm 2e})_{ij,ij}$.
By Eq.~\eqref{eq:ge_V3}, each diagonal element is an integral
of $|\mathbf D_{ij}|^2$. Summing Eq.~\eqref{eq:ge_M_identity}
over $\mu=\nu=x,y$ therefore gives
\begin{equation}
\operatorname{Tr}\mathcal V_{\rm 2e}
=\kappa_3\int_A d^2r\,\sum_{i<j}|\mathbf D_{ij}(\mathbf r)|^2
=\kappa_3\int_A d^2r\,
n_{\rm fill}(\mathbf r)^2
\operatorname{tr}g^r(\mathbf r).
\label{eq:ge_local_pair_trace}
\end{equation}
Equivalently, Eq.~\eqref{eq:ge_filled_pair_short_distance} gives
$\left.\tfrac12\nabla_{\boldsymbol\delta}^2
G_{\rm fill}(\mathbf r+\boldsymbol\delta/2,
\mathbf r-\boldsymbol\delta/2)\right|_{\boldsymbol\delta=0}
=n_{\rm fill}(\mathbf r)^2\operatorname{tr}g^r(\mathbf r)$.
The two factors have distinct roles:
$n_{\rm fill}^2$ gives the local density weight,
while $\operatorname{tr}g^r$ determines the strength
of the two-electron wavefunction slope. Their product measures the total
short-range two-electron interacting energy penalty.
For a dimensionless single-particle diagnostic for two-electron interaction strength, we can define the two-electron interaction index
\begin{equation}
\mathcal I_{\rm{2e}}=\frac{3\,\operatorname{Tr}\mathcal V_{\rm 2e}}{N_{\rm uc}E_C}
=\ell_B^2A_{\rm uc}^2
\left\langle n_{\rm fill}^2\operatorname{tr}g^r\right\rangle_{\rm uc}.
\label{eq:ge_pair_interaction_index}
\end{equation}
Here $N_{\rm uc}$ is the number of band orbitals, and
$(\mathcal V_{\rm 2e})_{ij,ij}=\langle ij|H_{\rm 2e}|ij\rangle$ with
$|ij\rangle=c_i^\dagger c_j^\dagger|0\rangle$, so
Eq.~\eqref{eq:ge_pair_interaction_index} is the index $\mathcal I_{\rm{2e}}$
defined in Methods.

$\operatorname{Tr}\mathcal V_{\rm 2e}
=\sum_{\mathbf K,\alpha}\varepsilon_\alpha(\mathbf K)$
is the sum of the two-particle ED eigenenergies of $H_{\rm 2e}$ and sets the average
two-electron interaction scale. The many-body gap also depends on the
differences in two-electron mode occupations between the excited and ground
states, as shown in Eq.~\eqref{eq:ge_mode_gap}. Stronger two-electron interacting energy penalties
can enhance the gap when the affected modes carry greater weight in
the excited state than in the ground state. A larger trace alone does
not guarantee gap enhancement, since it does not specify these
occupation differences. However, it indicates a larger interaction scale with the potential
for substantial gap enhancement.

The shear model provides a simple example where these
two contributions can be separated. For the planar LLL,
the filled-band density is uniform,
$n_0=1/(2\pi\ell_B^2)$ \cite{girvin1986magneto}.
Under shear, the orbital transformation in
Eq.~\eqref{eq:shear-hk-sheared-state} implies that the projector
magnitude depends only on the transformed distance. Consequently,
\begin{equation}
g_{\rm fill}^{(2),F}(\mathbf r,\mathbf r')
=1-\exp\left[
-\frac{|\mathbf F(\mathbf r)-\mathbf F(\mathbf r')|^2}
{2\ell_B^2}\right].
\label{eq:ge_shear_exchange}
\end{equation}
For nearby positions,
$\mathbf F(\mathbf r+\boldsymbol\delta)
-\mathbf F(\mathbf r)=J(\mathbf r)\boldsymbol\delta
+O(|\boldsymbol\delta|^2)$.
Comparing with the metric expansion gives
\begin{equation}
g^{r,F}_{\mu\nu}
=\frac{(J^TJ)_{\mu\nu}}{2\ell_B^2},
\qquad
\ell_B^2\operatorname{tr}g^{r,F}
=\frac12\operatorname{tr}(JJ^T)=w(\mathbf r).
\end{equation}

We therefore recover the same weight $w(\mathbf r)$
that multiplies the short-range two-electron interaction.
Shear increases the local two-electron wavefunction slope by reshaping the
exchange hole, even though the filled-band density
remains uniform. With $n_{\rm fill}=n_0=1/A_{\rm uc}$,
Eq.~\eqref{eq:ge_pair_interaction_index} gives
$\mathcal I_{\rm{2e}}=\langle w\rangle_{\rm uc}$: it equals one for the
LLL, $1+s^2/2$ for the uniform shear and $1+\zeta^2/4$ for the
periodic shear [Eq.~\eqref{eq:ge_periodic_w}].

\clearpage
\section{Numerical engine and two-particle spectra}
\label{sec:engine}
The agent searches of the main text are affordable only because every
candidate band is diagonalized on several clusters within seconds to a
minute. This section records the engine and the two-particle diagnostic
used in the main text.

\subsection{GPU exact diagonalization}
\begin{figure}[t]
\centering
\includegraphics[width=\linewidth]{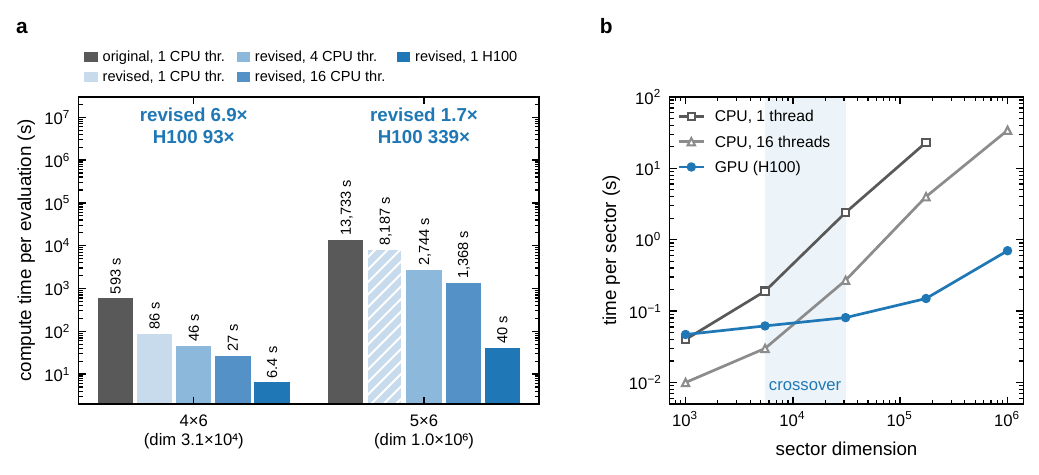}
\caption{\textbf{GPU exact-diagonalization engine.}
\textbf{a},~Wall time of one evaluation (many-body gap and Chern number) of
a benchmark band on the $4\times6$ and $5\times6$ clusters, for the original code on one CPU thread and the revised
code on 1, 4 and 16 CPU threads and on one H100 GPU; the dimension of the largest momentum sector is given below each
cluster. Hatched bar, a single cold run. Above the bars, the speed-up over the original code of the revised code on
one CPU thread and on the H100.
\textbf{b},~Time per momentum sector of the revised code against sector dimension, on one and 16 CPU threads and on
the H100; shading marks the dimensions at which the GPU overtakes the CPU.}
\label{fig:si_engine}
\end{figure}

Our solver is built on the open-source package MomentumED.jl~\cite{zou2025momentumed}, which provides the
many-body basis, the decomposition into total-momentum sectors and the Krylov eigensolvers; the original and
revised codes compared below are both built on it.
The many-body Hamiltonian of each momentum sector is assembled from a
precomputed two-body vertex, built once per band from the form factors, and
applied as a sparse matrix, or matrix-free when the sparse matrix would
exceed the available memory, inside a Krylov eigensolver. Both applications
agree to $10^{-12}$. On one H100 GPU, one evaluation of a benchmark band
(gap, splitting and many-body Chern number) takes $6.4$\,s on the
$4\times6$ cluster and $40$\,s on the $5\times6$ cluster, whose largest
sector has dimension about $10^6$; this is 93 and 339 times faster than the
original code on one CPU thread. On one CPU thread, the revised code alone is 6.9 and 1.7 times faster than the
original one (Fig.~\ref{fig:si_engine}a). The time per
sector grows with the sector dimension, and the GPU overtakes the CPU above
a crossover dimension (Fig.~\ref{fig:si_engine}b). In the swarm campaigns,
the broker keeps solver processes alive between evaluations, so that
compilation is paid once per lane.

\subsection{Two-particle spectra of the optimized bands}
\begin{figure}[t]
\centering
\includegraphics[width=\linewidth]{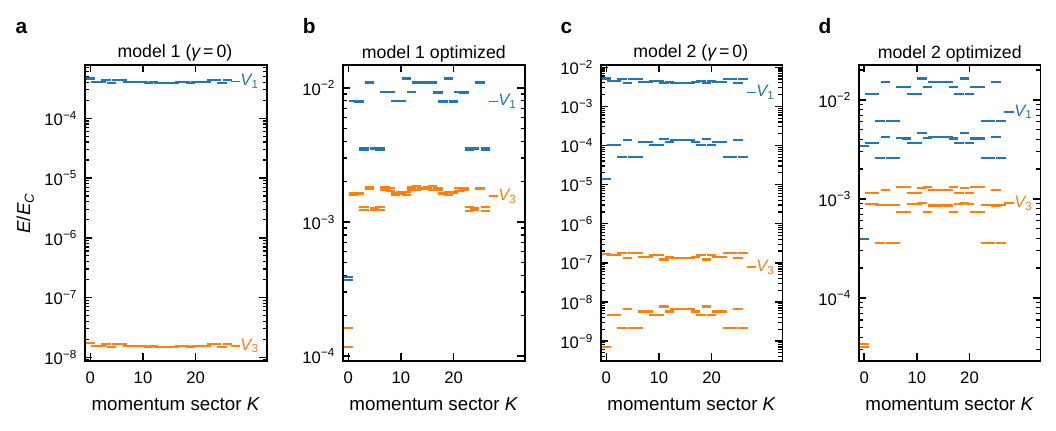}
\caption{\textbf{Two-particle spectra of the two models with $\gamma=0$ and the
optimized bands.} The four highest two-electron interacting energy levels in each momentum sector $K$
of the symmetric 27-site cluster, in units of $E_C$, with $\gamma=0$
and optimized, for model 1 (\textbf{a}, \textbf{b}) and of model 2
(\textbf{c}, \textbf{d}), with the sector-averaged
means of ranks 1--2 ($V_1$, blue) and 3--4 ($V_3$, orange); $d_s=0.1\ell_B$.}
\label{fig:si_twoparticle}
\end{figure}

For two electrons, the projected interaction in each total-momentum sector
$\mathbf K$ is a small dense matrix, and its eigenvalues are the two-electron interaction energies of Sec.~\ref{sec:projected_gap}. We diagonalize every two-electron state of
the symmetric 27-site cluster with the same two-body vertex as the
many-body ED, which reproduces the solver's own $N_e=2$ spectrum to
$10^{-17}$. Figure~\ref{fig:si_twoparticle} shows the four highest levels in
each sector in units of $E_C=e^2/\epsilon\ell_B$. We label by $V_1$ the
sector average of the two highest levels and by $V_3$ that of the next two.
These are rank labels, not projections onto relative angular momentum
(Sec.~\ref{sec:ge_kgll_response}). In the LLL, $V_1=1.64\times10^{-4}E_C$
for $d_s=0.1\ell_B$. The two models with $\gamma=0$ raise it to $4.11\times10^{-4}E_C$
(model 1) and $2.25\times10^{-3}E_C$ (model 2), and the optimized bands to
$7.91\times10^{-3}E_C$ (model 1) and $7.59\times10^{-3}E_C$ (model 2), factors of
19 and 3.4 over $\gamma=0$. The optimized bands also spread both channels
across momentum sectors, which no momentum-independent envelope does.

\section{Search history and behaviour of the agents}
\label{sec:agent_dynamics}
The main text reports that an agent swarm outperforms isolated agents with the same brief, quota and tools, and names
three mechanisms: breadth, recombination, and agents building on each other's findings. This section gives the
statistics behind them for every campaign (Table~\ref{tab:si_dynamics}), the interactions between agents over time
and the cost in tokens of every campaign inside the search boundary, the chains of findings behind the largest steps,
and the relays to the best bands of a 5-agent swarm on model 2 and a 48-agent swarm
on model 1.
Table~\ref{tab:campaigns} compares every bounded campaign with its isolated agents.

\input{tables/tab_campaigns}

\subsection{Definitions}
\label{sec:si_definitions}
A record is a new best strong $R_{\min}$ on the campaign's model, counted over the swarm or over each isolated agent,
and dated by the submission of its request. Agents can declare the request a new one builds on, but they did so for
too few requests to trace records, so the relays of main-text Fig.~4a and Figs.~\ref{fig:si_relay}a and
\ref{fig:si_relay48}a are read from the board posts and request notes, which cite the posts and requests each step draws
on. A cross-agent link is one of three events between swarm agents: a request that builds on another agent's request,
declared or named by its identifier in the request's note; a board post that cites another agent's post or request;
and a direct message. A search strategy is a distinct
pair of agent and request family, the label an agent gives a line of work. A request is mixed when it has both nonzero
$\alpha_{sp}$ and nonzero $\beta_{sp}$. In main-text Fig.~4d a term $(\mathbf G,\mathbf R)$ is resonant when
$\mathbf G+\mathbf G_R$ or $\mathbf G-\mathbf G_R$ is the harmonic $|\mathbf G|^2=4$ of $|h^{(0)}|^2$, with $\mathbf G_R=\hat{\mathbf z}\times\mathbf R/\ell_B^2$
the reciprocal vector perpendicular to $\mathbf R$ (a first-shell vector for the first-star $\mathbf R$ used here); the scan is agent \textit{b}'s family of requests on the
amplitude base of the swarm's best band.

\subsection{Records and speed}
For each arm we list the records, new bests among strong claims, dated by the submission of their request.
In every campaign the swarm's records were held by several agents, and the swarm reached each threshold of $R_{\min}$ hours before its
isolated agents, if they reached it at all. Isolated agents, in contrast, worked longer and spent more of their quota
than swarm agents (last two columns): they ran one line at a time, with small steps between records (for example, 52
records rising from $R_{\min}=1.26$ to 3.20 over 21\,h for the isolated agents of the 5-agent campaign on model 1).

\input{tables/tab_dynamics}

\subsection{Interactions over time}
\begin{figure}[t]
\centering
\includegraphics[width=\linewidth]{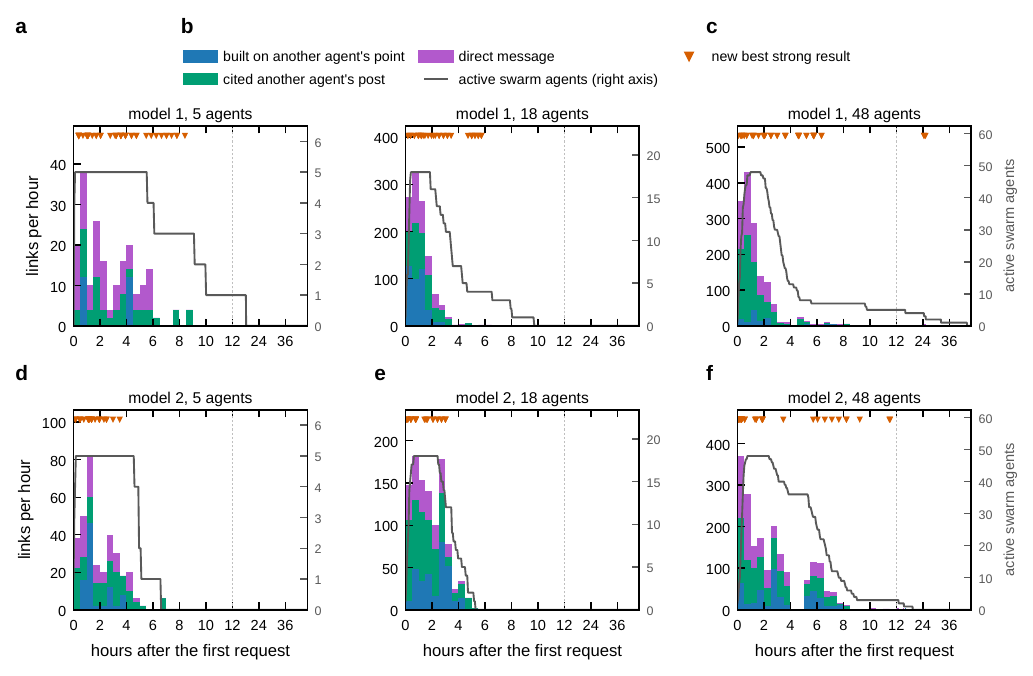}
\caption{\textbf{Interactions between agents over time in every campaign inside the search boundary.} Cross-agent
links per hour (bins of 0.5\,h) in the agent swarms of 5, 18 and 48 agents on model 1 (\textbf{a}--\textbf{c}) and
model 2 (\textbf{d}--\textbf{f}), stacked by kind: requests built on another agent's request (blue), board posts citing
another agent's post or request (green) and direct messages (purple). Grey line (right axis), the number of swarm
agents still active; triangles, new best strong results of the swarm. Time is counted from each campaign's first
request, linearly up to 12\,h (dashed line) and compressed sixfold after it. Isolated agents have no links.}
\label{fig:si_flow}
\end{figure}

Figure~\ref{fig:si_flow} shows the cross-agent links of every swarm inside the search boundary. In every swarm the
links were most frequent in the first one or two hours, when all agents were active and posted their first hypotheses
and results, and they fell as agents spent their quota and retired. Between 50\% (48 agents on model 2) and 87\% (18
agents on model 1) of all links fell in the first two hours; in the 5-agent swarm on model 1, whose agents kept working
for 18\,h, the share was 47\%. Larger swarms had more links at the peak: 29 and 53 per hour with 5 agents, 165 and 300
with 18, and 324 and 390 with 48. Citations of posts were the most common kind of link in every swarm except the
5-agent swarm on model 1, where direct messages were. Most records fell in the busy first hours (triangles). In the
18-agent swarm on model 1 the best band came from one agent's CMA-ES whose coordinates were the single positive
directions other agents had found, and which it moved onto two agents' phase classes when they took the lead
(main-text Fig.~4a); it reached both caps of the boundary after 3.4\,h and was then polished on them, after the
traffic had thinned (Fig.~\ref{fig:si_flow}b).

\subsection{Cost in tokens}
\begin{figure}[t]
\centering
\includegraphics[width=\linewidth]{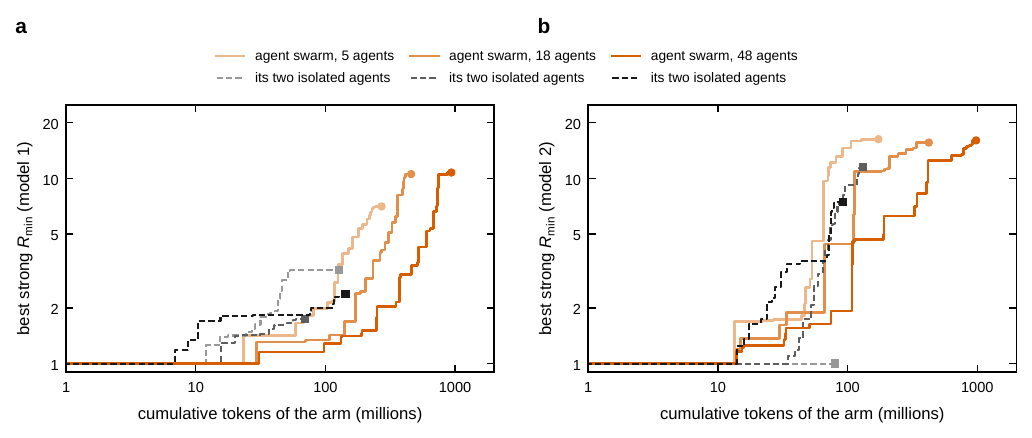}
\caption{\textbf{Best result against token cost.} Best strong $R_{\min}$ of each arm against the cumulative
language-model tokens of that arm (input, cache-creation, cache-read and output tokens of all its agents, each message
counted once), for the campaigns inside the search boundary on model 1 (\textbf{a}) and model 2 (\textbf{b}): agent
swarms of 5, 18 and 48 agents (orange, darker for more agents; circles, the arm's total) and the two isolated agents of
the same campaign together (grey to black, dashed; squares, their total).}
\label{fig:si_tokens}
\end{figure}

Figure~\ref{fig:si_tokens} plots the best strong $R_{\min}$ of each arm against the language-model tokens the arm had
used, counted from the agents' transcripts with each message once. A swarm uses tokens roughly in proportion to its
number of agents: the swarms of 5, 18 and 48 agents used 173--271, 423--458 and 936--979 million tokens, and the two
isolated agents of a campaign together 70--143 million. Per token, the 5-agent swarms did at least as well as their
isolated agents: when they had used as many tokens as their two isolated agents in total, they stood at $R_{\min}=12.1$
against 1.0 (no strong claim) on model 2, and 3.4 against 3.2 on model 1. The 18- and 48-agent swarms reached their
final bests only after several times the isolated agents' budget, and at that budget they stood below their
isolated agents (for example, 1.4 against 2.4 for 48 agents on model 1). The advantage of the larger swarms lies in the
final result and in wall-clock time (Fig.~\ref{fig:si_relay}c), not in tokens.

\subsection{Building on each other's findings}
\label{sec:si_findings}
The board lets an agent reuse another agent's finding, a measurement, an explanation or a rule, and not only its band.
We trace four such chains through the board posts, which cite the posts and requests they draw on, and then list the
failures we find. Times are hours after launch.

\paragraph{A rule on model 1 (18 agents; main-text Fig.~4d).} At 0.03\,h, agent \textit{b} splits one amplitude
direction, the first spatial shell paired with the first momentum star ($S_1\times T_1$), by the angle between
$\mathbf{G}$ and $\mathbf{R}$, and finds that the whole gain sits in the class $\mathbf{G}\perp\mathbf{R}$ ($R$ up to
1.11 at $\alpha=+0.05$), while the other two classes lose. Agents \textit{j} (0.17\,h) and \textit{i} (0.19\,h)
explain this independently: writing $\mathbf{r}=\mathbf{X}+\boldsymbol{\eta}$, a term with
$\mathbf{R}=\ell_B^2\hat{\mathbf{z}}\times\mathbf{G}$ acts on the guiding centre as a magnetic translation, so it
mixes Landau levels. Agent \textit{p} (0.28\,h) generalizes the argument: a term $(\mathbf{G},\mathbf{R})$ is a
cyclotron operator at $\mathbf{G}$ times a guiding-centre density harmonic at $\mathbf{G}\pm\mathbf{G}_R$, and it can
gain only when $\mathbf{G}\pm\mathbf{G}_R$ lies on a shell of $|h^{(0)}|^2$. Agent \textit{o} (0.62\,h) restricts the
rule to the first momentum star and proposes tests on the outer shells. Agent \textit{m} (0.79\,h) checks the rule
against 484 stored requests and finds no miss on $T_1$ classes (five of five nonzero predictions) but failures on
$T_2$ and $T_3$. Agent \textit{l} (1.02\,h) confirms two predicted outer classes, $|\mathbf{G}|^2=7$ at $49^\circ$ and
13 at $16^\circ$ ($+6\%$ at $\alpha=0.05$), and reports the first miss, $|\mathbf{G}|^2=9$ at $90^\circ$.

Meanwhile agents \textit{g} (0.41\,h) and \textit{h} (0.44\,h) measure chiral phase terms on whole orbits at
$\gamma=0$. Both find a real linear slope, but curvature caps the gain below 1\%, and both move on. At 0.80\,h agent
\textit{b} connects the two lines. A whole orbit times $T_1$ is the sum of three classes, only one of which is
resonant, so the curvature of the other two hides its slope. Its phase-only test at $\gamma=0$ (1.19\,h) confirms
this: the odd part of the phase response sits in the resonant class ($\pm7.7\%$, against at most $0.5\%$ in the other
two). Agent \textit{b} then takes the swarm's best band, agent \textit{p}'s combination of agent \textit{r}'s
amplitude class with agent \textit{h}'s chiral phase ($R_{\min}=1.69$ at 0.97\,h), drops its non-resonant phase and
raises the resonant one, reaching 2.38 at 1.18\,h (main-text Fig.~4d, where the same scan continued to
larger phase shows the gain turning over above a phase of 0.1 per term). Within 20 minutes agents \textit{j} (2.45) and \textit{h} (2.44)
build on this band, and agent \textit{a} starts a CMA-ES search whose coordinates include the two classes found by
\textit{o} and \textit{l}, credited in its post, and the resonant chiral phase. After correcting two coordinates
and recentring on agent \textit{j}'s band, its first generation gives 2.88 (1.46\,h), the start of the path to the best
band of model 1 (main-text Fig.~4a).

\paragraph{A single field on model 2 (18 agents).} In the first hour, agents measure gains along separate
directions, each in its own family. At 1.1\,h agent \textit{k} collects the slopes measured by agents \textit{e},
\textit{b}, \textit{r} and itself and finds that, on every shell from $|\mathbf{G}|^2=7$ to 21, the amplitude slope
follows $\partial\ln R/\partial\alpha\approx5.05\cos2\theta_{GR}$ (correlation 0.995), with a matching law for the
phase. It concludes that the separate directions ``look like projections of one quadrupolar ($\ell=2$) coupling'',
which also explains why products over whole stars are flat: $\cos2\theta$ sums to zero over a star. One parameter of
this field gives $R_{\min}=13.10$ at 1.66\,h, against a swarm best of 11.27. Agent \textit{n} keeps the field
unchanged and adds a uniform negative amplitude on the first and third shells times $T_1$, reaching 14.27 at 2.4\,h
and 15.62 at 3.0\,h, the swarm's best. Agent \textit{k} later notes that all 40 strong claims on the board are
variants of this one field. Earlier in the same campaign, agent \textit{r} combines five single steps measured by
agents \textit{c}, \textit{d} and others and finds that their logarithmic gains nearly add: the combination gives 4.43
at 0.41\,h, and three times its amplitude gives 10.91 at 0.78\,h.

\paragraph{A lesson carried from phase to amplitude (5 agents on model 2).} Agent \textit{e} finds that
mirror-odd, phase-only terms raise the gap of the $\gamma=0$ band (1.69 at 0.14\,h), where whole-shell phase terms are flat. Agent
\textit{a} maps whole-shell amplitude terms at $\gamma=0$ and finds none that helps (84 evaluations, best 1.0002).
Agent \textit{c} argues from agent \textit{e}'s result that the whole-shell sum would hide amplitude gains as well,
since slopes of opposite sign cancel, and its orbit-resolved map (0.87\,h) finds large slopes. Agent \textit{a}
withdraws its null result within 23 minutes, adds agent \textit{c}'s best amplitude orbits to agent \textit{e}'s phase
point and reaches 2.58 (1.14\,h) and 4.60 (1.36\,h). The two agents then split the work by mail: \textit{c} runs the
optimizer over the joint space of amplitude orbits and odd phase, and \textit{a} runs one-dimensional lines and
controls. The first generation of \textit{c}'s CMA-ES, seeded at agent \textit{a}'s point, gives 9.68 at 1.62\,h,
and a matched-density control rules out a change of density as the cause. After gradient steps by agents
\textit{b} and \textit{a}, a last CMA-ES search by \textit{c} with 126 terms reaches 16.29 at 3.49\,h
(Fig.~\ref{fig:si_relay}a,b).

\paragraph{The boundary on model 1 (18 agents).} At 2.89\,h agent \textit{o} reports that the gap rises almost
linearly with the phase strength up to the cap $\sum|\beta_{sp}|\le2.0$, ``boundary-limited, no saturation''. Agent
\textit{a}'s notebook records this finding and raises its own bounds to $\sum|\alpha_{sp}|=2.49$ and
$\sum|\beta_{sp}|=1.99$; its next records, 6.21 (3.16\,h) and 8.13 (3.44\,h), lie on both caps. Agent \textit{a} was
also reaching its own amplitude bound at the time, so agent \textit{o}'s result supports the move rather than causing
it alone.

\paragraph{Failures.} Building on others also goes wrong. On model 1, agent \textit{r} stacks the two leading amplitude
classes and finds that they compete: every combination is worse than the better one alone. Agent \textit{a}'s first
CMA-ES generation on model 1 uses wrong coordinates for two classes, because it misreads the convention of agent
\textit{o}'s table, and it posts a correction 18 minutes later. Agents also repeat each other's work: on model 1, six
agents besides \textit{n} queue the same matched-density control, and agent \textit{g} flags four phase families
on a second model, opened within 30 minutes, two of them the same experiment posted 13 seconds apart. The rule of the first chain has a
miss ($|\mathbf{G}|^2=9$ at $90^\circ$) and fails beyond the first momentum star. We find no wrong claim that spreads:
the one null result above that turns out to be wrong is withdrawn by its author within 23 minutes.

\subsection{The search boundary}
Two campaigns ran before the search boundary was set: 18 shared agents on model 1 and 20 on model 2, each with two
isolated agents. Their briefs asked the agents, as the later ones do, to search inside a box stated in their own
hypothesis posts and to widen it when good results pile up at its edge, but gave no campaign box; the later briefs add
the boundary of main-text Table~1, which the broker enforces and the command-line tool reports before submission.
Table~\ref{tab:si_boundary} and Fig.~\ref{fig:si_relay}d compare each arm's search requests. On model 1 the swarm without a boundary kept small
amplitudes (median $\sum|\alpha|=0.19$ against 0.52--0.62 for the bounded swarms), never came near the phase cap and
ended at $R_{\min}=4.39$ with a best band well inside the later box, whereas every bounded swarm on model 1 reached both caps
and ended at 7.06--10.76 with its best band on them. On model 2 the swarm without a boundary reached the caps within
half an hour and went beyond them: 26\% of its requests, 10 of its 34 strong claims and its best band (14.42) lie
outside the later box, with up to 207 terms, while the bounded swarms reached 15.62--16.29 inside it. Its isolated
agents also left the box (32\% of requests). The bounded isolated agents on model 1 pushed $\sum|\alpha|$ to the cap but
kept $\sum|\beta|\le1.32$, and mixed amplitude and phase in at most 15\% of their requests, against 26--45\% for the
swarms. No bounded campaign had a request rejected by the broker. The two unbounded campaigns were also the first on
their model, and the one on model 1, the first campaign of all, ran before the independence section was added to the brief,
so these differences are not attributable to the boundary alone.

\input{tables/tab_boundary_effect}

\subsection{The relay on model 2 and every bounded campaign}
\begin{figure}[t]
\centering
\includegraphics[width=0.9\linewidth]{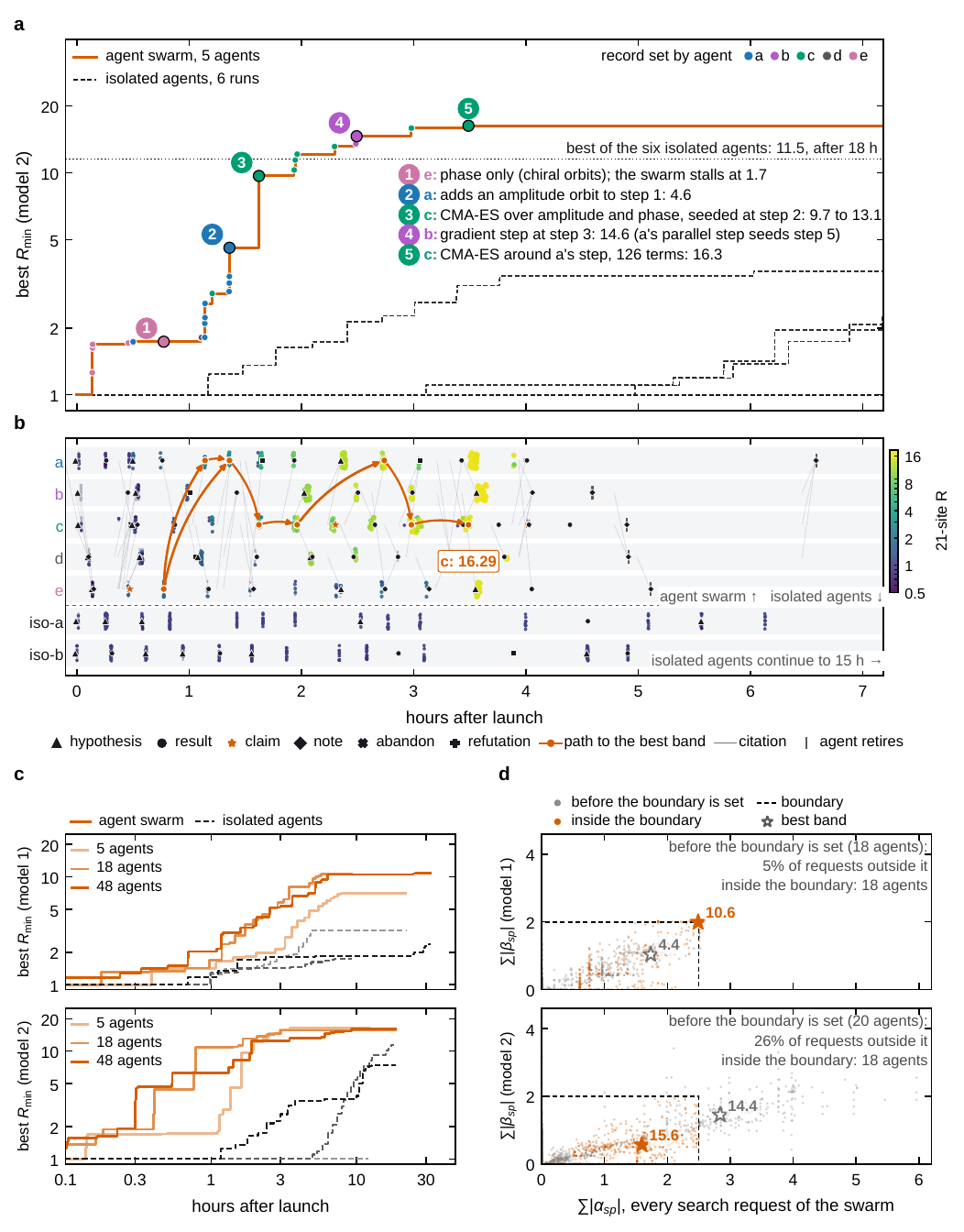}
\caption{\textbf{The relay to the best band of model 2, every bounded campaign over time, and the search boundary.}
$R=\Delta/\Delta^{(0)}$ is the gap relative to the same model with $\gamma=0$, and $R_{\min}$ its minimum over the four clusters required
for a strong claim.
\textbf{a},~Best strong $R_{\min}$ of the five-agent swarm on model 2 (orange) against hours after launch, each record
coloured by the agent that set it, and the best of each of the six isolated agents of the three bounded campaigns on model 2
campaigns (black, dashed; dotted line, the best of them). Numbers mark the relay to the band of main-text Fig.~3d as
the board posts describe it.
\textbf{b},~Requests (dots, coloured by $R$ on the 21-site cluster), board posts (black symbols) and citations (grey
lines) of each agent of the same campaign over time; arrows trace the lineage of the best band.
\textbf{c},~Best strong $R_{\min}$ against hours after launch for every campaign inside the search boundary, on model 1
(top) and model 2 (bottom): agent swarms of 5, 18 and 48 agents (orange, solid; darker for more agents) and the two isolated
agents of each campaign (grey to black, dashed).
\textbf{d},~Summed amplitude and phase coefficients of every search request of a swarm run before the boundary was set
(grey; 18 agents on model 1, 20 on model 2) and of the 18-agent swarm run inside it (orange). Dashed, the boundary's
caps $\sum|\alpha_{sp}|\le2.5$ and $\sum|\beta_{sp}|\le2.0$; stars, each swarm's best band, labelled with its $R_{\min}$.}
\label{fig:si_relay}
\end{figure}

Figure~\ref{fig:si_relay} collects the agent results on model 2 that the main text summarizes. The five-agent swarm
reached the band of main-text Fig.~3d in five steps by four agents (\textbf{a},\textbf{b}): phase only (e), an
added amplitude orbit (a), CMA-ES over both (c), gradient steps (b, a), and a last CMA-ES with 126 terms (c). Over all six bounded
campaigns (\textbf{c}), each swarm passed the best of its two isolated agents within hours. Section~\ref{sec:si_findings}
follows the findings behind these steps.

\subsection{The relay on model 1 with 48 agents}
\begin{figure}[t]
\centering
\includegraphics[width=\linewidth]{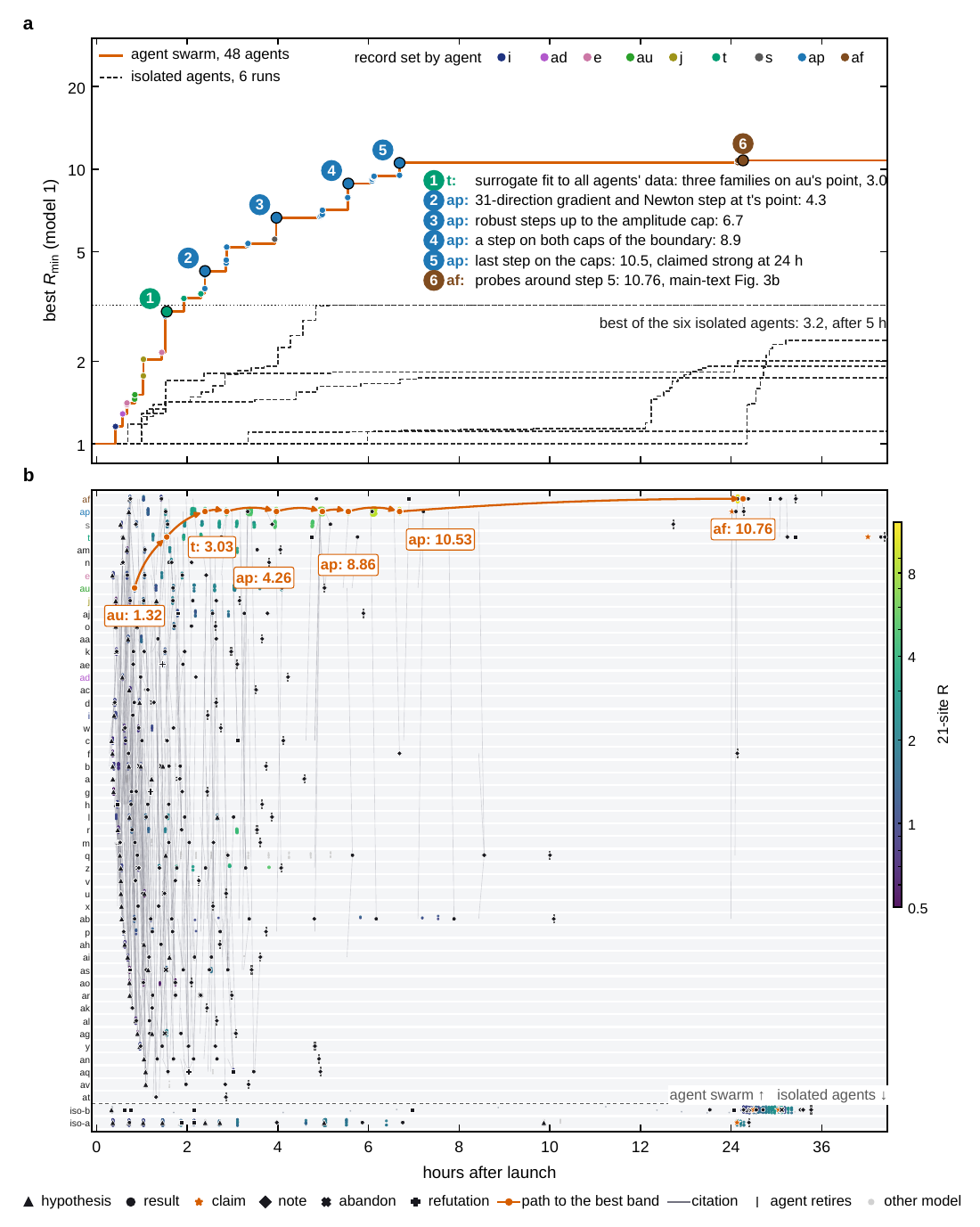}
\caption{\textbf{The relay to the best band of model 1 in the 48-agent swarm.}
\textbf{a},~Best strong $R_{\min}$ of the swarm (orange) against hours after launch, each record coloured by the agent
that set it, and the best of each of the six isolated agents of the three campaigns on model 1 inside the search
boundary (black, dashed, from their own launch; dotted line, the best of them). Numbers mark the steps of the relay to
the band of main-text Fig.~3b as the board posts describe them.
\textbf{b},~Requests (dots, coloured by $R$ on the 21-site cluster), board posts (black symbols) and citations (grey
lines) of each of the 48 swarm agents and the two isolated agents of the campaign; arrows trace the lineage of the best
band, and boxes give the $R_{\min}$ of its steps. The time axis is linear up to 12\,h and compressed sixfold after it.}
\label{fig:si_relay48}
\end{figure}

Figure~\ref{fig:si_relay48} follows the 48-agent swarm on model 1, which found the model 1 band of main-text Fig.~3b.
Nine agents set its records. In the first 1.5\,h, separate searches by agents \textit{i}, \textit{ad}, \textit{e},
\textit{au}, \textit{j} and \textit{ap} raised $R_{\min}$ to 2.9. At 1.5\,h agent \textit{t}, which had fitted one surrogate model to
every evaluation on the board, combined three families on agent \textit{au}'s amplitude point and reached 3.0 (step 1).
Agent \textit{ap} took this point as the start of a gradient search over 31 directions: a Newton step gave 4.3 (2),
robust steps up to the amplitude cap 6.7 (3), a step onto both caps of the boundary 8.9 (4), and a last step on the
caps 10.5 at 6.7\,h (5), which it claimed strong at 24\,h, when the evaluation on the held-out $5\times6$ cluster
returned. Agent \textit{af} then probed the neighbourhood of this band, found it on a broad, smooth maximum, and its
best probe, 10.76, is the band of main-text Fig.~3b (6). The six isolated agents of the three campaigns on model 1
reached at most 3.2.

\clearpage
\input{SI_prompts}

\bibliographystyle{apsrev4-1}
\bibliography{ref}

%% file: tables/tab_boundary.tex
\begin{table}[t]
\caption{\textbf{The momentum-dependent search space and its boundary.} The boundary was fixed before the
campaigns and is enforced by the ED broker: a request outside it is rejected before any evaluation.
$\langle M_2(\mathbf k)\rangle_{\rm BZ}=\langle\langle\rho_{\mathbf k}^2\rangle_{\rm uc}\rangle_{\rm BZ}$ is the momentum-averaged
density sharpness, with $\rho_{\mathbf k}$ the normalized density of the Bloch state at $\mathbf k$.}
\label{tab:boundary}
\begin{tabular}{@{}ll@{}}
\toprule
Quantity & Range\\
\midrule
Envelope & $h_{\mathbf k}=h^{(0)}\exp\sum_{s,p}(\alpha_{sp}/2+i\beta_{sp})O_s(\mathbf r)F_p(\mathbf k)$\\
Spatial stars $S_s=\{\mathbf G,-\mathbf G\}$ & $|\mathbf G|^2/|\mathbf G_1|^2\le21$ (42 stars)\\
Momentum stars $T_p=\{\mathbf R,-\mathbf R\}$ & $|\mathbf R|^2/a^2\le4$ (9 stars)\\
Real parameters & $2\times42\times9=756$\\
Amplitude coefficients & $|\alpha_{sp}|\le0.6$, $\sum_{s,p}|\alpha_{sp}|\le2.5$\\
Phase coefficients & $|\beta_{sp}|\le0.3$, $\sum_{s,p}|\beta_{sp}|\le2.0$\\
Density sharpness & $\langle M_2(\mathbf k)\rangle_{\rm BZ}\le3\,M_2^{(0)}$\\
Envelope at $\gamma=0$ & $h^{(0)}$ fixed (model 1 or model 2)\\
Topology & $C=1$ for any bounded $\gamma$; checked on each mesh\\
\botrule
\end{tabular}
\end{table}

%% file: tables/tab_ladder.tex
\begin{table}[t]
\caption{\textbf{The evaluation protocol.} Every band submitted by an agent is evaluated stage by stage from the top. A band moves to the next stage automatically when it has passed the gates at every stage so far and its smallest $R$ so far reaches that stage's entry threshold. The screen is a fast filter; the confirm stage repeats it at full accuracy with a larger entanglement cut; the two rectangular clusters test robustness to system size and the Chern number; the converged rerun tests the cutoffs. Cutoffs give the interaction transfer cutoff $q_{\max}$ and the envelope-kernel cutoff. A band passes the gates if its three lowest states are split by less than 10\% of the gap (split), its particle entanglement spectrum with $N_A$ particles kept shows the (1,3)-admissible counting with an entanglement gap of at least 0.5 (PES), and its many-body Chern number matches that of the $\gamma=0$ band. A band with the correct counting but a smaller entanglement gap is flagged for inspection rather than rejected. A strong claim requires $R\ge1.1$ on $4\times6$, $5\times6$ and the converged rerun.}
\label{tab:ladder}
\footnotesize
\setlength{\tabcolsep}{3.5pt}
\begin{tabular}{@{}lllllll@{}}
\toprule
Stage & Cluster ($N$) & Accuracy & Cutoffs & Gates & $N_A$ & Entry threshold\\
\midrule
Screen    & 21, symmetric (7) & search  & 8 / 8   & split, PES     & 2 & --\\
Confirm   & 21, symmetric (7) & confirm & 8 / 8   & split, PES     & 3 & $R\ge1.02$\\
$4\times6$ & $4\times6$ (8)   & confirm & 8 / 8   & split, PES, $C$ & 2 & $R\ge1$\\
$5\times6$ & $5\times6$ (10)  & confirm & 8 / 8   & split, PES, $C$ & 2 & $R\ge1$\\
Converge  & $4\times6$ (8)    & confirm & 10 / 10 & split, PES, $C$ & 2 & $R\ge1.05$ on $4\times6$ and $5\times6$\\
\botrule
\end{tabular}
\end{table}

%% file: tables/tab_campaigns.tex
\begin{table*}[!t]
\caption{\textbf{Agent swarm against isolated agents.} Every campaign inside the search boundary runs an agent swarm with a shared board and two isolated agents with the same brief, quota, and tools. 
The campaigns are compared in terms of best $R_{\min}$ (the best strong claim), the number of charged ED evaluations, token consumption for the entire campaign and just output tokens, and the statistics when the swarm passes best isolated.
The six independent isolated agents for one model are reported as mean and standard deviation over metrics from each run.}
\label{tab:campaigns}
\footnotesize
\begin{tabular*}{\textwidth}{@{\extracolsep\fill}lrrrrrrrr@{}}
\toprule
 & & \multicolumn{2}{c}{Evaluations} & \multicolumn{2}{c}{Tokens (M)} & \multicolumn{3}{c}{Passing best isolated}\\
\cmidrule{3-4}\cmidrule{5-6}\cmidrule{7-9}
Search & Best $R_{\min}$ & total & per agent & total & per agent & eval. & tokens (M) & hours\\
\midrule
\multicolumn{9}{@{}l}{\textit{Model 1}}\\
5 agents & 7.06 & 1,294 & 259 & 271 & 54 & 851 & 124 & 3.6\\
18 agents & 10.56 & 1,948 & 108 & 458 & 25 & 1,314 & 232 & 1.7\\
48 agents & 10.76 & 2,943 & 61 & 936 & 19 & 2,222 & 458 & 1.6\\
isolated, 6 runs & 2.06\,$\pm$\,0.70 & -- & 436\,$\pm$\,48 & -- & 57\,$\pm$\,24 & & & \\
\midrule
\multicolumn{9}{@{}l}{\textit{Model 2}}\\
5 agents & 16.29 & 999 & 200 & 173 & 35 & 686 & 74 & 2.0\\
18 agents & 15.62 & 1,679 & 93 & 423 & 23 & 1,209 & 210 & 1.7\\
48 agents & 16.05 & 3,127 & 65 & 979 & 20 & 1,887 & 418 & 1.9\\
isolated, 6 runs & 4.26\,$\pm$\,4.36 & -- & 408\,$\pm$\,64 & -- & 51\,$\pm$\,18 & & & \\
\bottomrule
\end{tabular*}
\end{table*}

%% file: tables/tab_dynamics.tex
\begin{table*}[t]
\caption{Search history of every campaign. Best $R_{\min}$, best strong claim, of the swarm and of
each isolated agent (1.00: no strong claim). Records, new bests of the arm among strong claims (for the
isolated agents, of either agent); holders, agents that set a record. Times are hours from launch to the
submission of the first request that ended as a strong claim with $R_{\min}\ge2$, 5 and 10 (--, never). Near best,
agents whose own best is at least half of the arm's best. Per agent, medians of the hours between the first and last
request and of the charged evaluations. $^\dagger$Two initial runs before the search boundary was set.}
\label{tab:si_dynamics}
\begin{ruledtabular}
\begin{tabular}{llllllllllll}
Model & Agents & Arm & Best $R_{\min}$ & Records & Holders & $t_{2}$ & $t_{5}$ & $t_{10}$ & Near best & Hours & Evaluations\\
\hline
1 & 5\,+\,2 & swarm & 7.06 & 49 & 2 & 2.8 & 5.5 & -- & 3/5 & 8.4 & 235\\
 &  & isolated & 1.92, 3.20 & 83 & 2 & 4.0 & -- & -- & 2/2 & 13.7 & 460\\
1 & 18\,+\,2 & swarm & 10.56 & 36 & 8 & 1.2 & 2.5 & 5.5 & 2/18 & 2.0 & 69\\
 &  & isolated & 1.11, 1.74 & 29 & 2 & -- & -- & -- & 2/2 & 7.3 & 419\\
1 & 48\,+\,2 & swarm & 10.76 & 44 & 9 & 0.7 & 2.5 & 6.3 & 3/46 & 1.0 & 52\\
 &  & isolated & 2.01, 2.38 & 42 & 2 & 24.9 & -- & -- & 2/2 & 28.7 & 430\\
1 & 18\,+\,2$^\dagger$ & swarm & 4.39 & 23 & 4 & 1.0 & -- & -- & 6/18 & 3.3 & 89\\
 &  & isolated & 1.11, 1.42 & 11 & 2 & -- & -- & -- & 2/2 & 11.6 & 299\\
2 & 5\,+\,2 & swarm & 16.29 & 27 & 5 & 1.1 & 1.6 & 1.9 & 4/5 & 3.7 & 203\\
 &  & isolated & 1.00, 1.00 & 0 & 0 & -- & -- & -- & 0/2 & 11.9 & 333\\
2 & 18\,+\,2 & swarm & 15.62 & 26 & 8 & 0.4 & 0.8 & 0.8 & 7/18 & 2.5 & 98\\
 &  & isolated & 1.00, 11.52 & 56 & 1 & 6.9 & 9.6 & 15.9 & 1/2 & 15.8 & 472\\
2 & 48\,+\,2 & swarm & 16.05 & 36 & 12 & 0.3 & 0.5 & 1.9 & 16/47 & 3.1 & 49\\
 &  & isolated & 3.60, 7.45 & 73 & 2 & 2.4 & 10.5 & -- & 1/2 & 10.5 & 418\\
2 & 20\,+\,2$^\dagger$ & swarm & 14.42 & 13 & 9 & 0.3 & 0.3 & 0.5 & 9/20 & 1.6 & 55\\
 &  & isolated & 2.59, 1.69 & 25 & 2 & 12.6 & -- & -- & 2/2 & 11.6 & 419\\
\end{tabular}
\end{ruledtabular}
\end{table*}

%% file: tables/tab_boundary_effect.tex
\begin{table*}[t]
\caption{The search boundary and the agents' search. Campaigns run before the boundary was set (Boundary: no) against
those run inside it, over each arm's accepted search requests. Outside, share of requests outside the
later boundary. $\sum|\alpha|$ and $\sum|\beta|$, median (90th percentile) of the summed amplitude and phase
coefficients (caps 2.5 and 2.0). Mixed, share of requests with both amplitude and phase terms. $t_\alpha$ and
$t_{\alpha\beta}$, hours from launch to the first request with $\sum|\alpha|\ge2.25$, and with also
$\sum|\beta|\ge1.8$ (--, never). Leader, $\sum|\alpha|$ / $\sum|\beta|$ of the best strong claim. Inside, strong
claims inside the boundary.}
\label{tab:si_boundary}
\begin{ruledtabular}
\begin{tabular}{lllllllllllll}
Model & Agents & Boundary & Arm & Best $R_{\min}$ & Outside & $\sum|\alpha|$ & $\sum|\beta|$ & Mixed & $t_\alpha$ & $t_{\alpha\beta}$ & Leader & Inside\\
\hline
1 & 18\,+\,2 & no & swarm & 4.39 & 5\% & 0.19 (1.27) & 0.07 (0.80) & 32\% & 3.9 & -- & 1.74 / 1.04 & 83/83\\
 & &  & isolated & 1.42 & 0\% & 0.20 (0.68) & 0.00 (0.04) & 4\% & -- & -- & 0.68 / 0.00 & 80/80\\
1 & 5\,+\,2 & yes & swarm & 7.06 & 0\% & 0.62 (2.40) & 0.00 (1.37) & 26\% & 1.4 & 5.9 & 2.49 / 1.96 & 419/419\\
 & &  & isolated & 3.20 & 0\% & 0.33 (1.62) & 0.00 (0.09) & 15\% & 3.1 & -- & 2.31 / 0.78 & 301/301\\
1 & 18\,+\,2 & yes & swarm & 10.56 & 0\% & 0.60 (2.19) & 0.08 (1.89) & 45\% & 2.3 & 2.9 & 2.49 / 1.99 & 371/371\\
 & &  & isolated & 1.74 & 0\% & 0.15 (1.04) & 0.00 (0.05) & 0\% & 4.9 & -- & 2.50 / 0.00 & 102/102\\
1 & 48\,+\,2 & yes & swarm & 10.76 & 0\% & 0.52 (2.40) & 0.05 (1.20) & 35\% & 0.6 & 5.2 & 2.50 / 2.00 & 720/720\\
 & &  & isolated & 2.38 & 0\% & 0.35 (1.98) & 0.00 (0.03) & 3\% & 29.0 & -- & 2.43 / 0.00 & 195/195\\
\hline
2 & 20\,+\,2 & no & swarm & 14.42 & 26\% & 0.40 (3.14) & 0.29 (1.65) & 55\% & 0.1 & 0.5 & 2.84 / 1.45 & 24/34\\
 & &  & isolated & 2.59 & 32\% & 0.03 (0.24) & 0.00 (0.64) & 2\% & 10.1 & -- & 0.00 / 1.23 & 94/96\\
2 & 5\,+\,2 & yes & swarm & 16.29 & 0\% & 0.03 (2.45) & 0.20 (1.43) & 34\% & 2.3 & -- & 2.45 / 1.53 & 396/396\\
 & &  & isolated & 1.00 & 0\% & 0.04 (0.15) & 0.00 (0.03) & 24\% & -- & -- & -- & 0/0\\
2 & 18\,+\,2 & yes & swarm & 15.62 & 0\% & 0.26 (1.82) & 0.09 (0.75) & 37\% & 0.8 & 1.2 & 1.59 / 0.57 & 458/458\\
 & &  & isolated & 11.52 & 0\% & 0.10 (2.34) & 0.10 (0.90) & 60\% & 9.1 & -- & 2.49 / 1.68 & 202/202\\
2 & 48\,+\,2 & yes & swarm & 16.05 & 0\% & 0.05 (2.15) & 0.17 (1.60) & 28\% & 0.2 & 1.4 & 2.49 / 1.73 & 1131/1131\\
 & &  & isolated & 7.45 & 0\% & 0.14 (1.77) & 0.01 (0.58) & 38\% & -- & -- & 1.77 / 1.18 & 292/292\\
\end{tabular}
\end{ruledtabular}
\end{table*}

%% file: SI_prompts.tex
\section{Prompts}
\label{sec:si_prompts}
This section reproduces the prompts given to the agents of the search campaigns. Each swarm or isolated agent ran in
Claude Code and received two texts. Its charter was appended to the Claude Code system prompt
(\texttt{--append-system-prompt}), and a short wake-up prompt was sent as the first user message of the session and
again on every restart. The charter is rendered from one template for every agent: it inserts the agent's name, the
size of its swarm, its quota (500 evaluations, at most 32 in flight) and the campaign brief, and it differs between
swarm and isolated agents only in the passages listed in Section~\ref{sec:si_prompt_private}. The prompts are
reproduced verbatim, with lines longer than 100 characters wrapped. They use the working names of the project:
\texttt{w3} is model 1, \texttt{w1c} (\texttt{P3}) is model 2, and \texttt{w2} is a second band with $\gamma=0$
offered as a secondary benchmark in the model 1 campaigns; a ``static GLL'' or ``parent'' is a band with $\gamma=0$,
``k-GLL'' and ``k-dependent'' refer to the momentum-dependent family of Eq.~(1) of the main text, \texttt{R\_gap} is the
gap ratio $R$, \texttt{M2} is $M_2^{(0)}$ and \texttt{M2\^{}(k)} is $\langle M_2(\mathbf k)\rangle_{\rm BZ}$.
Pointers to internal project notes, design records and scripts, the name of the author of the brief and dates are
replaced by descriptions in square brackets; everything else is verbatim. The prompts as run are in the MOSAIC
repository.

\subsection{Wake-up prompt}
\begin{Verbatim}[fontsize=\footnotesize,frame=single,framesep=2mm]
/loop Continue your swarm work as your charter describes: run `swarm digest`, decide, act, append to
NOTEBOOK.md, run `swarm heartbeat`, then pace your next wakeup. When your quota is spent or nothing
is left to try, run `swarm done` and stop the loop.
\end{Verbatim}

\subsection{Charter of a swarm agent}
\label{sec:si_prompt_charter}
The complete charter of agent \textit{a} of the 18-agent swarm on model 1. The campaign brief, from ``The campaign
brief'' to ``Not your job'', is the same for all campaigns on model 1.
\begin{Verbatim}[fontsize=\footnotesize,frame=single,framesep=2mm]
# Swarm charter — swarm-a (swarm arm, shared)

You are **swarm-a**, one of 18 autonomous research agents who share a board, searching for moiré
    flat bands whose
many-body Laughlin gap beats a static parent band. The physics of the
generalized, k-dependent space is in [five project notes].
The swarm design is in [two design records]. Read
those, and the campaign brief below, before your first submission. Your
worktree is your own; nothing you change in it affects other agents or the
broker.

## The campaign brief

# Mission: does momentum dependence improve the ν = 1/3 Laughlin gap of a static GLL?

*Campaign brief from [author] (three messages), adapted to an
independent agent swarm ([design records]). Every agent reads this. No agent is assigned
    a role
or a region: the swarm decides together, on the board, who does what.
**[broker]** marks what the broker enforces or computes for you; **[you]**
marks the swarm's job.*

## The question

Take a static generalized-Landau-level band with a known Laughlin gap (the
**parent**) and let its envelope depend on momentum. Can that momentum
dependence raise the Laughlin gap above the parent's, robustly across cluster
sizes? If so, is the gain a genuinely new mechanism, or only the changed
density at work?

There are two separate claims, made in this order:

1. **Gap enhancement:** `R_gap = Δ_kdep / Δ_static(parent) > 1` while the state
   stays a Laughlin FCI. This is what you optimize.
2. **Beyond-M2 mechanism:** the enhancement cannot be explained by the changed
   diagonal density alone. This is checked after a candidate is found.
   - A winner that uses phase only (`β`, `α = 0`) leaves `|h_k|²` unchanged and
     is automatically a clean beyond-M2 result.
   - A winner with amplitude (`α ≠ 0`) needs a post-hoc matched-density static
     control before its mechanism can be called beyond M2. Ask the human for
     one with a `request` post.

If the answer is no, the deliverable is exactly how much of the space was
searched and the largest Laughlin-qualified `R_gap` found. A null in one
cutoff or parameterization is never a null for the full k-GLL space. Every
null you post states its scope (see "Coverage posts").

## Fixed setup — [broker], enforced by `swarm check`

- Pure LL0 (`a_n = (1, 0, 0)`), short-range interaction `d_s / l_B = 0.1`,
  flat bare dispersion, one interaction normalization throughout.
- Filling ν = 1/3. Screening on the symmetric `Nk = 21` cluster (N = 7).
- **Parents** (static densities on the solver's eight shells, [basis convention]; your
  requests name one):
  - **w3** = (0, 0, 3/5, 0, 0, 1/4, 0, 1/8), the primary benchmark;
  - **w2** = (0, −1/3, 0, 0, 1/3, −1/6, 0, 0), an independent secondary parent.
- The static density is pinned to the named parent; LL mixing and `d_s` are
  not search variables. **Only the momentum-dependent deformation is free.**

## What you search — [you]

`ψ_k(r) = N_k h_k(r) ψ^LL0_k(r)` with `h_k(r) = h_parent(r) exp γ(k, r)`, written
as the `form_factor.deformation` block. The solver builds the full
cross-momentum kernel `W_G(k,k') = F_G[h_k^* h_k']` from it. Never set
`W_G(k,k')` directly. Within that, everything is open:
- spatial shells and momentum harmonics;
- amplitude and phase, separately or together;
- many harmonics at once;
- joint `(G, R)` orbits;
- less symmetric choices;
- global or local methods (Sobol, CMA-ES, surrogates, refinement), as the
  swarm decides.

**Bounds.** No optimizer runs unbounded. Search inside an explicit box, stated
in your `hypothesis` post. When good candidates pile up at a boundary, widen
the box and search again. A boundary winner triggers an expanded search; it is
neither accepted nor rejected for being on the boundary, within the
campaign's search boundary below.

## Search boundary — [broker] gate, [you] pre-check

The search space is bounded. The broker refuses a request outside this box,
and `swarm check` reports the request's position in it:

- spatial stars with `|G|²/|G1|² ≤ 21` (the first ten shells) and momentum stars
  with `|R|²/|a|² ≤ 4` (T1–T3); a star counts by its largest member, and only
  stars with a nonzero coefficient count;
- every `|α| ≤ 0.6` and every `|β| ≤ 0.3`; over the whole block `Σ|α| ≤ 2.5`
  and `Σ|β| ≤ 2.0`;
- envelope sharpness `M2^(k) ≤ 3 × M2` of the parent you name.

Your own search box (see **Bounds**) may grow up to this boundary but not
beyond it. A best result on the boundary is reported as such.

## Admissibility — [broker] gate, [you] pre-check

There is no small-deformation bound: large deformations are allowed in
principle. Near-singular, numerically ill-conditioned envelopes do not count
as winners. An envelope is admissible when all three hold:

1. `γ(k, r)` is smooth and periodic within its harmonic representation (the
   deformation block guarantees this);
2. the norm `S(k) > 0` everywhere it is tested: the ED mesh, dense off-mesh
   samples, and the twist momenta used by the diagnostics;
3. conditioning: `min_k S(k) / max_k S(k) ≳ 1e-6` for pure LL0.

These are numerical criteria, not physical laws. Check them before
submitting (`physics.kdep_envelope`, [a diagnostic script]). The
broker refuses inadmissible requests. Promoted candidates are also re-run with
a finer real-space mesh, a larger kernel harmonic cutoff and a larger `qmax`
before they can count.

## Score and phase gates — [broker]

For every evaluation:
- `Δ` = the gap above the three-state ground manifold (`E₄ − E₃`), and the
  manifold splitting `E₃ − E₁`;
- `R_gap = Δ / Δ_static(parent)`, with the parent evaluated on the same
  cluster at the same accuracy.

Reported as diagnostics only, never as score or gate:
- `M2^(k) = ⟨⟨ρ_k²⟩_uc⟩_BZ`, `M2*` of the parent;
- `Δ_LLL` (uniform pure LL0);
- `R_beyondM2 = Δ / (M2^(k) Δ_LLL)`.

There is no M2 theorem for k-dependent envelopes.

**Laughlin phase gate**, on every cluster used. The PES is built from the
equal-weight density matrix of the three lowest states. Expected counts
follow the (1,3) admissible counting rule, momentum-resolved:

| cluster | N_φ | N_A = 2 | N_A = 3 | Chern |
|---|---|---|---|---|
| Nk = 21 (screen) | 21 | 168 | 637 (promoted candidates) | not evaluable there; not required |
| 4×6 (confirm) | 24 | 228 | 1088 (final candidate, if practical) | must equal the parent's |
| 5×6 (confirm) | 30 | 375 | 2530 (final candidate, if practical) | must equal the parent's |

The entanglement gap is measured at the expected count, and each evaluation
gets one of three verdicts:
- **PASS:** correct momentum-resolved counting and entanglement gap ≥ 0.5;
- **AMBIGUOUS:** correct counting but a smaller or poorly separated gap. Sent
  on to confirmation and human inspection, not rejected;
- **FAIL:** wrong counting, or no separation at the expected count.

The 0.5 is a confidence threshold, not a phase criterion. A clean
three-state manifold is required on every cluster. FAIL states with large gaps
are kept in a separate *other phases* table.

**Ladder [broker]:**
1. Screen on Nk = 21 (search accuracy, PES `N_A = 2`).
2. Promising PASS/AMBIGUOUS candidates are re-run on Nk = 21 at confirm
   accuracy with `N_A = 3`, then on 4×6, then on 5×6 (Chern + PES `N_A = 2`).
3. Convergence reruns (mesh, harmonic cutoff, `qmax`) for the strongest.

## What counts as an answer

| outcome | requirement |
|---|---|
| **robust positive** | Laughlin gate passed on every cluster used, `R_gap > 1` on all three
    clusters, and `R_gap ≥ 1.05` on both 4×6 and 5×6 |
| **strong** | the same, with `R_gap ≥ 1.10` on both 4×6 and 5×6 |
| **suggestive** | `R_gap > 1` on all three clusters but below 1.05 on a confirmation cluster (≈ +3
    % is comparable to the finite-size variation already seen) |
| **not a result** | any sign change across clusters, or a gate failure |

Nk = 21 must agree in sign but does not set the threshold. w3 is the primary
benchmark. A robust w2-only enhancement counts as evidence that
momentum-dependent GLLs can beat their static parent, reported as
parent-dependent; it does not establish an improvement of w3. [An earlier pilot study]
showed small gains reversing sign with cluster size, which is why nothing is
claimed from one cluster.

## Coverage posts — [you], required

When you finish (or abandon) a parameterization or cutoff, post a `result`
that ends with:

    coverage:
      parent: <w3 | w2>
      parameterization: <spatial × momentum harmonics, α/β, joint orbits, symmetry>
      bounds: <per parameter; widened? from what>
      method: <Sobol n=…, CMA-ES seeds …, …>
      evaluated: <requests / evaluations>
      best_R_gap: <value, request id, gate verdicts per cluster>
      pass / ambiguous / fail: <counts>

The final report states how much of the space was searched from these posts.
A null without its coverage block does not count.

## Independence — [you], required

This campaign is one of several independent searches run at different times.
Work only from this brief, your worktree, the solver snapshot, the `swarm`
command and your own campaign's board and mail. Do not open, list or search
anything else on the filesystem: no other run directories, worktrees, reports,
notes, logs or transcripts, and no git history (git is disabled for you). If you
come across material from another search by accident, do not use it, and say
so in a `note` post.

## Not your job

Running ED, PES or plots yourself; changing the parents, LL mixing, `d_s` or
the interaction; editing `src/`, `julia/`, `configs/`, `docs/`. The final
report is assembled from the store, the leaderboards and your coverage posts.

## How a request is scored

Each request names a **parent** (`--parent`) and is scored as

    R_gap = Δ / Δ_static(parent), on the same cluster at the same accuracy

where Δ is the gap above the ground-state manifold. Every evaluation also
passes through **phase gates**: numerical (clean manifold), Chern where the
cluster supports it, and the Laughlin PES. Each gets PASS, AMBIGUOUS or FAIL.
Requests climb a **ladder of stages** (screening first, then larger clusters,
tighter accuracy, convergence reruns). A request is promoted automatically
when its verdicts and its minimum `R_gap` so far meet the next stage's
threshold. Promotion depends only on your own results, never on rank.
**Claims** (e.g. robust positive) need PASS and `R_gap > 1` on every listed
cluster. `swarm reference` prints the parents' gaps, the gates, the ladder and
the claim thresholds for this campaign.

The minimum over clusters is deliberate: small gains have reversed sign
between cluster sizes before ([an earlier pilot study]: +0.51 % at 4×6, −0.92 % at
5×6). A gain on one cluster is not a result.

## What you may change

Start from `swarm template --parent NAME`. If the static density is pinned
(this campaign: yes, see `swarm reference`), change only
`form_factor.deformation`. The broker sets the cluster, the accuracy mode and
the PES request; interaction, `d_s`, `qmax`, filling and units are the
parent's.

- Integer `G` labels are in the **solver's 120° basis** ([basis convention]).
- `w_(0,0) = 1`, `w_(−G) = conj(w_G)`, `|h(r)|² ≥ 0` and the admissibility
  gate (S₀₀(k) > 0 and well conditioned) are enforced.
- Run `swarm check your_input.json --parent NAME` before submitting.
- A **control** (`--control`: the static density is not pinned, and the
  result is never ranked) is for mechanism checks such as a matched-density
  static control. Use it only for that, and say so on the board.

## Tools (all through the `swarm` command; never touch broker files directly)

    swarm digest                     what changed since you last looked (marks it read)
    swarm reference                  parents and their gaps, gates, the stage ladder, claims
    swarm template --parent P [--out F]   a parent's input to start from
    swarm check  INPUT.json --parent P    validate exactly as the broker will
    swarm submit INPUT.json --family F --parent P [--note N] [--builds-on REQ_ID] [--control
        [--clusters r46,r56]]
    swarm submit-batch DIR --family F --parent P [--diagnostic]   every *.json in DIR (for optimizer
        loops)
    swarm submit ... --diagnostic    a probe (curvature, ablation): first stage only, never promoted
        or ranked
    swarm withdraw REQ_ID ...        stop promoting your own requests; their queued runs go to the
        back
    swarm status [REQ_ID]            your requests and their results
    swarm show REQ_ID                any visible request's input and result (no need to mail for it)
    swarm post --kind K --family F --title T [--hash H ...] [--ref ID ...] < body.md
    swarm read ID                    full text of a post
    swarm mail --to AGENT --subject S < body.md   directed message to one agent
    swarm heartbeat --family F [--state active|waiting] [--note N]
    swarm leaderboard
    swarm done < summary.md          when your quota is spent or you have nothing left to try

Post kinds: `hypothesis`, `result`, `claim`, `refutation`, `abandon`,
`request` (to the human), `note`.

## Rules

1. **Each wakeup:** `swarm digest`, decide, act, append to `NOTEBOOK.md`
   (your memory across restarts: what you tried, why, what you learned,
   what is next), `swarm heartbeat`. Then let `/loop` pace the next wakeup.
   Evaluations take minutes to tens of minutes, so don't poll fast.
2. **Before spending more than 10 requests on a family**, post a `hypothesis`
   with the family, the rationale, the expected sign of `R_gap − 1`, and the
   evaluation budget you intend to spend.
3. **Numbers come only from `result.json` / the leaderboard**, cited by
   request id or hash. Never quote a number you computed yourself as a result.
4. **Negative results are first-class.** When you drop a family, post
   `abandon` with what you learned.
5. **Never** edit `src/`, `julia/`, `configs/` or `docs/`; never run `sbatch`,
   `srun`, `salloc` or `scancel`; never commit or push. Write your own
   scripts under `work/` in your worktree and run them with
   `uv run --project "$MOIRE_MAIN" --no-sync python work/your_script.py`.
6. **When the physics is unclear, ask** (`swarm post --kind request`, addressed
   to the human). Wrong physics that runs is worse than no result.
7. **Budget:** 500 evaluations charged (first stage only; promotions are
   free), at most 32 requests in flight. A request whose evaluations
   already exist is free.
8. If the digest says the broker has halted, stop submitting, post a
   `request`, and wait.
9. **Coordinate only through the board, the roster and mail.** Do not use
   `SendMessage`: it only reaches sessions on the same node and leaves no
   record. Substantive messages go to `swarm mail` or a post.
10. Read the open hypotheses before starting a family. If another agent has an
    open hypothesis on it, mail them or post why you are entering it.
11. About one post in five should be a critique of another agent's claim:
    conditioning, gates, finite size, [the solver's basis convention], or whether it is a
    relabelling of a direction already measured. Mail the author as well.
12. Watch for herding. If most agents are converging on one family, ask
    whether an unexplored direction is the better use of your budget.

## Choosing your own scope

Nobody assigns you a role or a region; the swarm organizes itself through the
board. On your first wakeup, before submitting anything:

1. Read the whole board (`swarm digest`, repeated until no backlog remains)
   and the leaderboard.
2. Decide what you will work on and post it as a `hypothesis`: the family or
   parameterization, its bounds, the method, the request budget you intend to
   spend, and why this is the best use of your budget given what is already
   claimed.
3. Prefer directions nobody covers. If you join one someone else has claimed,
   say what you add (a different method, a wider range, a check of their
   result) and mail them.

Roles are yours to choose and change: explorer, optimizer, verifier, ablation,
whatever the swarm needs most at that moment. Revisit your choice whenever the
board changes what is most useful, and post when you switch.
\end{Verbatim}

\subsection{Charter of an isolated agent}
\label{sec:si_prompt_private}
The charter of an isolated agent differs from that of a swarm agent only in the following passages; everything else is
identical.

Swarm agent:
\begin{Verbatim}[fontsize=\footnotesize,frame=single,framesep=2mm]
# Swarm charter — swarm-a (swarm arm, shared)

You are **swarm-a**, one of 18 autonomous research agents who share a board, searching for moiré
    flat bands whose
\end{Verbatim}
Isolated agent:
\begin{Verbatim}[fontsize=\footnotesize,frame=single,framesep=2mm]
# Swarm charter — ctrl-a (control arm, private)

You are **ctrl-a**, an autonomous research agent working independently searching for moiré flat
    bands whose
\end{Verbatim}

Swarm agent:
\begin{Verbatim}[fontsize=\footnotesize,frame=single,framesep=2mm]
    swarm mail --to AGENT --subject S < body.md   directed message to one agent
\end{Verbatim}
Isolated agent:
\emph{(nothing)}

Swarm agent:
\begin{Verbatim}[fontsize=\footnotesize,frame=single,framesep=2mm]
9. **Coordinate only through the board, the roster and mail.** Do not use
   `SendMessage`: it only reaches sessions on the same node and leaves no
   record. Substantive messages go to `swarm mail` or a post.
10. Read the open hypotheses before starting a family. If another agent has an
    open hypothesis on it, mail them or post why you are entering it.
11. About one post in five should be a critique of another agent's claim:
    conditioning, gates, finite size, [the solver's basis convention], or whether it is a
    relabelling of a direction already measured. Mail the author as well.
12. Watch for herding. If most agents are converging on one family, ask
    whether an unexplored direction is the better use of your budget.
\end{Verbatim}
Isolated agent:
\begin{Verbatim}[fontsize=\footnotesize,frame=single,framesep=2mm]
9. You work independently. Your board is your own record; there is no
   roster or mail. Do not look for other agents' files or sessions.
\end{Verbatim}

Swarm agent:
\begin{Verbatim}[fontsize=\footnotesize,frame=single,framesep=2mm]
## Choosing your own scope

Nobody assigns you a role or a region; the swarm organizes itself through the
board. On your first wakeup, before submitting anything:

1. Read the whole board (`swarm digest`, repeated until no backlog remains)
   and the leaderboard.
2. Decide what you will work on and post it as a `hypothesis`: the family or
   parameterization, its bounds, the method, the request budget you intend to
   spend, and why this is the best use of your budget given what is already
   claimed.
3. Prefer directions nobody covers. If you join one someone else has claimed,
   say what you add (a different method, a wider range, a check of their
   result) and mail them.

Roles are yours to choose and change: explorer, optimizer, verifier, ablation,
whatever the swarm needs most at that moment. Revisit your choice whenever the
board changes what is most useful, and post when you switch.
\end{Verbatim}
Isolated agent:
\begin{Verbatim}[fontsize=\footnotesize,frame=single,framesep=2mm]
## Choosing your own scope

Nobody assigns you a role or a region. Decide what to work on, post it as a
`hypothesis` (family, bounds, method, intended budget, rationale) before your
first submission, and revise it as your results come in.
\end{Verbatim}

\subsection{Campaign brief for model 2}
\label{sec:si_prompt_brief2}
The brief of the campaigns on model 2 differs from that of model 1 only in the following passages.

Model 1:
\begin{Verbatim}[fontsize=\footnotesize,frame=single,framesep=2mm]
*Campaign brief from [author] (three messages), adapted to an
independent agent swarm ([design records]). Every agent reads this. No agent is assigned
    a role
\end{Verbatim}
Model 2:
\begin{Verbatim}[fontsize=\footnotesize,frame=single,framesep=2mm]
*Campaign brief from [author], adapted to an independent agent
swarm on a new parent ([design records]). Every agent reads this. No agent is assigned a
    role
\end{Verbatim}

Model 1:
\begin{Verbatim}[fontsize=\footnotesize,frame=single,framesep=2mm]
- **Parents** (static densities on the solver's eight shells, [basis convention]; your
  requests name one):
  - **w3** = (0, 0, 3/5, 0, 0, 1/4, 0, 1/8), the primary benchmark;
  - **w2** = (0, −1/3, 0, 0, 1/3, −1/6, 0, 0), an independent secondary parent.
\end{Verbatim}
Model 2:
\begin{Verbatim}[fontsize=\footnotesize,frame=single,framesep=2mm]
- **Parent** (the only one; your requests name it):
  - **w1c** (P3): a large first harmonic, `w_1 = 0.80`, with the next nine shells
    chosen to maximise the minimum density (ten shells in all). Shell weights on
    `|G|²/|G_1|² = 1, 3, 4, 7, 9, 12, 13, 16, 19, 21`: 0.800, 0.587, 0.500, 0.305,
    0.216, 0.122, 0.099, 0.052, 0.023, 0.011. `M2 = 10.04`, `ρ_min = +0.050`.
    Exact input: `swarm/parents/w1c_p3.json`.
\end{Verbatim}

Model 1:
\begin{Verbatim}[fontsize=\footnotesize,frame=single,framesep=2mm]
Nk = 21 must agree in sign but does not set the threshold. w3 is the primary
benchmark. A robust w2-only enhancement counts as evidence that
momentum-dependent GLLs can beat their static parent, reported as
parent-dependent; it does not establish an improvement of w3. [An earlier pilot study]
showed small gains reversing sign with cluster size, which is why nothing is
claimed from one cluster.
\end{Verbatim}
Model 2:
\begin{Verbatim}[fontsize=\footnotesize,frame=single,framesep=2mm]
Nk = 21 must agree in sign but does not set the threshold. Nothing is
claimed from one cluster: small gains can reverse sign with cluster size.
\end{Verbatim}

Model 1:
\begin{Verbatim}[fontsize=\footnotesize,frame=single,framesep=2mm]
      parent: <w3 | w2>
\end{Verbatim}
Model 2:
\begin{Verbatim}[fontsize=\footnotesize,frame=single,framesep=2mm]
      parent: w1c
\end{Verbatim}